\documentclass{aa}

\newcommand\Sctilde{\stackrel{\sim}{\smash{\mathrm{Sc}}\rule{0pt}{1.1ex}}}
\usepackage{graphicx}
	\graphicspath{{./img/},{./}}
\usepackage{txfonts}
\usepackage{hyperref}
\usepackage{subcaption}
\usepackage{xcolor}
\usepackage{textcomp,mathabx}
\usepackage{siunitx}
    \DeclareSIUnit{\astronomicalunit}{AU}
	\DeclareSIUnit{\parsec}{pc}
	\DeclareSIUnit{\earthmass}{M_\Earth}
	\DeclareSIUnit{\solarmass}{M_\Sun}
	\DeclareSIUnit{\jupitermass}{M_J}
	\DeclareSIUnit{\year}{yr}

\begin{document} 

    \bibliographystyle{aa}
    \title{Planetesimal formation via the streaming instability in simulations of infall dominated young disks}
    \titlerunning{Planetesimal formation in Class 0/I disks}

    \author{L.-A. H\"uhn\inst{1}
        \and
        C. P. Dullemond\inst{1}
        \and
        U. Lebreuilly\inst{2}
        \and
        R. S. Klessen\inst{1,3,7,8}
        \and
        A. Maury\inst{2,9,10}
        \and
        G. P. Rosotti\inst{4}
        \and
        P. Hennebelle\inst{2}
        \and
        E. Pacetti\inst{5}
        \and
        L. Testi\inst{6}
        \and
        S. Molinari\inst{5}
        }

    \institute{Institut für Theoretische Astrophysik, Zentrum für Astronomie der Universität Heidelberg, Albert-Ueberle-Str. 2, 69120 Heidelberg, Germany\\
            \email{huehn@uni-heidelberg.de}
        \and
               Université Paris-Saclay, Université Paris-Cité, CEA, CNRS, AIM, 91191 Gif-sur-Yvette, France
        \and
               Interdisziplinäres Zentrum für Wissenschaftliches Rechnen, Universität Heidelberg, Im Neuenheimer Feld 205, 69120 Heidelberg, Germany
        \and
               Dipartimento di Fisica, Universit\`{a} degli Studi di Milano, Via Celoria, 16, I-20133 Milano, Italy
        \and
               INAF -- Istituto di Astrofisica e Planetologia Spaziali (INAF-IAPS), Via Fosso del Cavaliere 100, 00133 Roma, Italy
        \and
               Dipartimento di Fisica e Astronomia “Augusto Righi”, ALMA Mater Studiorum - Universit\a di Bologna, via Gobetti 93/2, I-40190 Bologna, Italy
        \and
               Harvard-Smithsonian Center for Astrophysics, 60 Garden Street, Cambridge, MA 02138, U.S.A.
        \and 
               Elizabeth S. and Richard M. Cashin Fellow at the Radcliffe Institute for Advanced Studies at Harvard University, 10 Garden Street, Cambridge, MA 02138, U.S.A.
        \and
               Institute of Space Sciences (ICE), CSIC, Campus UAB, Carrer de Can Magrans s/n, E-08193 Barcelona, Spain
        \and
               ICREA, Pg. Lluís Companys 23, Barcelona, Spain
            }

    \date{\today}
 
    \abstract
    {Protoplanetary disks naturally emerge during protostellar core-collapse. In their early evolutionary stages, infalling material dominates their dynamical evolution. In the context of planet formation, this means that the conditions in young disks are different from the typically considered disks where infall has subsided. High inward velocities are caused by the advection of accreted material which is deficient in angular momentum, rather than being set by viscous spreading, and accretion gives rise to strong velocity fluctuations. Therefore, we aim to investigate when it is possible for the first planetesimals to form and subsequent planet formation to commence. We analyze the disks obtained in numerical 3D nonideal magnetohydrodynamical simulations, which serve as a basis for 1D models representing the conditions during the Class 0/I evolutionary stages. We integrate the 1D models with an adapted version of the \texttt{TwoPopPy} code to investigate the formation of the first planetesimals via the streaming instability. In disks with temperatures such that the snow line is located at ${\sim}\SI{10}{\astronomicalunit}$ and where it is assumed that velocity fluctuations felt by the dust are reduced by a factor of 10 compared to the gas, ${\sim}\SI{e-3}{\solarmass}$ of planetesimals may be formed already during the first $\SI{100}{\kilo\year}$ after disk formation, implying the possible early formation of giant planet cores. The cold-finger effect at the snow line is the dominant driver of planetesimal formation, which occurs in episodes and utilizes solids supplied directly from the envelope, leaving the disk solid reservoir intact. However, if the cold-finger effect is suppressed, early planetesimal formation is limited to cold disks with efficient dust settling whose dust-to-gas ratio is initially enriched to $\epsilon_0\geq 0.03$.}


    \maketitle
%
\section{Introduction}
Forming planets from dust grains present in protoplanetary disks surrounding young stars remains an uncertain process, where grains have to grow over many orders of magnitude and accumulate to eventually form planetary cores (for a review, see \citealt{drazkowsa2023}). It is currently believed that a complex string of processes needs to occur in order to achieve the formation of a planetary system akin to those observed. Protoplanetary disks need to maintain a sufficient solid mass for the formation of these planets and are required to meet specific conditions that are believed to be favorable for these processes. Fundamental aspects of the dynamics of protoplanetary disks remain subject to debate (for a review, see \citealt{lesur2023}), and progress on the understanding of one aspect of planet formation is usually made while assuming a reasonable understanding of the others. Such approaches are reasonable given the vast range of spatial and time scales involved. However, they have the downside of losing self-consistency regarding the initial conditions of protoplanetary disks, such as the total available solid mass or the elapsed time since the initial formation of the star and disk system from the parent molecular cloud.

Observations have revealed the presence of circumstellar disks all along the star formation sequence, from young, deeply embedded protostars (Class 0 systems, \citealt{maury2019,sheehan2022}) to pre main-sequence stars aged a few million years (Class II young stellar objects (YSOs), \citealt{andrews2018}). Observations reveal a large diversity in the properties of these disks, from being very compact and potentially massive around the youngest protostars (for a review, see \citealt{tsukamoto2023}) to being less gas rich and very structured when they are revealed around T Tauri stars (for a review, see \citealt{miotello2023}). Interestingly, recent observations of protoplanetary disks have challenged the existing model of planet formation. A large fraction of Class II disks may have difficulties forming planetary systems similar to those observed, as their solid budget available for planet formation may be too low \citep{manara2018} to form, for example, the giant planets present around 10\%-20\% of stars \citep{cumming2008,mayor2011}. Even under the optimistic assumption of a high degree of efficiency when converting available solids to giant planet cores, only ${\sim}16\%$ of observed Class II disks have a dust disk mass of ${\gtrsim}\SI{10}{\earthmass}$ \citep{vandermarel2021}, commonly cited to be the minimal core mass of giant planets in the core accretion scenario. Additionally, observations of dust masses in $1$ -- $\SI{2}{\mega\year}$-old disks around brown dwarfs reveal through direct measurements that they contain only small amounts of dust, with $M_\mathrm{dust}\lessapprox\SI{1}{\earthmass}$ \citep{testi2016,sanchis2020}. This is considerably lower than the total mass of the planetary systems found around Trappist-1 ($7$ -- $\SI{10}{\earthmass}$) and Proxima Centauri ($>1$ -- $\SI{2}{\earthmass}$), further suggesting that insufficient amounts of dust are available to form known planetary systems.

Furthermore, substructures are abundant phenomena in Class II disks, as large observational programs obtaining images with the Atacama Large Millimeter/submillimeter Array (ALMA) like "DSHARP" \citep{andrews2018} have shown. They are believed to be related to planet formation, either by serving as their birthplace (e.g., \citealt{testi2014}) or by being caused by the presence of a planet in the disk (see, e.g., \citealt{teague2018,bae2023}). In any case, the formation of these substructures would take time, but they are observed even in young disks like HL Tau \citep{hltau}, which is a Class I/II disk. In addition, the dust trapped in these substructures could be secondary dust produced from fragmenting collisions of planetesimals stirred-up by a planet \citep{turrini2019,testi2022}, which would require giant planets to be present in young disks already, in turn requiring an early onset of planet formation. Convincing indications of annular structures have been reported only toward a couple Class I disks \citep{seguracox2020, flores2023}, however it may be that substructures are common in young protostellar disks but obscured by the high optical depth of the dust emission. Altogether, these observations question the current planet formation paradigm whose starting point is a ${\sim}\SI{2}{\mega\year}$-old fully formed disk with negligible amounts of mass being accreted from the molecular cloud.

It is therefore imperative to put more emphasize on investigating the conditions reigning during the earlier evolutionary stages of protoplanetary disks, and the implications for planet formation. Indeed, during the protostellar Class 0 phase, the budget in mass to form planetary bodies is naturally more favorable, as a large amount of material, up to a few times the final stellar mass, is available in the dense envelope surrounding the protostellar embryo and its disk \citep{lada1987, andre2000}. Solid masses of young disks obtained from observations \citep{tychoniec2020, sheehan2022} are generally uncertain \citep{tung2024}, but observations of protostellar cores show signatures of vigorous mass transfer, up to ${\sim}\SI{e-5}{\solarmass\per\year}$ \citep{evans2015}, from the surrounding envelope down to disk scales. Sometimes observed to be episodic \citep{bjerkeli2023}, these infall/accretion events may allow replenishing the disk scales with both gas and dust efficiently, coming either from the inner envelope \citep{sai2022}, or from the mass reservoir at larger scales \citep{pineda2023,cacciapuoti2023}. Despite its short duration ($<\SI{0.1}{\mega\year}$, \citealt{maury2011,kristensen2018}), the Class 0 phase thus could be a key in evolving the solid particles and start assembling the building blocks of planetary cores during the early protoplanetary disk formation stages.

In the current most promising picture of planet formation, the first stage is the formation of kilometer sized objects, so-called planetesimals. An often employed idea is that the so-called streaming instability facilitates the formation of planetesimals through the direct gravitational collapse of accumulations of dust grains (\citealt{yg2005}, see also reviews by \citealt{johansen2014,lesur2023}). In previous works, planetesimal formation during the infall stage was investigated by considering the drift of large dust grains about \SI{10}{\micro\meter} in size in the envelope of young, embedded disks \citep{bate2017,lebreuilly2020}. While theoretical details of grain growth in the envelope and the achievable maximum grain size are highly debated \citep{ormel2009,guillet2020,silsbee2022,bate2022,lebreuilly2023}, observational evidence indicates the presence of large grains around young, embedded disks at a distance of $\SI{500}{\astronomicalunit}$ \citep{galametz2019, valdivia2019, cacciapuoti2023a}. \citet{cridland2022} find that, in some cases, the drift of large grains in the envelope can increase the dust-to-gas ratio of infalling material to the point where conditions for planetesimal formation are met without the need for any processes accumulating dust within the disk. While it remains unclear how efficiently dust can grow in the envelope, disks with a dust-to-gas ratio that is initially higher than 1\% could form in this scenario, exceeding what is found in the solar neighborhood. In this work, however, we put the focus on disk processes facilitating the onset of the streaming instability inside the disk.

Theoretical studies have been made in the past about the possibility of planetesimal formation in the disk build-up stage via disk processes. \citet{vorobyov2024} found using 3D hydrodynamics that dust density enhancements can already occur in the midplane in the first $\SI{20}{\kilo\year}$ after disk formation, neglecting the magnetic field. \citet{dd2018} have modeled the buildup and evolution of protoplanetary disks all the way to the end of Class II in a 1D viscous evolution scenario, employing an analytical prescription of the infalling material \citep{hg2005,shu1977,ulrich1976}. However, their setup was strongly simplified, neglecting the magnetic field. It would have a significant impact on the properties of a young disk, because the infall profile and disk size is expected to be different in a magnetized scenario. They also did not consider the impact infall has on the dynamics of early disks, which is substantially different from Class II disk dynamics. Improvements have been made by \citet{morbidelli2022} and \citet{mm2023} by employing more realistic infall and dynamical prescriptions based on results from \citet{lee2021}, while remaining focused on the Solar System and sticking to semi-analytical descriptions.

\begin{figure*}[htp]
    \centering\includegraphics[width=\linewidth]{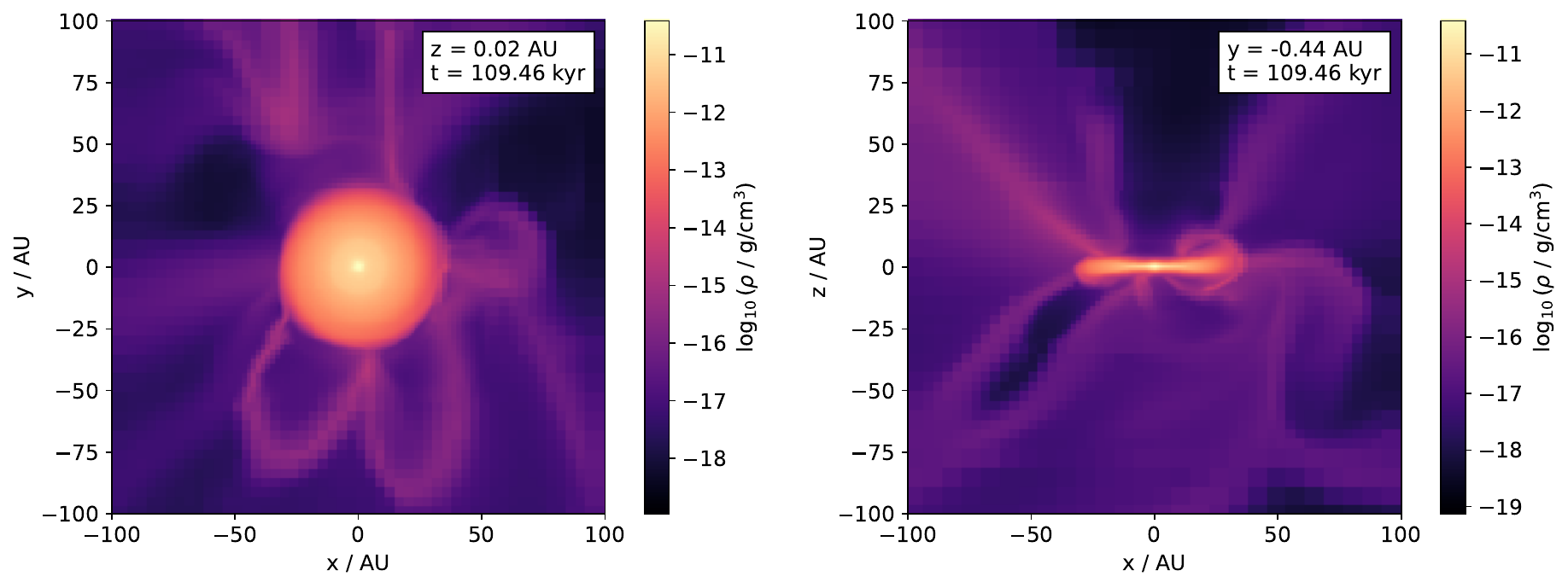}
    \caption{Snapshot of the protoplanetary disk from the R2 run of H20 at $t=\SI{109.46}{\kilo\year}$. \textit{Left}: Slice in the $x$-$y$ plane with $z=\SI{0.02}{\astronomicalunit}$. \textit{Right}: Slice in the $x$-$z$ plane with $y=\SI{-0.44}{\astronomicalunit}$. The color denotes the density on a logarithmic scale.}
    \label{fig:3D_snapshot}
\end{figure*}%
Given these simplifying assumptions about the conditions present in early disks made in the past, we investigate the possibility of an early onset of planet formation under more realistic conditions. We base our model on 3D nonideal magnetohydrodynamical (MHD) models of core-collapse from \citet[H20]{hennebelle2020a} where disks emerge naturally as the central star forms. An example of such a disk is shown in Fig. \ref{fig:3D_snapshot}, which shows one snapshot of the core-collapse simulations that will be used as the main reference in this work. We use these simulations to directly compute parameter profiles for the use in a 1D disk evolution framework, describing the infall and resulting time-evolution of the gas disk, which we combine with a two-population approximate dust evolution \citep{birnstiel2012} to model the accumulation of solids to the point where the streaming instability could potentially facilitate the formation of planetesimals.

In Section \ref{sec:methods}, we describe how we use the quantities from the H20 core-collapse simulations as a basis for a 1D disk evolution model. After that, we discuss key aspects and resulting model quantities of one simulation from H20 in Section \ref{sec:disk_quantities}, which serves as our reference simulation for the discussion of planetesimal formation. The key characteristics are presented in Section \ref{sec:ref_sim}, the disk parameters we obtain in Sections \ref{sec:basic_parameters}, \ref{sec:torque_parameters} and \ref{sec:dust_evolution}, and the resulting disk evolution in Section \ref{sec:1D_comparison}. Based on that, we discuss the possibility of planetesimal formation in Section \ref{sec:results}. Subsequently, we consider additional core-collapse simulations in Section \ref{sec:othermhd} and how the choice of planetesimal formation criterion influences our results in Section \ref{sec:pl_form_crit}. Finally, we discuss the limitations and implications of our work in Section \ref{sec:discussion} and conclude our paper in Section \ref{sec:conclusions}.

\section{Methods}\label{sec:methods}
Our model comprises two components. First, we obtain azimuthally and vertically averaged profiles for protoplanetary disk quantities from 3D non-ideal MHD simulations of isolated core-collapse performed by H20 which have a resolution of $\Delta x = \SI{1}{\astronomicalunit}$ at the highest level of refinement. These profiles are used as a basis for our modified version of the \texttt{TwoPopPy} 1D evolution code, which was used to model the dust and gas evolution in the two population approximation \citep{birnstiel2012}. We consider two species of solids, silicates and water ice, as well as two gaseous species, the background gas consisting of hydrogen and helium, and water vapor.

\subsection{Gas disk model}\label{sec:gas_disk_model}
\begin{table}[htp]
    \centering
    \caption{Parameters used for the calculation of radial resolution and vertical extent.}
    \begin{tabular}{lcccc}
        \hline
        \hline
        Name & $\Delta r_0$ & $H_0$ / \si{\astronomicalunit} & $q$\\
        \hline
        R1 & 0.3 & $\num{2.0}$ & $1.3$\\
        R2 & 0.15 & $\num{1.0}$ & $1.2$\\
        R7 & 0.2 & $\num{1.9}$ & $1.2$\\
        \hline
    \end{tabular}
    \label{tab:extraction_params}
\end{table}
The H20 simulations were performed on a Cartesian grid with adaptive mesh refinement using \texttt{RAMSES} \citep{teyssier2002}. We use the physical quantities found on that grid to infer disk quantities for use in a 1D evolution model context. To achieve this, we first calculate the relevant quantity on the Cartesian grid for every cell, and then perform an average over those cells that lie within cylindrical shells with radial and vertical widths $\Delta r$ and $\Delta z$, respectively. To account for different levels of refinement and the Cartesian nature of the grid, we choose the radial width to be
\begin{equation}
    \Delta r(r) = \max(\Delta r_\mathrm{min}, \Delta r_0r),
\end{equation}
where $\Delta r_\mathrm{min} = \SI{2}{\astronomicalunit}$. The values for $\Delta r_0$ are chosen individually for each simulation according to Tab. \ref{tab:extraction_params}. Furthermore, we choose the vertical extent $\Delta z$ motivated by the vertical extent of three pressure scale heights, resulting in
\begin{equation}
    \Delta z = 3H_0\left(\frac{r}{\SI{10}{\astronomicalunit}}\right)^q,
\end{equation}
where $H_0$ and $q$ are also chosen individually according to Tab. \ref{tab:extraction_params} in an effort to maintain a good balance between the physical extent of three pressure scale heights varying over time and numerical stability. When performing the average, cell quantities are weighted either by cell volume or mass. If a quantity is related to material vertically falling onto the disk, it is evaluated for cells with $z=\pm \Delta z(r)$ within one resolution element, and the average is performed only in vertical and azimuthal direction.

The surface density is calculated as
\begin{equation}
    \Sigma(r) = \frac{M_{\Delta r}(r)}{2\pi r\Delta r},
\end{equation}
with $M_{\Delta r}(r)$ being the mass contained in the cylindrical shell of thickness $\Delta r$ centered at $r$. The temperature is taken as a volume-weighted average,
\begin{equation}
    T_\mathrm{MHD}(r) = \frac{\sum_i T_idx_i^3}{\sum dx_i^3},\label{eq:T_mhd}
\end{equation}
with $dx_i$ the size of an individual cell and the sum being taken over all cells within the cylindrical shell centered around $r$ with thickness $\Delta r$.

In the core-collapse simulations performed by H20, the temperature is treated using a strongly simplified approximation. They follow a simple prescription as a function of the particle density $n$ mimicking radiative transfer and the thermal properties of the gas,
\begin{equation}
    T_\mathrm{pres} = T_0\left(1+\frac{(n/n_1)^{\gamma_1-1}}{1+(n/n_2)^{\gamma_1-\gamma_2}}\right),
\end{equation}
employing the parameters $T_0=\SI{10}{\kelvin}$, $n_1=\SI{e10}{\per\cubic\centi\meter}$, $n_2=\SI{3e11}{\per\cubic\centi\meter}$, $\gamma_1=5/3$, $\gamma_2=7/5$. This prescription does not contain any heating sources related to stellar irradiation or the accretion shock, so the disk temperature is likely underestimated. In fact, the accretion luminosity from the stellar surface in particular is believed to be a significant, albeit poorly constrained, heating source. Therefore, we artificially raise the temperature and perform a parameter study to investigate the consequences in later sections. We raise the temperature computed via Eq. \eqref{eq:T_mhd} by a multiplication with a constant factor $\xi$,
\begin{equation}
    T(r) = \xi\ T_\mathrm{MHD}(r),\label{eq:T_scale}
\end{equation}
while simultaneously reducing model parameters scaled by the temperature by the same factor as shown in subsequent paragraphs. By employing this choice, we ensure that the underlying dynamics remains unchanged by the temperature scaling. We note that, while a change in the disk temperature affects the disk scale height, which is in turn used to determine the extent of the vertical averaging, the scaling is applied only after the averaging has taken place. Applying the scaling first would not change the results, as the extent of the vertical region would increase together with the scale height, yielding the same result.

An external source term is calculated to account for the significant mass infall during the disk buildup stage,
\begin{equation}
    \dot\Sigma_\mathrm{ext}(r) = \left[-\rho(r,z)(v_z(r,z)-v_r(r,z)\tan(\theta(r)))\right]_{-\Delta z}^{\Delta z},
\end{equation} 
where $\theta(r)$ denotes angle of the disk surface described by the vertical extent, $\tan\theta(r)=\frac{d(\Delta z)}{dr}$.

To model the gas evolution of the disk, we take inspiration from the prescription of \citet{tabone2022},
\begin{equation}
    \frac{\partial \Sigma}{\partial t} - \frac{3}{r}\frac{\partial}{\partial r}\left( \frac{1}{r\Omega_K}\frac{\partial}{\partial r}\left(r^2\alpha_\mathrm{SS}\Sigma c_s^2\right)\right) - \frac{3}{2r}\frac{\partial}{\partial r}\left(\frac{\alpha_\mathrm{EA}\Sigma c_s^2}{\Omega_K}\right) = \dot \Sigma_\mathrm{ext},\label{eq:1D_adv_diff}
\end{equation}
where $\Omega_K$ is the Keplerian frequency, $c_s=\sqrt{k_BT/\mu}$ is the isothermal sound speed, $\mu$ is the mean molecular weight, $k_B$ is the Boltzmann constant and $\alpha_\mathrm{SS}$ is the viscosity parameter, $\alpha_\mathrm{SS}=\nu\Omega_K/c_s^2$ \citep{ss1973}. The model assumes that $v_\phi\approx\Omega_K r$, where $v_\phi$ is the azimuthal velocity of the disk, which holds true for the disks we consider in this work. The equation has been modified from its original form to account for material infall. Instead of an outflow caused by disk winds exerting an external torque characterized by $\alpha_\mathrm{DW}$, the accretion of envelope material causes an external torque. Combined with the magnetic component of the stress tensor \citep{bh1998}, the envelope accretion parameter $\alpha_\mathrm{EA}$ is given by
\begin{equation}
    \alpha_\mathrm{EA} = -\frac{4r}{3\Sigma c_s^2}\left[\dot\Sigma\delta v_\phi^\mathrm{infall}+\frac{B_zB_\phi}{4\pi}\right]^{\Delta z}_{-\Delta z}\xi^{-1},\label{eq:alpha_EA}
\end{equation}
with the magnetic field components $B_{r,z,\phi}$ and $\delta v_\phi^\mathrm{infall}=v_\phi^\mathrm{infall}-r\Omega_K$. The infalling material constitutes a positive contribution to $\alpha_\mathrm{EA}$ due to it being deficient in angular momentum, i.e., $\delta v_\phi^\mathrm{infall}<0$. The deficiency in angular momentum is a consequence of magnetic braking in the envelope. The infall contribution to $\alpha_\mathrm{EA}$ therefore has the same sign as in a scenario of disk-wind-driven evolution of Class II disks, where outflowing material has an excess of angular momentum \citep{bp1982}.

Furthermore, $\alpha_\mathrm{SS}$ can be calculated as
\begin{equation}
    \alpha_\mathrm{SS} = \frac{2}{3c_s^2}\left(\langle\delta v_r\delta v_\phi\rangle-\frac{\langle B_rB_\phi\rangle}{4\pi}\right)\xi^{-1}.\label{eq:alpha_SS}
\end{equation}
Here, $\langle\delta v_r\delta v_\phi\rangle$ is obtained as the mass-weighted average of the product of the individual cells' deviation from the mean flow $\langle v_{r,\phi}\rangle$,
\begin{equation}
    \delta v_{r,\phi;i} = v_{r,\phi;i} - \langle v_{r,\phi}\rangle,
\end{equation}
where the mean flow is in turn also calculated as the mass-weighted average,
\begin{equation}
    \langle v_{r,\phi}\rangle = \frac{\sum_i v_{r,\phi;i}\rho_i dx_i^3}{\sum_i \rho_i dx_i^3}.
\end{equation}
A mass-weighted average is also applied to compute the product of the vertical and azimuthal magnetic field components.

\subsection{Dust model}\label{sec:dust_model}
We use the two population approximation to model the dust evolution with \texttt{TwoPopPy} \citep{birnstiel2012}. For details about the model, we refer to the corresponding publication. In the following, our adaptations to that model are highlighted.

Two solid components, silicates and water ice, are considered and evolved separately, such that
\begin{equation}
    \frac{\partial \Sigma_i}{\partial t} + \frac{1}{r}\frac{\partial}{\partial r}\left[r\left(\Sigma_i\bar v_d-D_d\Sigma_g\frac{\partial}{\partial r}\left(\frac{\Sigma_i}{\Sigma_g}\right)\right)\right]=\dot\Sigma_\mathrm{ext,i},\label{eq:dust_evol}
\end{equation}
where the index $i$ denotes the respective solid species. Equation \eqref{eq:dust_evol} uses the mass-weighted velocity of the two population approximation (see \citealt{birnstiel2012}), the total gas surface density composed of hydrogen, helium and water vapor, $\Sigma_g = \Sigma_\mathrm{H+He}+\Sigma_\mathrm{vap}$ and the dust diffusivity,
\begin{equation}
    D_d = \frac{\delta_rc_s^2}{\Omega_K},\label{eq:dust_D}
\end{equation}
where we make an assumption for the Stokes number $\mathrm{St}=t_\mathrm{stop}\Omega_K$ of $\mathrm{St}\ll 1$, where $t_\mathrm{stop}$ is the particle stopping time. Furthermore, following the assumption of \citet{dd2018} that 50\% of the infalling solids are water ice and silicates, respectively, the external source term is given by
\begin{equation}
    \dot\Sigma_\mathrm{ext,i} = \frac{1}{2}\epsilon_0\Sigma_\mathrm{ext},\label{eq:infall_dtg}
\end{equation}
with the dust-to-gas ratio of the infalling material, $\epsilon_0$. In Eq. \eqref{eq:dust_D}, the parameter $\alpha_\mathrm{SS}$ used in the original model has been replaced by a direct characterization of the turbulent root-mean-square (RMS) velocities. As the diffusion term in Eq. \eqref{eq:dust_evol} describes mixing in the radial direction, we only consider velocities in the radial direction,
\begin{equation}
    \delta_r = \frac{\langle\delta v_r^2\rangle}{c_s^2}\xi^{-1}\chi^{-1},\label{eq:delta_r}
\end{equation}
to separate turbulence-induced angular momentum transport and passive tracer diffusion caused by velocity fluctuations. We introduce an additional scaling factor $\chi$ in Eq. \eqref{eq:delta_r}, which we use in the following sections to investigate the influence of the turbulent diffusion strength on planetesimal formation. Because water vapor acts as a tracer in the background gas, we also describe the water vapor surface density evolution according to Eq. \ref{eq:dust_evol}, employing $\Sigma_i = \Sigma_\mathrm{vap}$, setting $\dot\Sigma_\mathrm{ext,i}=0$ and replacing $\bar v_d$ with the gas radial velocity, $v_\mathrm{gas}$.

In accordance with the aforementioned separation between angular momentum transport and tracer diffusion, an analogous separation between diffusion in different spatial direction is made, by defining
\begin{equation}
    \delta_z = \frac{\langle\delta v_z^2\rangle}{c_s^2}\xi^{-1}\chi^{-1},\label{eq:delta_z}\\
\end{equation}
to describe mixing in the vertical direction, counteracting the settling of dust grains. Furthermore, we define a parameter for mixing that is not direction-dependent, 
\begin{equation}
    \delta_t = \frac{\langle\delta v_r^2\rangle+\langle\delta v_\phi^2\rangle+\langle\delta v_z^2\rangle}{3c_s^2}\xi^{-1}\chi^{-1}.\label{eq:delta_t}
\end{equation}
The relative turbulent velocity for similar-sized grains is then given by
\begin{equation}
    \Delta v_\mathrm{turb} = \sqrt{3\delta_t\mathrm{St}}c_s,\label{eq:oc_vrel}
\end{equation}
which results from adapting the expression for the Reynolds number used in the derivation of this quantity by \citet{oc2007} to be
\begin{equation}
    \mathrm{Re} = \frac{UL}{\nu_\mathrm{mol}} = \frac{v_\mathrm{RMS}^2}{\Omega_K\nu_\mathrm{mol}} = \frac{\delta_tc_s^2}{\Omega_K\nu_\mathrm{mol}},
\end{equation}
where $U$ and $L$ are the length scale and velocity of the largest eddies, respectively, $\nu_\mathrm{mol}$ is the molecular kinematic viscosity, and assuming $\frac{U}{L} = \Omega_K$. The dust scale height is in turn adapted from \citet{dubrulle1995},
\begin{equation}
    H_d = H_g\sqrt{\frac{\delta_z}{\delta_z+\mathrm{St}}},\label{eq:dust_scaleheight}
\end{equation}
with the gas pressure scale height $H_g=c_s/\Omega_K$. Equation \eqref{eq:dust_scaleheight} can be interpreted as a prescription for dust grain settling, where it is assumed that the settling efficiency instantaneously adapts to the strength of vertical mixing and the particle Stokes number. We note that introducing the temperature scaling factor $\xi$ to $\delta_z$ (see Eq. \eqref{eq:delta_z}) leads to a small artificial decrease in the dust scale height, because the Stokes number of a grain of a given size does not scale with the temperature, so that for $\mathrm{St}\gg\delta_z$, $H_g/H_d\ \propto\ \sqrt{\xi}$ (cfr. Eq. \ref{eq:dust_scaleheight}). However, the dominant parameter affecting settling in our model is $\chi$, as $H_g/H_d\ \propto\ \chi$ for $\mathrm{St}\gg\delta_z$ and if the grain size is limited by fragmentation, as is the case for our models.

By applying Eqs. \eqref{eq:oc_vrel} and \eqref{eq:dust_scaleheight} to the growth rate of monodisperse coagulation $\dot a = \rho_g/\rho_\bullet\Delta v_\mathrm{turb}$ \citep{kornet2001}, the growth timescale of the dust grains can be written as
\begin{equation}
    \tau_\mathrm{growth} = \frac{a}{\dot a} \approx \frac{1}{\Omega_K\epsilon}\sqrt{\frac{\delta_z\mathrm{St}}{\delta_t(\delta_z+\mathrm{St})}},\label{eq:tau_growth}
\end{equation}
with $\rho_\bullet$ the dust grain internal density and $a$ the dust grain size. Equation \eqref{eq:tau_growth} introduces an additional factor $\sqrt{\delta_z\mathrm{St}/(\delta_t(\delta_z+\mathrm{St}))}$ compared to \citet{birnstiel2012}. In particular, the growth timescale depends on the local dust grain size via the Stokes number in our model, which is approximated as the current maximal size of the large grain population in the two-population model for convenience. Our results are not affected greatly by this choice, as the dust grain size is limited by fragmentation for the majority of the simulation.

Furthermore, the limiting Stokes number due to fragmentation is now given by
\begin{equation}
    \mathrm{St}_\mathrm{frag} = \frac{1}{3}\frac{v_\mathrm{frag}^2}{\delta_t c_s^2}.\label{eq:st_frag}
\end{equation}
Here, the fragmentation velocity $v_\mathrm{frag}$ is set based on the water content in each radial grid cell, such that $v_\mathrm{frag}=\SI{10}{\meter\per\second}$ where $\Sigma_\mathrm{ice}/(\Sigma_\mathrm{ice}+\Sigma_\mathrm{sil})>0.01$, and $v_\mathrm{frag} = \SI{1}{\meter\per\second}$ otherwise \citep{gb2015,aw2014}, with a smooth transition for numerical stability. In total, the fragmentation velocity is therefore given by
\begin{equation}
    \frac{v_\mathrm{frag}}{\SI{1}{\meter\per\second}} = 1 + \frac{9}{1+\exp\left(-1000\left(\frac{\Sigma_\mathrm{ice}}{\Sigma_\mathrm{ice}+\Sigma_\mathrm{sil}}-0.01\right)\right)}.\label{eq:vfrag}
\end{equation}

The dust grain internal density and gas mean molecular weight are calculated dynamically following \citet{da2018},
\begin{align}
    \mu &= \left(\Sigma_\mathrm{H+He}+\Sigma_\mathrm{vap}\right)\left(\frac{\Sigma_\mathrm{H+He}}{\mu_\mathrm{H+He}}+\frac{\Sigma_\mathrm{vap}}{\mu_\mathrm{vap}}\right)^{-1},\\
    \rho_\bullet &= \left(\Sigma_\mathrm{sil}+\Sigma_\mathrm{ice}\right)\left(\frac{\Sigma_\mathrm{sil}}{\rho_{\bullet,\mathrm{sil}}}+\frac{\Sigma_\mathrm{ice}}{\rho_{\bullet,\mathrm{ice}}}\right)^{-1},
\end{align}
with $\mu_\mathrm{H+He}=2.34m_p$, $\mu_\mathrm{vap}=18m_p$, $\rho_{\bullet,\mathrm{sil}}=\SI{3}{\gram\per\cubic\centi\meter}$, $\rho_{\bullet,\mathrm{ice}}=\SI{1}{\gram\per\cubic\centi\meter}$ and $m_p$ the proton mass.

Envelope material entering the disk is assumed to exhibit the same dust-to-gas ratio as the initial dust-to-gas ratio of the disk, which is varied in the following sections. Additionally, when new dust enters a radial grid cell due to infall, the new pebble size $a_1$ used by the two population model is adjusted following \citet{morbidelli2022}. It is set to be the mass-weighted average of the previous size $a_\mathrm{1,prev}$ and the size of the newly injected material, given by the monomer size $a_0$,
\begin{equation}
    a_1 = \frac{a_0\dot\Sigma_\mathrm{dust}dt+(\Sigma_\mathrm{sil}+\Sigma_\mathrm{ice})a_\mathrm{1,prev}}{\dot\Sigma_\mathrm{dust}+\Sigma_\mathrm{sil}+\Sigma_\mathrm{ice}}.\label{eq:dust_infall_size}
\end{equation}

\subsection{Solid composition}
We employ the prescription by \citet{so2017} to model the time-dependent evaporation and condensation of water ice and vapor, respectively\footnote{For consistency with the literature, we refer to the chemical process of sublimation as "evaporation" and to deposition as "condensation".}. After every integration time step, we calculate the equilibrium vapor pressure according to the Clausius-Clapeyron equation,
\begin{equation}
    P_\mathrm{eq} = P_\mathrm{eq,0}\exp\left(-\frac{A}{T}\right),
\end{equation}
with $P_\mathrm{eq,0} = \SI{1.14e13}{\gram\per\centi\meter\per\square\second}$ and $A=\SI{6062}{\kelvin}$ \citep{lk1991}. The condensation and evaporation rates are then given by
\begin{align}
    R_c &= 8\sqrt{\frac{k_BT}{\mu_\mathrm{vap}}}\frac{3}{4\pi\rho_\bullet\bar a}\frac{\Omega_K}{c_s},\\
    R_e &= 8\sqrt{2\pi}\frac{3}{4\pi\rho_\bullet\bar a}\sqrt{\frac{\mu_\mathrm{vap}}{k_BT}}P_\mathrm{eq},
\end{align}
respectively. Here, we assume a typical particle size $\bar a$ given by the particle-number-weighted average,
\begin{equation}
    \bar a = \frac{N_0a_0+N_1a_1}{N_0+N_1} = \frac{(1-f_m)a_0a_1^3+f_ma_1a_0^3}{(1-f_m)a_1^3+f_ma_0^3},\label{eq:abar}
\end{equation}
with the number of particles belonging to the small and large population $N_0$ and $N_1$, respectively, as well as their corresponding particles sizes $a_0=\SI{1}{\micro\meter}$ and $a_1$. The particle size of the large population $a_1$ is a function of time and radial distance and given by the two population model. In later sections, we investigate the impact of our choice of $a_0$ and the corresponding mean particle size on our results. Consequently, the time derivatives of vapor and ice surface density read
\begin{align}
    \dot\Sigma_\mathrm{ice} &= R_c\Sigma_\mathrm{ice}\Sigma_\mathrm{vap}-R_e\Sigma_\mathrm{ice},\\
    \dot\Sigma_\mathrm{vap} &= -R_c\Sigma_\mathrm{ice}\Sigma_\mathrm{vap}+R_e\Sigma_\mathrm{ice}.
\end{align}
In our dust model, we rely on the assumption that the dust composition is homogenous at all times due to instantaneous vertical mixing and establishment of the coagulation-fragmentation equilibrium. In particular, this assumption means that the ice primarily condensing on the small grains of the distribution is instantaneously also distributed to the large grains of the distribution. This approach is reasonable given the high magnitude of the vertical mixing parameter $\delta_z$ we find in the core-collapse simulations (see Section \ref{sec:dust_evolution}).

\subsection{Planetesimal formation}\label{sec:methods_pltm_form}
When employing the model for planetesimal formation via the streaming instability by \citet{da2018}, we require that two criteria are met to consider the conditions favorable for the formation of planetesimals via streaming instability to occur. First, the Stokes number of the large population, $\mathrm{St}_1$, needs to be larger than a threshold value, $\mathrm{St}_1\geq\num{e-2}$ \citep{bs2010}. Second, the mass ratio of the large dust to the gas needs to be large enough, $\rho_\mathrm{dust,1}/\rho_\mathrm{gas}\geq \epsilon_{mid,crit}$. If a radial grid cell satisfies this criterion, planetesimals are formed from a fraction of the large dust $\Sigma_\mathrm{dust,1}=f_m\Sigma_\mathrm{dust}$,
\begin{equation}
    \dot\Sigma_\mathrm{pltm} = \zeta\Sigma_\mathrm{dust,1}\Omega_K,
\end{equation}
employing a planetesimal formation efficiency of $\zeta=\num{e-3}$ \citep{simon2016,drazkowsa2016}.

For a given vertically integrated dust-to-gas ratio $\Sigma_\mathrm{dust}/\Sigma_\mathrm{gas}$, the threshold value $\epsilon_\mathrm{mid,crit}$ can be seen as a limit to the vertical mixing (see Eq. \ref{eq:dust_scaleheight}). A minimal value for the bulk dust-to-gas ratio in the absence of turbulence also exists \citep{carrera2015}, but given the strong turbulent mixing found in the core-collapse simulations (see Fig. \ref{fig:mhd_delta}), requiring a minimal midplane dust-to-gas ratio of $\epsilon_\mathrm{mid,crit}$ always serves as the stronger condition. Requiring $\mathrm{St}\geq\num{e-2}$ is restrictive compared to $\mathrm{St}\geq\num{3e-3}$ found by \citet{carrera2015}, and more recent works indicate that the streaming instability might lead to strong clumping for even smaller dust grains given a high enough bulk dust-to-gas ratio \citep{yang2017,ly2021}. Despite that, we still employ the more restrictive criterion for two reasons. First, the vertical stirring parameter we use in our 1D model is substantially higher than what is typically used in the aforementioned studies ($\delta_z>\num{e-2}$). Even though vertical stirring is encapsulated as a limiting mechanism in the requirement of $\epsilon_\mathrm{mid,crit}$, it remains to be investigated whether small grains can also be clumped under these conditions. Second, realistic systems, especially those that experience many fragmenting collisions like in our system, exhibit a grain size distribution, and it remains unclear how the critical condition for streaming instability behaves in such cases \citep{bs2010,krapp2019,paardekooper2020,schaffner2021,yang2021}. In our simulations, the condition on the midplane dust-to-gas ratio is stronger than the one on size in most cases.

Consequently, a sink term for ices and silicates is introduced,
\begin{align}
    \dot\Sigma_\mathrm{ice,pltm} &= -\frac{\Sigma_\mathrm{ice}}{\Sigma_\mathrm{ice}+\Sigma_\mathrm{sil}}\dot\Sigma_\mathrm{pltm},\\
    \dot\Sigma_\mathrm{sil,pltm} &= -\frac{\Sigma_\mathrm{sil}}{\Sigma_\mathrm{ice}+\Sigma_\mathrm{sil}}\dot\Sigma_\mathrm{pltm}.
\end{align}
In the following sections, we employ two different threshold values. First, a more strict criterion of $\epsilon_\mathrm{mid,crit}=1$ \citep{drazkowsa2016,yg2005}, and second, a weaker criterion of $\epsilon_\mathrm{mid,crit}=0.5$ \citep{gole2020}.

\section{Reference core-collapse simulation}\label{sec:disk_quantities}

We compute parameters that govern the surface density time evolution of disks arising in the 3D core-collapse simulations. They employ nonideal MHD on a Cartesian grid with adaptive mesh refinement. In particular, the model "R2" from H20 will be used as a reference, with a comparison to models "R1" and "R7" in later sections. This simulation was run for a physical time of ~\SI{100}{\kilo\year}, with a resolution of $dx=\SI{1}{\astronomicalunit}$ at the highest AMR refinement level. Only the gas disk was modeled.

\begin{figure}[htp]
    \centering\includegraphics[width=\linewidth]{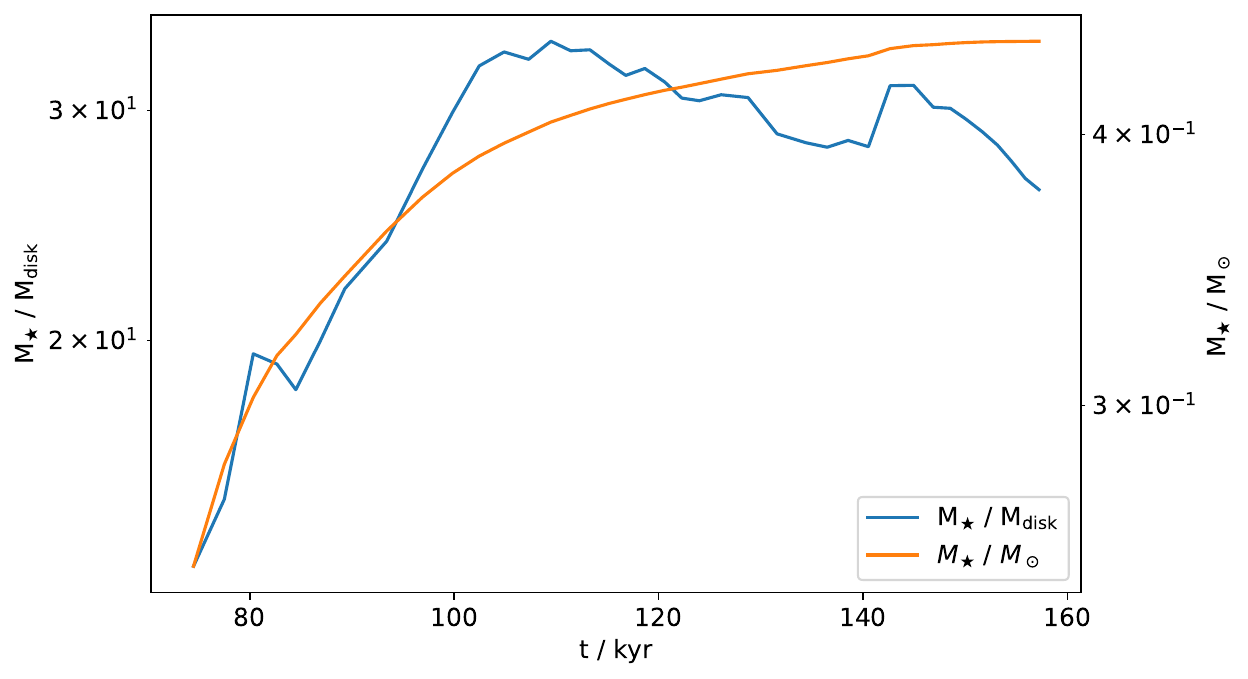}
    \caption{Mass evolution over time of the sink particle at the center of the R2 simulation from H20. The blue line shows the fraction $M_\bigstar/M_\mathrm{disk}$ with values indicated on the left axis, whereas the orange line shows the sink particle mass in units of solar masses, with values indicated on the right axis.}
    \label{fig:sink_evol}
\end{figure}%
The initial core mass of the models in H20 is $\SI{1}{\solarmass}$. We start considering snapshots once the central sink particle, mimicking the star, has formed and the initially violent dynamics has subsided. In the case of R2, this corresponds to $t_0=\SI{74.49}{\kilo\year}$. At this stage, the sink particle has a mass of $M_\bigstar = 13\ \mathrm{M}_\mathrm{disk} = \SI{0.25}{\solarmass}$. The evolution of the sink particle mass is shown in Fig. \ref{fig:sink_evol}. At the end of the R2 simulation, the mass has progressed to $M_\bigstar = 26\ \mathrm{M}_\mathrm{disk} = \SI{0.44}{\solarmass}$.

\subsection{Simulation resolution}\label{sec:ref_sim}

\begin{figure}[htp]
    \centering\includegraphics[width=\linewidth]{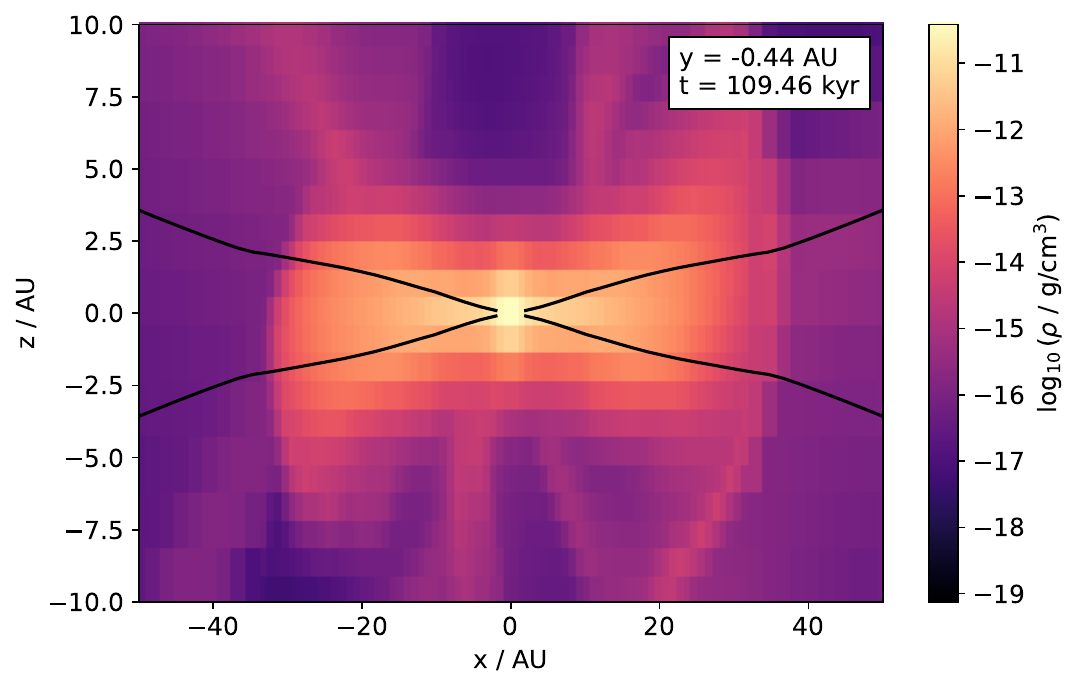}
    \caption{Zoom-in on the disk from R2 of H20 in the $x$-$z$ plane at $t=\SI{109.46}{\kilo\year}$, showing the density on a logarithmic scale. To accommodate for the disk's shape, $x$ and $z$ axes are scaled differently. The gas pressure scale height $H_g$ is marked by black lines.}
    \label{fig:3D_hp}
\end{figure}%
A zoom-in of the disk that is formed in the reference model R2 (see Fig. \ref{fig:3D_snapshot}) is shown in Fig. \ref{fig:3D_hp}, where the x-z plane is shown. The height corresponding to one pressure scale height $H_g$ is marked. The snapshot corresponds to a physical time of $\SI{109.46}{\kilo\year}$ after the beginning of core-collapse. At this time, $M_\star=\SI{0.41}{\solarmass}=34\ \mathrm{M}_\mathrm{disk}$. It can be seen that the disk itself is approximately axisymmetric, while the same does not hold true for the infall. However, for simplicity, this is neglected here so that we can model the disk only in the radial dimension, averaging in the azimuthal direction and vertically when applicable. As higher spatial resolution is computationally expensive, simulations running for ${\sim}\SI{100}{\kilo\year}$ are currently not feasible to be run with higher resolution. Thus, we are restricted to models with a limited resolution of at most three cells per scale height in the outer regions of the disk, with the resolution being worse in the inner disk. Therefore, we do not make a distinction between the bulk of the disk and the disk midplane at this time, even though the dynamics caused by the strong infall likely has a less severe impact on the midplane, which is the region large dust grains settle to and thus most relevant for planetesimal formation.

\subsection{Disk parameters}\label{sec:basic_parameters}
\begin{figure*}[htp]
    \centering\includegraphics[width=\linewidth]{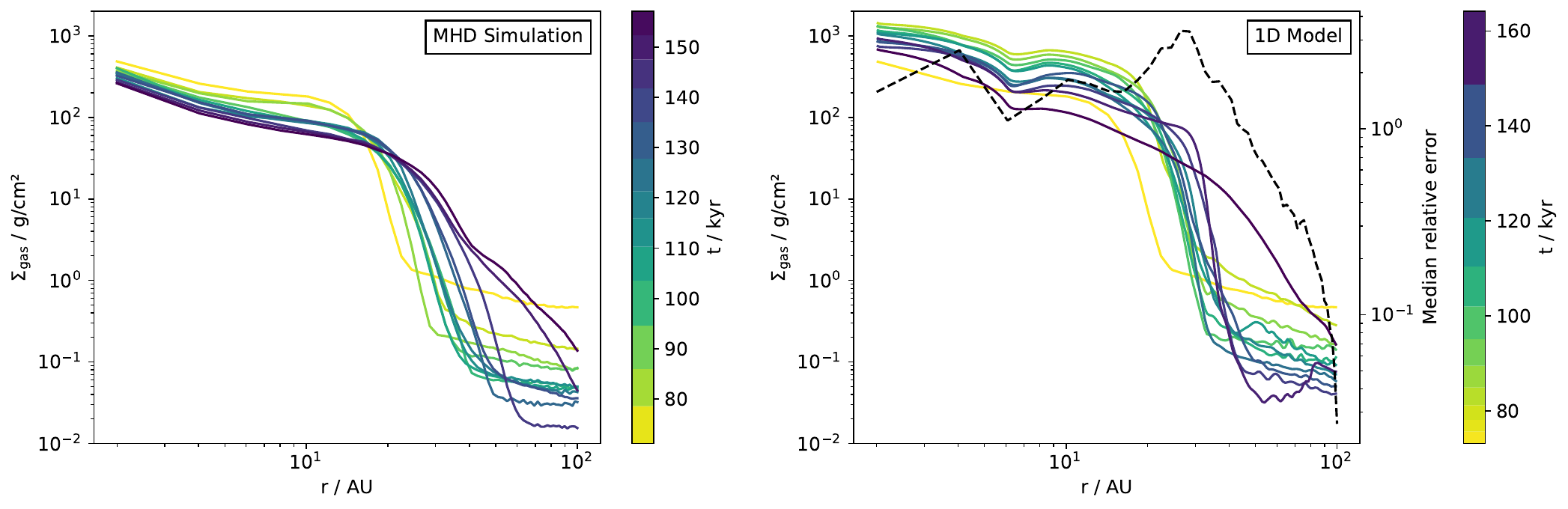}
    \caption{Gas surface density $\Sigma_\mathrm{gas}$ as a function of radial distance. The line color denotes the elapsed simulation time as indicated by the color bar, where a darker color indicates later times. \textit{Left}: Calculated directly from the R2 core-collapse simulation. \textit{Right}: Obtained in our 1D model when starting from the first core-collapse snapshot and employing the previously extracted model parameters. The black dashed line indicates the median relative error of $\Sigma_\mathrm{gas}$ obtained from the 1D model compared to the R2 simulation, with values indicated on the right ordinate.}
    \label{fig:mhd_sigma}
\end{figure*}
The left panel of Fig. \ref{fig:mhd_sigma} shows the azimuthally averaged surface density for all available snapshots. The goal of this section is to reconstruct the time evolution shown in this figure by integrating a 1D advection-diffusion equation. The employed parameter profiles are then used to model dust and gas evolution simultaneously in a 1D framework. In particular, the profiles of the torque-related parameters, which will be discussed in Section \ref{sec:torque_parameters}, show that the time evolution of the disk during the infall stage is best described by an infall induced external torque, rather than a viscosity-induced internal one, which is a fundamental difference to previous 1D models and may not be immediately obvious when just considering Fig. \ref{fig:mhd_sigma}. A disk driven by an external torque is similar to a wind-driven accretion case for Class II disks \citep{tabone2022}, but the fundamental physical difference to the disk shown in Fig. \ref{fig:mhd_sigma} is that the torque is a result of the accretion of angular momentum-deficient material, which has undergone magnetic braking, rather than the ejection of material whose angular momentum is increased due to the magnetic lever arm.

\begin{figure}[htp]
    \centering\includegraphics[width=\linewidth]{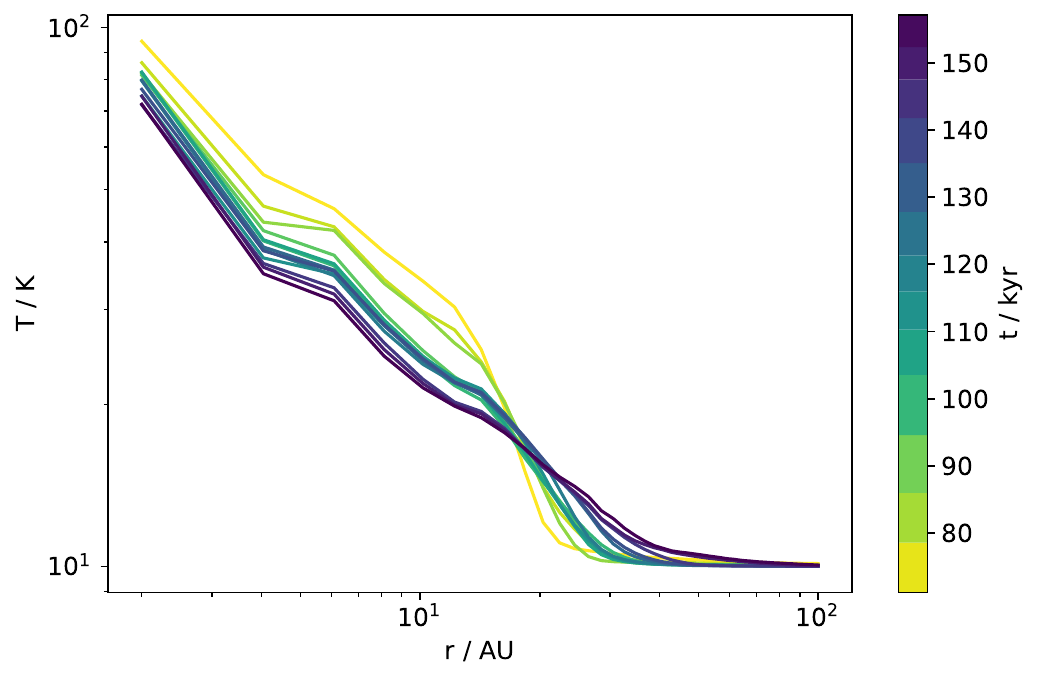}
    \caption{Average temperature as a function of radial distance found in R2. Different line colors indicate different simulation times according to the color bar, like Fig. \ref{fig:mhd_sigma}.}
    \label{fig:mhd_T}
\end{figure}%
Figure \ref{fig:mhd_T} shows the azimuthally and vertically averaged disk temperature without an additional scaling factor $\xi$ discussed in Section \ref{sec:gas_disk_model}. It is apparent that in this case, the disk is cooler than what is typically assumed for protoplanetary disks in their early evolutionary stages. In fact, since $T<\SI{150}{\kelvin}$ everywhere in the simulation grid, we cannot consider effects related to the snow line using this temperature profile, because the snow line is within the central resolution element around the star. Therefore, the snow line can only be considered as a potential planetesimal formation site for scaling factors $\xi>1$ that allow cells in the simulation grid to reach $T\geq\SI{150}{\kelvin}$.

Some disks found by H20 are gravitationally unstable, and modeling them would require approximating the impact of the long-range gravitational force as a local phenomenon to fit into the picture of a viscously evolving gas disk. Additionally, separate treatment for dust evolution is typically necessary. In order to constrain this work to disks whose surface density evolution can be approximated in a 1D framework to reasonable accuracy, we only consider core-collapse simulations with disks that remain stable during the runtime. The treatment of planetesimal formation in gravitationally unstable early-stage disks remains subject to future work. We verified that the disk in R2 is gravitationally stable throughout the simulations runtime, i.e. that the Toomre parameter,
\begin{equation}
Q = \frac{\Omega c_s}{\pi G\Sigma},\label{eq:toomre_Q}
\end{equation}
is larger than 1 for all snapshots. In fact, we find that $Q\gtrsim 10$ throughout the simulation. We also find the disks from other core-collapse simulations discussed in Section \ref{sec:othermhd} to be gravitationally stable according to this criterion.

\begin{figure*}[htp]
    \centering\includegraphics[width=\linewidth]{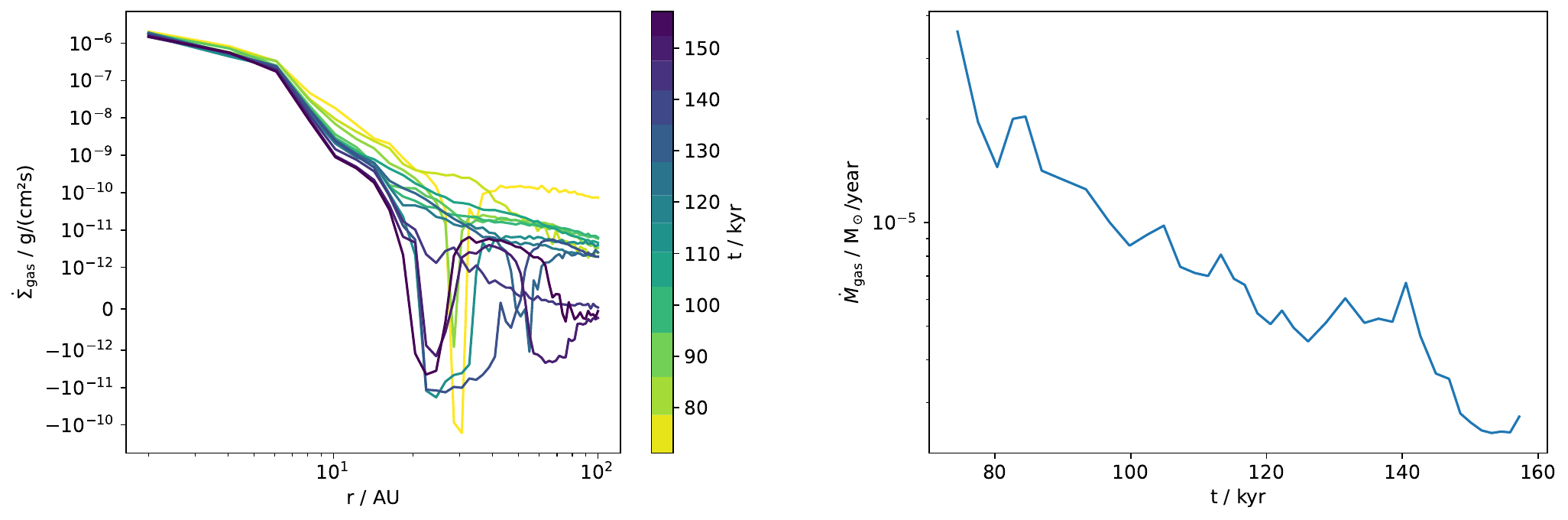}
    \caption{Source term of gas entering the disk bounds in R2. \textit{Left}: Time derivative of the surface density caused by mass entering the disk bounds, azimuthally averaged as a function of radial distance. The ordinate is scaled linearly for values with absolute value below \num{e-12}, but logarithmically otherwise. Line coloring like in Fig. \ref{fig:mhd_sigma}. \textit{Right:} Total instantaneously infalling mass as a function of time.}
    \label{fig:mhd_infall}
\end{figure*}%
Figure \ref{fig:mhd_infall} shows the radial profile of the azimuthally averaged infall from the envelope onto the disk. It is apparent that it is very centrally concentrated, with $\dot\Sigma$ dropping by four orders of magnitude before reaching the disk outer radius in the first simulation time step. Some outflow can be observed at the outer edge of the disk. However, in the interest of numerical stability, outflow is not considered, and the source term is set to zero at radii where it is negative. This simplification is also justified by the small magnitude and radial extend of the outflow compared to the infall.

The right-hand side of Fig. \ref{fig:mhd_infall} shows the total, i.e. radially integrated, mass infall rate. Over the course of the simulation, it drops by an order of magnitude as the initially vigorous infall starts subsiding. In order to compute the radially integrated infall rate, we exclude the region $r<\SI{6}{\astronomicalunit}$ as this region uses cells from within the sink particle accretion radius of $r_\mathrm{sink}=4dx$ for the calculation of $\dot\Sigma$. Numerical issues produce unphysically high values here, which was previously found by \citet{lee2021}. If the region was included, the magnitude of the infall rate would be $\dot M = \SI{e-4}{\solarmass\per\year}$, which is very large compared to the values of ${\sim}\SI{e-6}{\solarmass\per\year}$ typically assumed for 1D infall models.

\subsection{Torque parameters}\label{sec:torque_parameters}
Modeling the time evolution of the protoplanetary disk requires a characterization of both the internal torques, $\alpha_\mathrm{SS}$, and external torques, $\alpha_\mathrm{EA}$ (cfr. Eq. \ref{eq:1D_adv_diff}).

\begin{figure*}
    \centering\includegraphics[width=\linewidth]{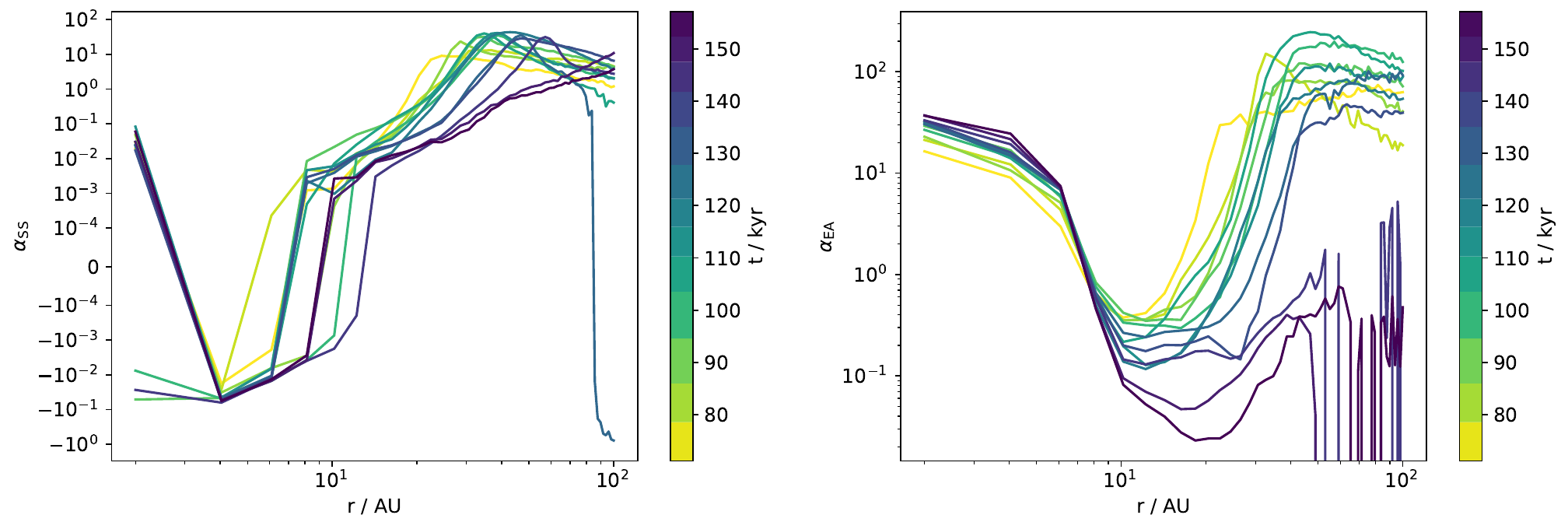}
    \caption{Torque parameters describing the dynamical evolution of the gas disk in R2, with line coloring like in Fig. \ref{fig:mhd_sigma}. \textit{Left}: Radial profile of the internal torque parameter $\alpha_\mathrm{SS}$ (cfr. Eq. \ref{eq:alpha_SS}), which is scaled linearly for values with absolute value less than \num{e-4}, but logarithmically otherwise. \textit{Right}: Radial profile of the external torque parameter $\alpha_\mathrm{EA}$ (cfr. Eq. \ref{eq:alpha_EA}) on a logarithmic scale.}
    \label{fig:mhd_alpha}
\end{figure*}
The radial profile of the azimuthally and vertically averaged $\alpha_\mathrm{SS}$, given by Eq. \eqref{eq:alpha_SS}, is shown in the left panel of Fig. \ref{fig:mhd_alpha}. It can be seen that $\alpha_\mathrm{SS}$ reaches negative values in most radial grid cells inside the disk radius, which is caused by $\langle\delta v_r\delta v_\phi\rangle<0$ dominating Eq. \eqref{eq:alpha_SS}. This is unlike the positive values used in commonly used 1D viscous evolution models for Class II and early-stage disks. In fact, $\alpha_\mathrm{SS}>0$ in Eq. \eqref{eq:1D_adv_diff} results in a diffusion of the gas surface density over time in accordance with the outward angular momentum transport induced by turbulent motion. Mathematically, there is no constraint on the sign of $\alpha_\mathrm{SS}$, and a negative value describes the transport of angular momentum toward the inner edge instead. In the context of a 1D radial diffusion model, $\alpha_\mathrm{SS}<0$ is not physical, though, as it leads to diffusion backward in time, removing entropy and violating thermodynamics. Instabilities at the root of turbulence in Class II disks, e.g. the Magnetorotational Instability (MRI), lead to the outward transport of angular momentum, but the MRI energy injection scales are much smaller than the resolution of the core-collapse simulations, as $L=\sqrt{\alpha_\mathrm{MRI}}H_p$.

We did not investigate whether the velocity fluctuations found in the core-collapse simulations are caused by disk turbulence or other processes like sound or shock waves from the infall, because such an investigation would be out of scope for this work. As the observed angular momentum transport is not in a direction where the commonly applied 1D viscous evolution equation is applicable, $\alpha_\mathrm{SS}$ as shown in Fig. \ref{fig:mhd_alpha} could not be used. Instead, we use a constant value of $\alpha_\mathrm{SS}=\num{e-1}$ and in doing so also neglect the very high values at $r\gtrsim\SI{20}{\astronomicalunit}$, as this region contains orders of magnitude less mass than the inner region (see Fig. \ref{fig:mhd_sigma}) due to it being outside the disk. While this values still implies considerable angular momentum transport, we find that this choice produces the closest match between the evolution of the disk in the core-collapse simulation and the 1D framework. We scale this constant value with $\xi$, analogously to $\alpha_\mathrm{EA}$ (cfr. Eq. \ref{eq:alpha_EA}). Other processes related to dust and vapor evolution that are often modeled using $\alpha_\mathrm{SS}$ were modeled with other diffusion parameters instead. They are related to the root-mean-square of the velocity fluctuations to uphold the best possible level of self-consistency (see Sect. \ref{sec:dust_evolution}).

Furthermore, the usage of this constant value for $\alpha_\mathrm{SS}$ can be further justified by the fact that the parameter characterizing the external torque, $\alpha_\mathrm{EA}$ (cfr. Eq. \ref{eq:alpha_EA}), caused mainly by the infall in this scenario, largely dominates the evolution. This is shown on the right-hand side of Fig. \ref{fig:mhd_alpha}. The angular momentum deficit of the infalling material causes values of $\alpha_\mathrm{EA}$ to reach a few 10 in the inner disk and of the order of one closer to the outer edge of the disk. The contribution of the magnetic field (cfr. Eq. \ref{eq:alpha_EA}) is negligible in comparison, only reaching values up to ${\sim}\num{e-1}$ at the disk edge. With the positive sign of $\alpha_\mathrm{EA}$ corresponding to inward advection of disk mass, the evolution is similar to a Class II disk whose evolution is dominated by disk winds, as discussed above. However, the different physical origin of the external torque results in higher values for $\alpha_\mathrm{EA}$ than found for $\alpha_\mathrm{DW}$ in disk-wind scenarios. Additionally, Fig. \ref{fig:mhd_alpha} shows that $\alpha_\mathrm{EA}$ reaches negative values for large radii late in the evolution. As this is outside the disk radius, and in the interest of numerical stability, we set $\alpha_\mathrm{EA}=0$ in the corresponding radial grid cells instead.

\subsection{Dust evolution parameters}\label{sec:dust_evolution}

\begin{figure*}[htp]
    \centering\includegraphics[width=\linewidth]{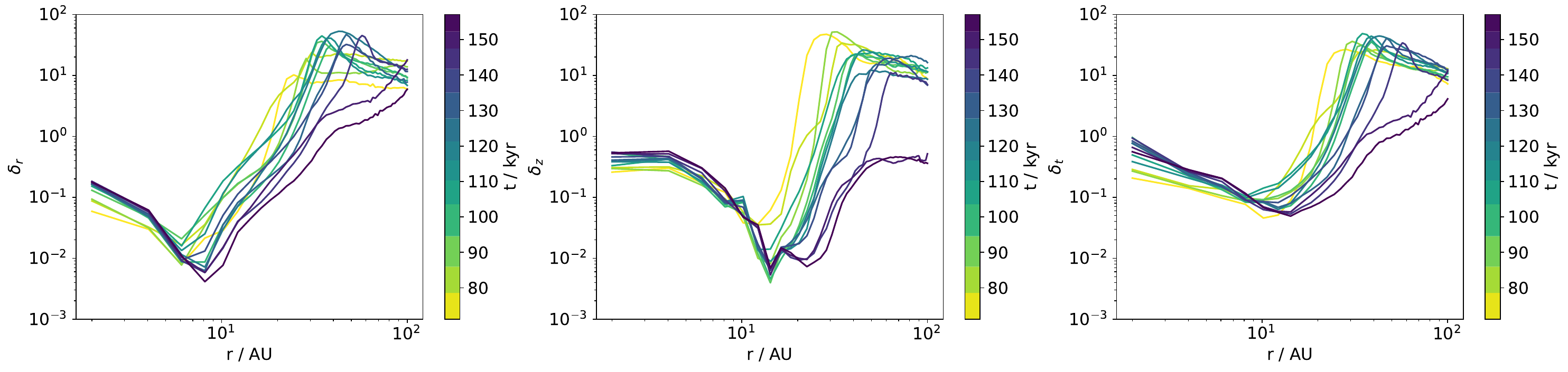}
    \caption{Radial profiles of model parameters describing dust properties and evolution, calculated from the disk in R2 and with line coloring like in Fig. \ref{fig:mhd_sigma}. \textit{Left:} Radial diffusion parameter $\delta_r$ (cfr. Eq. \ref{eq:delta_r}). \textit{Middle:} Vertical mixing parameter $\delta_z$ (cfr. Eq. \ref{eq:delta_z}). \textit{Right:} Turbulence parameter $\delta_t$ (cfr. Eq. \ref{eq:delta_t}).}
    \label{fig:mhd_delta}
\end{figure*}%
To describe the radial diffusion of dust and vapor, vertical settling of dust and the fragmentation limit in the two-population framework \citep{birnstiel2012}, $\alpha_\mathrm{SS}$ was replaced by $\delta_r$, $\delta_z$ and $\delta_t$, respectively (see Section \ref{sec:methods}). The profiles we calculate from the core-collapse simulation are shown in Fig. \ref{fig:mhd_delta}.

The first panel in Fig. \ref{fig:mhd_delta} shows $\delta_r$. It can be seen that values within the disk outer radius range between $\delta_r\sim\num{e-3}$ and $\delta_r\sim\num{e-2}$. While the magnitude is high for small radii, which is likely related to the sink accretion and overestimated mass infall as described above, the magnitude drops for increasing radius, but increases again for larger radii while still within the disk outer radius. In other words, there is a region in the disk where, for some snapshots, the radial diffusion parameter drops by an order of magnitude. The profile of $\delta_z$ is shaped similarly, but with a greater magnitude of up to several \num{e-1}. For some snapshots, the vertical settling parameter drops by up to two orders of magnitude at $r\sim\SI{10}{\astronomicalunit}$, which translates to more favorable conditions for planetesimal formation. Lastly, the third panel of Fig. \ref{fig:mhd_delta} shows that $\delta_t$ does not exhibit strong variation in its magnitude, staying at $\delta_t\sim\num{e-1}$ within the disk. 

\subsection{Gas evolution in the 1D framework}\label{sec:1D_comparison}

By employing the parameter profiles described in this section, the gas surface density evolution found in the core-collapse simulation is reconstructed from an initial snapshot, which is chosen such that initial strong changes in the surface density have subsided. Log interpolation is used to obtain the corresponding radial profiles on the 1D simulation grid. Furthermore, the parameter profile at any given moment in time is found by performing a linear interpolation at each time step. Therefore, the runtime of the 1D simulation is limited to the runtime of the original core-collapse simulation. The right panel of Fig. \ref{fig:mhd_sigma} shows the time evolution of the surface density as found in the 1D framework without including any dust evolution. It can be seen that it matches the original in order of magnitude, which we consider to be sufficient for the investigation of planetesimal formation. We prefer such an approximate model of the disk evolution and dynamics over the direct application of the disk surface density and gas velocities as found in the core-collapse simulations. The reason for this is that a direct application would not result in a self-consistent simultaneous description of the gas and solid surface density due to the usage of averages in azimuthal and vertical direction, as well as the mismatch of internal and output time step size of the core-collapse simulation.

\begin{figure}[htp]
    \centering\includegraphics[width=\linewidth]{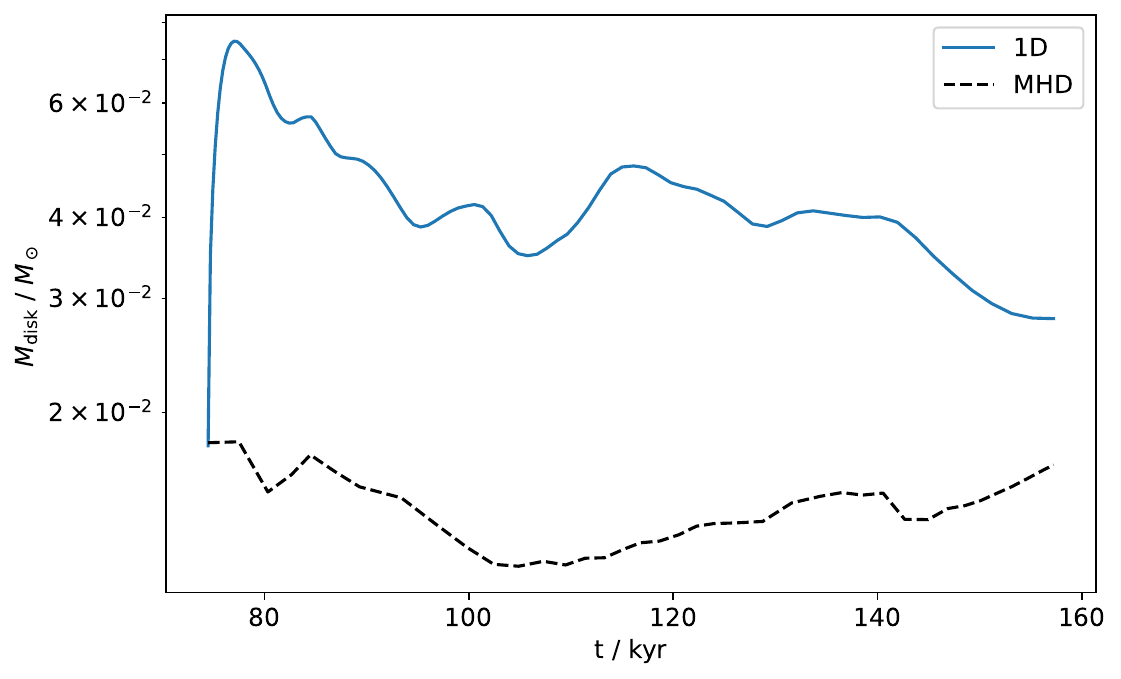}
    \caption{Disk mass as a function of time, calculated directly from the R2 core-collapse simulation (black dashed line) and from the surface density profiles obtained in our 1D model (blue line).}
    \label{fig:mhd_Mdisk}
\end{figure}
\begin{figure*}[htp]
    \centering\includegraphics[width=\linewidth]{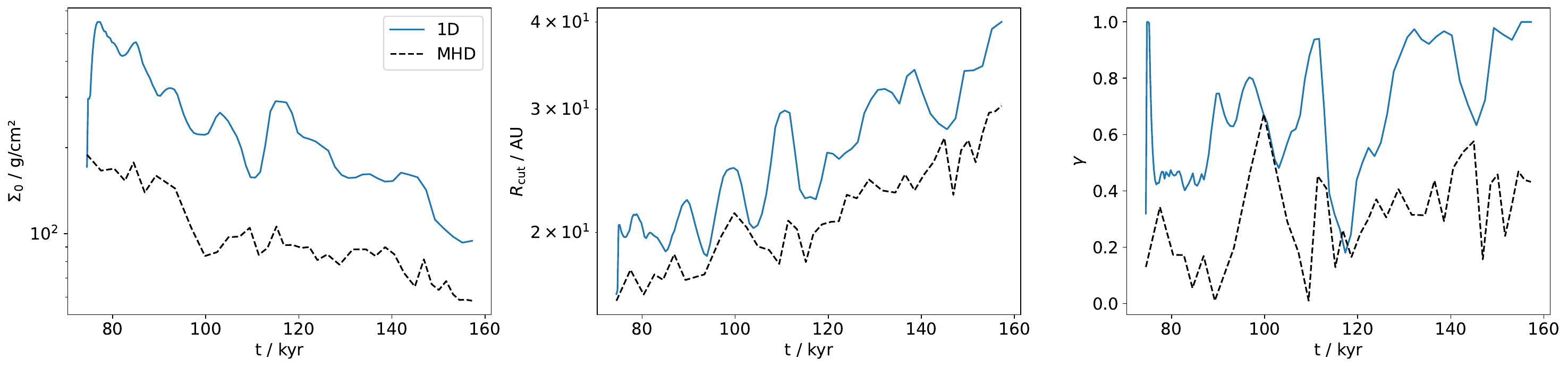}
    \caption{Parameters describing the disk surface density profile according to Eq. \eqref{eq:cutoff_disk} as a function of time. Fits to the profiles calculated directly from the R2 simulation are shown in black dashed lines, whereas fits to the profiles obtained in our 1D model are depicted with blue solid lines. \textit{Left}: Surface density at reference radial distance $r=\SI{12}{\astronomicalunit}$. \textit{Middle}: Disk cut-off radius, describing the radial distance where the surface density starts dropping exponentially. \textit{Right}: Slope of the surface density profile.}
    \label{fig:mhd_fitparams}
\end{figure*}

%
The kink seen in the 1D model's surface density profiles at $r=\SI{6}{\astronomicalunit}$, as well the increased total disk mass of the 1D model compared to the core-collapse simulation, are related to two effects. First, the calculation of the mass infall rate may result in unphysically high values within the sink particle accretion radius. Second, we do not consider the removal of mass exceeding the sink particle accretion threshold within the sink's accretion radius in the 1D model. In fact, we find that a change in the sink particle accretion mass threshold affects the slope and magnitude of the surface density profiles considerably beyond the sink particle accretion radius. Increasing the accretion threshold leads to a higher disk mass, which gives a better match to the disk mass found in our 1D model. However, in the interest of investigating simulations that cover the longest physical time period, we constrained our analysis to the R2 simulation of H20 where the threshold was not increased. This simulation employs a threshold density of $n_\mathrm{acc}=\SI{e13}{\per\centi\meter\cubed}$. While the magnitude of the surface density affects the Stokes number of the dust grains, this has no considerable effect on our results, as we find planetesimal formation to occur predominantly at $r\gtrsim\SI{10}{\astronomicalunit}$ in the following sections, whereas the mass mismatch occurs at $r\lesssim\SI{6}{\astronomicalunit}$. We note that the increase of the median relative error indicated in Fig. \ref{fig:mhd_sigma} at $\SI{20}{\astronomicalunit}$ is caused by the disk cutoff radius mismatch between the core-collapse simulation and the 1D model as shown in Fig. \ref{fig:mhd_fitparams}. This has no considerable effect on the planetesimal formation model. Furthermore, while a potential change in the pressure gradient impacts the drift speed of dust grains, dust drift is not impactful in our model as it occurs on time scales that exceed the simulated physical time. Because the surface density is overestimated in the inner disk regions, the total disk mass, shown in Fig. \ref{fig:mhd_Mdisk}, is also overestimated by our 1D model by a factor of ${\sim}3$. It exhibits only minor variations over time, which implies that the infalling mass and the strong advection caused by it create a balance where infalling material is not retained in the disk. Lastly, a detailed explanation of the boundary conditions is given in Appendix \ref{sec:appendix_boundaries}.

To characterize the disk that forms in the R2 run and the quality of our reconstruction in the 1D framework, a series of fits is applied to the surface density profile over time, with a fit function of
\begin{equation}
    \Sigma(r) = \Sigma_0\left(\frac{r}{R_0}\right)^{-\gamma}\exp\left(-\left(\frac{r}{R_\mathrm{cut}}\right)^\delta\right)+\Sigma_\mathrm{env},\label{eq:cutoff_disk}
\end{equation}
with $R_0=\SI{12}{\astronomicalunit}$. The resulting fit parameters are shown in Fig. \ref{fig:mhd_fitparams}. The magnitude of the surface density at a fixed location, $\Sigma_0$, as well as the disk cut-off radius $R_\mathrm{cut}$, match order-of-magnitude. While the cut-off radius increases over time with the disk growing, $\Sigma_0$ shows less variation over time. The slope of the surface density profile $\gamma$ exhibits greater deviation between MHD simulation and 1D model, as it is likely impacted by inaccurate physics near the sink particle accretion radius. We note that the quality of the fits varies between snapshots, causing fluctuations in the fit parameters.

\section{Planetesimal formation study}\label{sec:results}

\begin{table*}[htp]
    \centering
    \caption{Summary of parameters varied for the study presented in Sect. \ref{sec:results}. Square brackets denote intervals, whereas curly braces denote a set of values. The value $a_1$ represents the size of the large grain population in the two-population model.}
    \begin{tabular}{lllll}
        \hline
        \hline
        Label & Meaning & Values & Equation(s) & Section(s)\\
        \hline
        $\xi$ & Temperature scaling factor & $[1,9]$ & \eqref{eq:T_scale}, \eqref{eq:alpha_EA}, \eqref{eq:alpha_SS}, \eqref{eq:delta_r}, \eqref{eq:delta_z}, \eqref{eq:delta_t} & \ref{sec:xi} \\
        $\bar a$ & Average dust grain size & $\{0.1,1\}\si{\micro\meter}$, $a_1$ & \eqref{eq:abar} & \ref{sec:abar_model}, \ref{sec:a1}\\
        $\chi$ & Velocity fluctuation scaling factor & $[1,100]$ & \eqref{eq:delta_r}, \eqref{eq:delta_z}, \eqref{eq:delta_t} & \ref{sec:chi_a0} \\
        $\epsilon_0$ & Initial and infall dust-to-gas ratio & $[0.01,0.07]$ & \eqref{eq:infall_dtg} & \ref{sec:eps0}\\
        $\epsilon_\mathrm{mid,crit}$ & Critical midplane dust-to-gas ratio & $\{0.5,1\}$ & - & \ref{sec:eps0}\\
        \hline
    \end{tabular}
    \label{tab:study_params}
\end{table*}
In this section, we investigate the possibility of planetesimal formation taking place in the disk found in the R2 core-collapse simulation over the runtime of the simulation. We conduct a parameter study by varying the temperature scaling parameter $\xi\in[1,9]$, the initial vertically integrated and infall dust-to-gas ratio $\epsilon_0\in[0.01,0.07]$, the critical value of the midplane dust-to-gas ratio for the onset of the streaming instability $\epsilon_\mathrm{mid,crit}\in\{0.5,1\}$ and the velocity fluctuation reduction factor $\chi\in[1,100]$, as well as the average grain size $\bar a$ throughout the following subsections. For $\bar a$, we consider both $\SI{1}{\micro\meter}$ and $\SI{0.1}{\micro\meter}$, as well as a scenario without small grains, so that the average is given by the size of the large population, $\bar a=a_1$. The various parameters that are subject of this study are summarized in Table \ref{tab:study_params}.

Promising mechanisms that lead to an enrichment in dust that is sufficient to trigger planetesimal formation via the streaming instability are the traffic-jam effect \citep{da2018,so2017} and the cold-finger effect \citep{cuzzi2004}. The traffic-jam effect relies on a difference in radial drift speed between dust aggregates of different composition to create a local enhancement at the location of evaporation fronts. However, radial drift does not play a significant role during the short early infall phase, because the time it takes for pebbles in the outer disk regions to drift to the inner disk is larger than the simulation runtime of ${\sim}\SI{100}{\kilo\year}$. Therefore, we focus our study on the ability of the cold-finger effect to provide local enhancement in the dust-to-gas ratio. The cold-finger effect relies on the evaporation of solid material as dust grains cross evaporation fronts. In particular, we put our focus on the evaporation of water ice at the snow line. Here, new water vapor is supplied by the arrival of grains with ice mantles from the outer disk. This vapor is, in turn, redistributed both toward hotter and colder regions in the disk by turbulent diffusion. Consequently, a fraction of the water vapor created at the snow line reaches a location of low enough temperature to condense back on the solid grains. This cycle results in retaining a fraction of the solid flux through the snow line location, thereby enhancing the local dust-to-gas ratio. Unlike the traffic-jam effect, the enhancement caused by the cold-finger effect is not reliant on the differential drift of dust grains compared to the gas, as long as sufficient flux of solids passes the snow line and radial diffusion is strong enough to offset the missing speed difference. Because the local dust-to-gas ratio is enhanced by the retention of ice, planetesimals formed by the cold-finger effect have a large water fraction, differently from those formed by the traffic-jam effect, where the enhancement is driven by silicates.

\subsection{Impact of the disk temperature}\label{sec:xi}

\begin{figure}[htp]
    \centering\includegraphics[width=\linewidth]{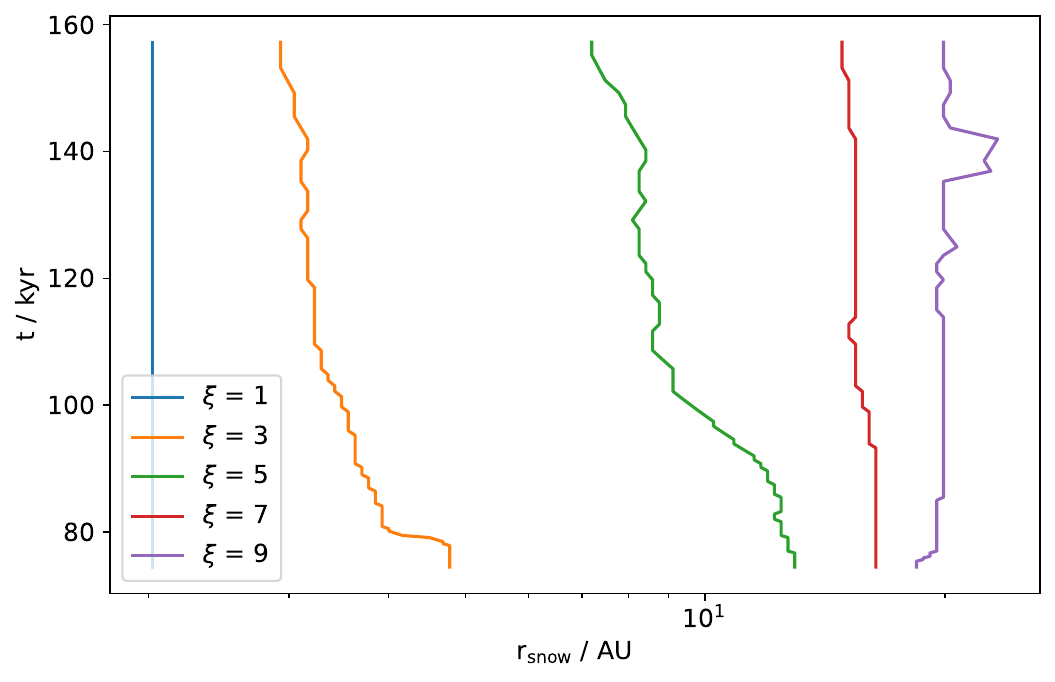}
    \caption{Radial location of the snow line as a function of time. The line color corresponds to the temperature scaling factor $\xi$.}
    \label{fig:R2_snowline}
\end{figure}%
First, we investigate the role of $\xi$, whose main effect on the simulations is to change the location of the snow line $r_\mathrm{snow}$ in the disk, defined as the radial location where the fragmentation velocity increases sharply (cfr. Eq. \ref{eq:vfrag}). Figure \ref{fig:R2_snowline} presents $r_\mathrm{snow}$ as a function of time for all investigated values of $\xi$. In the case of $\xi=1$, the snow line is not located within the simulation grid, i.e., it lies inside \SI{2}{\astronomicalunit}. Therefore, no snow line related effects can occur in runs with $\xi=1$, whereas the snow line is located within the simulation grid for higher values of $\xi$. For $\xi=3$ and $\xi=7$, the mean location of the snow line is $r_\mathrm{snow}\sim\SI{4}{\astronomicalunit}$ and $r_\mathrm{snow}\sim\SI{16}{\astronomicalunit}$, respectively, with limited movement covering ${\sim}\SI{2}{\astronomicalunit}$. Movement of the snow line is more significant for $\xi=5$ and $\xi=9$, covering ${\sim}\SI{6}{\astronomicalunit}$ and ${\sim}\SI{5}{\astronomicalunit}$, respectively. In those cases, the mean snow line locations are at \SI{11}{\astronomicalunit} and \SI{20}{\astronomicalunit}, respectively.

In simulations where the snow line is not found inside the simulation grid, i.e. $\xi=1$, material is not accumulated, due to the absence of any snow-line related effects. Therefore, the vertically integrated dust-to-gas ratio (also called metallicity in the literature), $(\Sigma_\mathrm{sil}+\Sigma_\mathrm{ice})/(\Sigma_\mathrm{vap}+\Sigma_\mathrm{H+He})$, does not exceed the initial value of 1\%. The high magnitude of the $\delta_z$ parameter (see Fig. \ref{fig:mhd_delta}) means that settling is inefficient. As a result, the midplane dust-to-gas ratio, the value indicative of the possibility of planetesimal formation, cannot reach the threshold value, and no planetesimals form in the $\xi=1$ scenario. This choice of $\xi$ corresponds to a cold disk which is not heated by the release of gravitational energy from the accretion.

\begin{figure*}[htp]
    \centering\includegraphics[width=\linewidth]{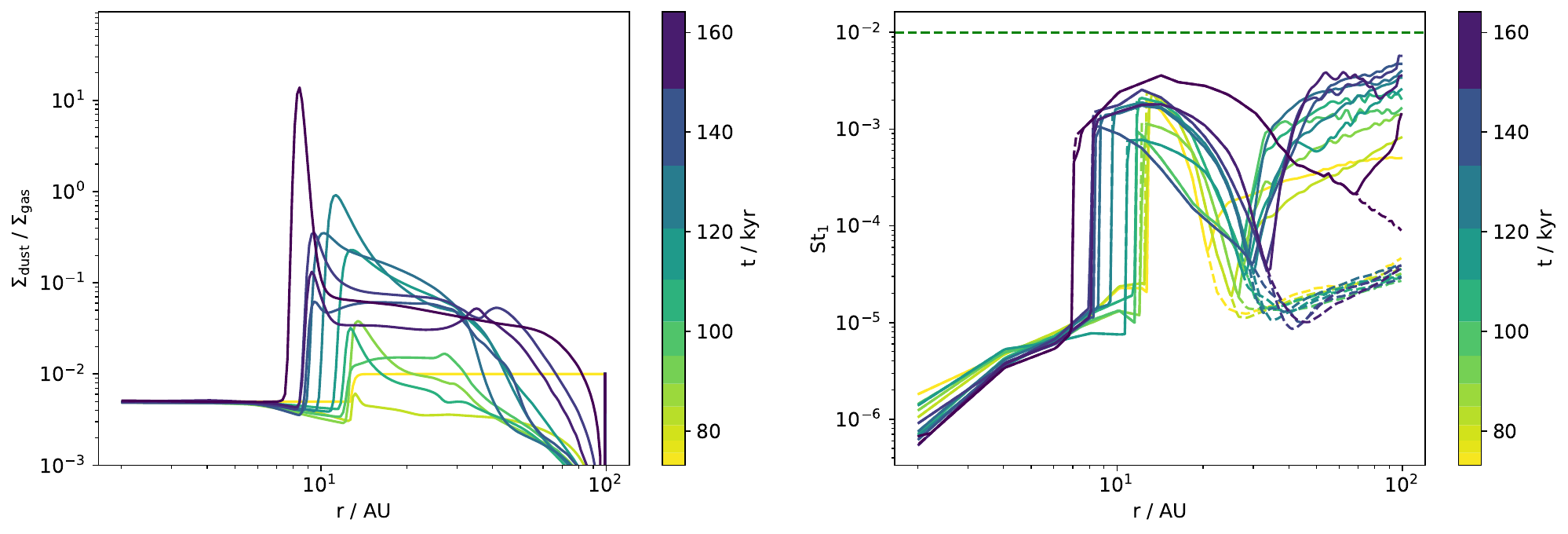}
    \caption{\texttt{TwoPopPy} simulation based on R2 quantities, but with the temperature scaled by $\xi=5$. The line color denotes the time in the simulation according to the color bar, with a darker color indicating a later time. \textit{Left}: Radial profile of the vertically integrated dust-to-gas ratio $\Sigma_\mathrm{dust}/\Sigma_\mathrm{gas}$. \textit{Right}: Radial profile of the pebble Stokes number $\mathrm{St}_1$, i.e., the Stokes number of the large grains in the two-population approximation. The dashed lines indicate the value expected from the two population model for this setup, $f_f\mathrm{St}_\mathrm{frag}$. The separate green dashed line at $\mathrm{St}_1=\num{e-2}$ indicates the minimal Stokes number required for strong clumping via the streaming instability.}
    \label{fig:R2_xi5_chi1}
\end{figure*}%
By comparing to core-collapse simulations with limited runtime, but that use a more detailed temperature model including radiative transfer and contributions of the accretion luminosity \citep{hennebelle2020b,lee2021,lebreuilly2024}, we find, using visual inspection, that applying a factor $\xi=5$ to $\xi=7$ to our temperature profile produces a good match to these models. On the left-hand side of Fig. \ref{fig:R2_xi5_chi1}, the vertically integrated dust-to-gas ratio of a run with $\xi=5$ is presented. With the snow line now being at a location within the radial grid of the 1D mode, the cold-finger effect leads to a very significant increase of the bulk dust-to-gas ratio, reaching a value of 1 after ${\sim}\SI{50}{\kilo\year}$ and even reaching 10 at the very end of the simulation. We note that, in principle, such high values of the bulk dust-to-gas ratio warrant a treatment of the back-reaction of the dust onto the gas, which we neglect here. We discuss this further in Section \ref{sec:discussion_1D_caveats}. Dust grains remain small, however, as they reach the low fragmentation limit set by high relative velocities, described by a high value of $\delta_t$ (see Fig. \ref{fig:mhd_delta}). As vertical settling remains inefficient, the midplane dust-to-gas ratio is not increased compared to the bulk dust-to-gas ratio, but the critical value is exceeded by orders of magnitude regardless.

The time evolution of the radial profile of $\mathrm{St}_1$ is shown on the right-hand side of Fig. \ref{fig:R2_xi5_chi1}. Due to the nature of the two population model, the Stokes number is a direct result of the limiting Stokes number due to drift or fragmentation, multiplied by a fudge factor $f_d$ or $f_f$, respectively. In this work, grains are always fragmentation limited, so that the Stokes number of the large grains is given by $\mathrm{St}_1=f_f\mathrm{St}_\mathrm{frag}=0.37\mathrm{St}_\mathrm{frag}$. In the region at $r>\SI{30}{\astronomicalunit}$, the Stokes number exceeds the limit given by fragmentation only because the lower limit given by the monomer size leads to larger Stokes numbers due to the very low density. While the Stokes number could in principle be lower than $f_f\mathrm{St}_\mathrm{frag}$ due to being limited by growth or the mixing with infalling small particles (cfr. Eq. \ref{eq:dust_infall_size}), these two effects are not significant for the run shown in Fig. \ref{fig:R2_xi5_chi1}. Therefore, despite the density criterion for planetesimal formation being met, the Stokes number of the large grains $\mathrm{St}_1$ never reaches the required value of $\mathrm{St}_1 \geq \num{e-2}$. Inside the disk, the maximal Stokes number only reaches values of ${\sim}\num{e-3}$. Therefore, planetesimals do not form for any run with $\chi=1$, i.e. the full turbulence as obtained directly from the 3D core-collapse simulation. While a relaxation of the size criterion discussed in Section \ref{sec:methods_pltm_form} would lead to planetesimal formation in this scenario, it also represents the case of the most violent dynamics, and it is unclear whether the streaming instability can produce strong clumping for very small grains under these conditions.

\begin{figure}[htp]
    \centering\includegraphics[width=\linewidth]{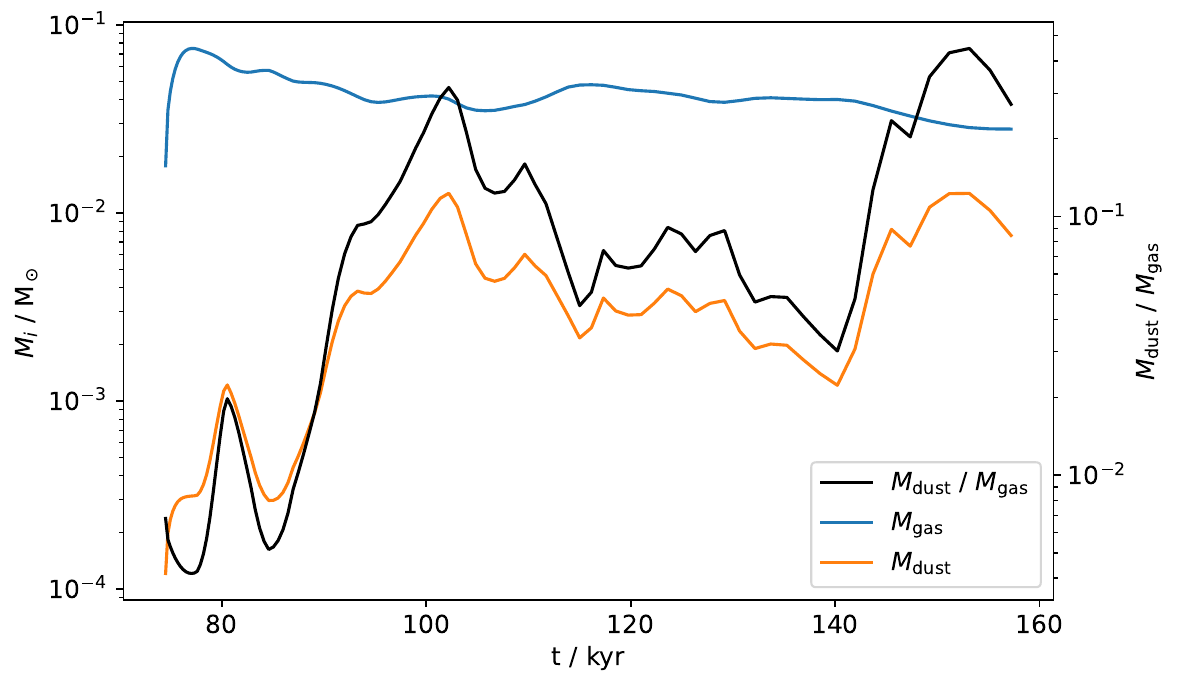}
    \caption{Total gas (orange line) and dust (blue line) masses as a function of time for a run with temperature scaling $\xi=5$. In addition, the total dust-to-gas ratio $M_\mathrm{dust}/M_\mathrm{gas}$ is shown on a separate axis using a black line.}
    \label{fig:R2_mass}
\end{figure}%
The reason for the strong enrichment of the vertically integrated dust-to-gas ratio even beyond a value of 1 is the large flux of material crossing the snow line, so that a large mass of ice can evaporate and condense, retaining a fraction of the material that was advected past the snow line. In fact, in contrast to Class II disks, the dust drift is not the dominant reason for a high radial speed of solids. Due to the high value of $\alpha_\mathrm{EA}$ caused by the strong infall, gas is advected with a high radial speed, between \SI{e3}{\centi\meter\per\second} in the outer disk and \SI{e5}{\centi\meter\per\second} in the inner disk. Dust drift only leads to an increase in speed by at most a factor of ${\sim}10$ for simulations with the largest grains ($\chi=100$). These high advection speeds in combination with a high infall rate (see Fig. \ref{fig:mhd_infall}) allow for high efficiency of the cold-finger effect, and thus a high bulk dust-to-gas ratio at the snow line. The flux passing through the snow line is naturally also related to its location. It is diminishing for increasing radial distance, as both the radial velocity and the dust surface density decrease with radial distance from the star. In a Class II disk with no infalling material, the total dust mass would be decreasing due to radial drift, and snow line related effects only concentrate solids present in the disk initially in order to achieve a higher local dust-to-gas ratio. In Class 0/I disks, the picture is different. Here, the large amount of mass accreted from the envelope is balanced by rapid accretion onto the star, so that the gas disk mass remains approximately constant over time. At the same time, the cold finger effect allows some accreted dust to be retained in the disk, increasing the total dust-to-gas ratio and thereby potentially establishing conditions favorable for planetesimals to form from the accreted solids. The time evolution of the total dust and gas masses, as well as the total dust-to-gas ratio, are shown in Fig. \ref{fig:R2_mass}.

\subsection{Relevance of the average dust grain size}\label{sec:abar_model}
\begin{figure}[htp]
    \centering\includegraphics[width=\linewidth]{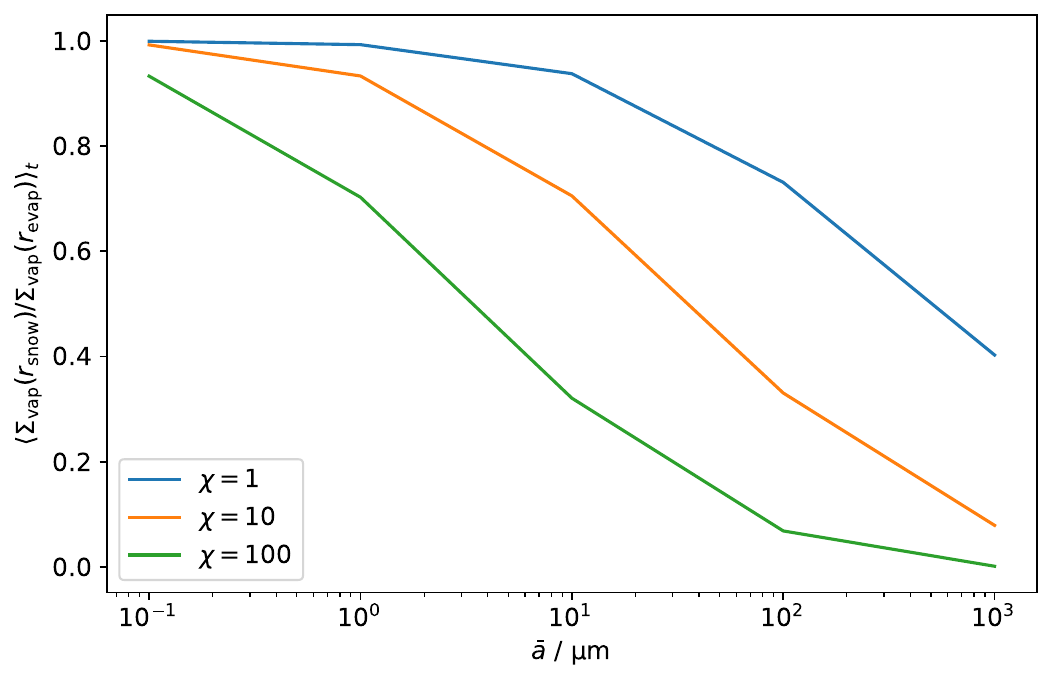}
    \caption{Time average of the mass fraction of water vapor reaching the snow line via backward radial diffusion from $r_\mathrm{evap}=r_\mathrm{snow}+v_\mathrm{gas}\tau_\mathrm{evap}$ (cfr. Eq. \ref{eq:backward_diff}). It is shown as a function of the mean grain size $\bar a$. The snow line location is chosen to match the $\xi=5$ case shown in Fig. \ref{fig:R2_snowline}. The line color indicates the factor $\chi$ by which the velocity fluctuations felt by the dust grains were reduced.}
    \label{fig:coldfinger_eff}
\end{figure}%
In our model, we find a strong dependence of the total mass of formed planetesimals on the average particle size. This can be understood by considering the timescale of evaporation of the ice mantle of a single dust grain and the distance covered by the particle during that time. Water vapor created during the evaporation must diffuse across the snow line to condense in order for the cold-finger effect to be efficient in retaining solids in the disk. If the distance covered by an evaporating particle increases, a smaller fraction of water vapor diffuses backward over the snow line and the cold-finger effect's effectiveness decreases. In order to gain a qualitative understanding of the importance of the grain size for the effectiveness of the cold-finger effect, we make the simplifying assumptions that the particles are fully made of ice and that the disk quantities stay constant during the evaporation process. Under these assumptions, the evaporation of the particle takes place on the timescale
\begin{equation}
    \tau_\mathrm{evap} = \bar a\left|\frac{da}{dt}\right|^{-1} = \bar a\rho_\bullet\sqrt{\frac{\pi k_BT}{8\mu_\mathrm{vap}}}P_\mathrm{eq}^{-1}.
\end{equation}
The radially outward mixing of a passive tracer in gas is described by \citep{pd2007}
\begin{equation}
    \frac{\Sigma_i}{\Sigma} = \sigma_i^{(0)} + \sigma_i^{(1)}\left(\frac{r_1}{r}\right)^{\frac{3}{2}\Sctilde},\label{eq:pd2007}
\end{equation}
with the equivalent Schmidt number $\Sctilde=\alpha_\mathrm{EA}/\delta_r$ describing the relative strength of advection and diffusion, $\Sigma_i$ being the tracer surface density, $\sigma_i^{(0)}$ the tracer background abundance and $\sigma_i^{(1)}$ the abundance of the tracer at an arbitrary location $r_1$. We assume that water vapor is created, on average, a certain distance inward of the snow line due to the finite time of the sublimation of water from the pebble as it drifts inward. This distance is given by $r_\mathrm{evap}=r_\mathrm{snow}+v_\mathrm{gas}\tau_\mathrm{evap}$ with $v_\mathrm{gas}<0$. By applying $\sigma_i^{(0)}=0$, $r_1=r_\mathrm{evap}$ and $\sigma_i^{(1)}=\Sigma_\mathrm{vap}(r_\mathrm{evap})/\Sigma(r_\mathrm{evap})$ to Eq. \eqref{eq:pd2007}, we find that the fraction of vapor that reaches the snow line is given by
\begin{equation}
    \frac{\Sigma_\mathrm{vap}(r_\mathrm{snow})}{\Sigma_\mathrm{vap}(r_\mathrm{evap})} = \left(1+\frac{v_\mathrm{gas}\tau_\mathrm{evap}}{r_\mathrm{snow}}\right)^{\frac{3}{2}\mathrm{\tilde{Sc}}}\frac{\Sigma(r_\mathrm{snow})}{\Sigma(r_\mathrm{evap})}.\label{eq:backward_diff}
\end{equation}
In the R2 core-collapse simulation, we typically have $\Sctilde\ \sim 100$, whereas the Schmidt number $\mathrm{Sc}=\nu/D$ is at least an order of magnitude smaller in Class II disks, reaching values of $\mathrm{Sc}\sim 10$ \citep{carballido2005}. Equation \eqref{eq:backward_diff} shows that $\frac{\Sigma_\mathrm{vap}(r_\mathrm{snow})}{\Sigma_\mathrm{vap}(r_\mathrm{evap})}\sim \tau_\mathrm{evap}^{1.5\mathrm{\tilde{Sc}}}$, with $\tau_\mathrm{evap}\sim \bar{a}$. Therefore, the Schmidt number being at least an order of magnitude larger makes the choice of the average particle size very impactful, in contrast to Class II models, where it has a negligible effect.

Figure \ref{fig:coldfinger_eff} shows the fraction of vapor reaching the snow line as calculated by Eq. \eqref{eq:backward_diff}, averaged over the simulation time and as a function of the average grain size for different values of $\chi$. The time-dependent results are discussed in Appendix \ref{sec:appendix_timedep_coldfinger}. As the reduction of velocity fluctuation strength reduces radial diffusion, a smaller mass fraction reaches the snow line for higher $\chi$. Therefore, even though a higher value of $\chi$ produces larger dust grains, the effectiveness of the cold-finger effect is reduced in turn. Furthermore, a larger average grain size implies a reduction in the mass fraction reaching the snow line, where less than 20\% is retained for $\bar a\geq\SI{e-2}{\centi\meter}$ ($\chi=100$) or $\bar a\geq\SI{e-1}{\centi\meter}$ ($\chi=10$). This greatly diminishes the effectiveness of the cold-finger effect to a point where it becomes negligible.

\subsection{Dependence on the magnitude of velocity fluctuations}\label{sec:chi_a0}

In Fig. \ref{fig:R2_xi5_chi1}, we show that solids cannot grow to a size sufficient to trigger the streaming instability and form planetesimals. The limiting factor for dust growth in our simulations is fragmentation due to high relative velocities. They are likely not of turbulent nature, but are rather induced by accretion-related shocks. Therefore, the high magnitude of those fluctuations might not protrude all the way to the midplane of the disk, which is the site of planetesimal formation. However, the resolution of the core-collapse simulations is not sufficient to resolve the scale height of the disk (see Fig. \ref{fig:3D_hp}), so that the impact of accretion on velocity fluctuations closer to the midplane remain subject to future investigations. It is likely that we overestimate their magnitude. Additionally, the limiting velocity for two grains to fragment, the so-called fragmentation velocity, is subject to debate, where laboratory experiments and theoretical calculations about the fragility of icy dust aggregates compared to silicate grains produce incompatible results. Different from our assumptions (see Section \ref{sec:methods}), higher fragmentation thresholds have been suggested in the past \citep{ormel2009}, whereas newer works indicate lower thresholds, challenging the assumed higher stickiness for icy dust aggregates \citep{mw2019,gundlach2018,steinpilz2019}. However, a sweep-up process during the early disk stages might allow grains to grow beyond the fragmentation barrier \citep{xa2023}. We also neglect how porosity could affect the maximum grain size and Stokes number \citep{okuzumi2012}.

In order to encapsulate all aforementioned arguments into this parameter study, we study the impact of reducing the turbulence dust grains are subject to, by adopting a factor $\chi$ (see Section \ref{sec:methods}). When considering the limiting Stokes number due to fragmentation, $\mathrm{St}_\mathrm{frag}$ (cfr. Eq. \ref{eq:st_frag}), assuming a lower $v_\mathrm{frag}$ yields an equivalent value to assuming an increased turbulent velocity reduction factor, i.e., $\mathrm{St}_\mathrm{frag}\propto\ \chi v_\mathrm{frag}^2$. On the other hand, the settling efficiency described by $H_g/H_d$ (cfr. Eq. \ref{eq:dust_scaleheight}) scales differently from the limiting Stokes number with $H_g/H_d\propto\ \chi v_\mathrm{frag}$ for $\mathrm{St}\gg\delta_z$. Therefore, comparing two runs where $\chi$ was decreased by a factor of 100 is not fully equivalent to comparing two runs where the fragmentation velocity was reduced by a factor of 10, because the settling efficiency is higher in the former case. However, because the settling efficiency is the most impactful factor for the possibility of planetesimal formation, a comparison can be made in that context. This requires the assumption that $\mathrm{St}\gg\delta_z$, which is fulfilled for large $\chi\gtrsim 10$, and that the size threshold for planetesimal formation is still met. In such a case, we expect that, if planetesimals can form for a given $\chi$ and $v_\mathrm{frag}=\SI{10}{\meter\per\second}$, they would also form for $v_\mathrm{frag}=\SI{1}{\meter\per\second}$ if they can form at $\chi/10$ and $v_\mathrm{frag}=\SI{10}{\meter\per\second}$. In the interest of simplicity, we limit this work to varying $\chi$ only, and we consider $\chi=10$ in the following.

\begin{figure*}[htp]
    \centering\includegraphics[width=\linewidth]{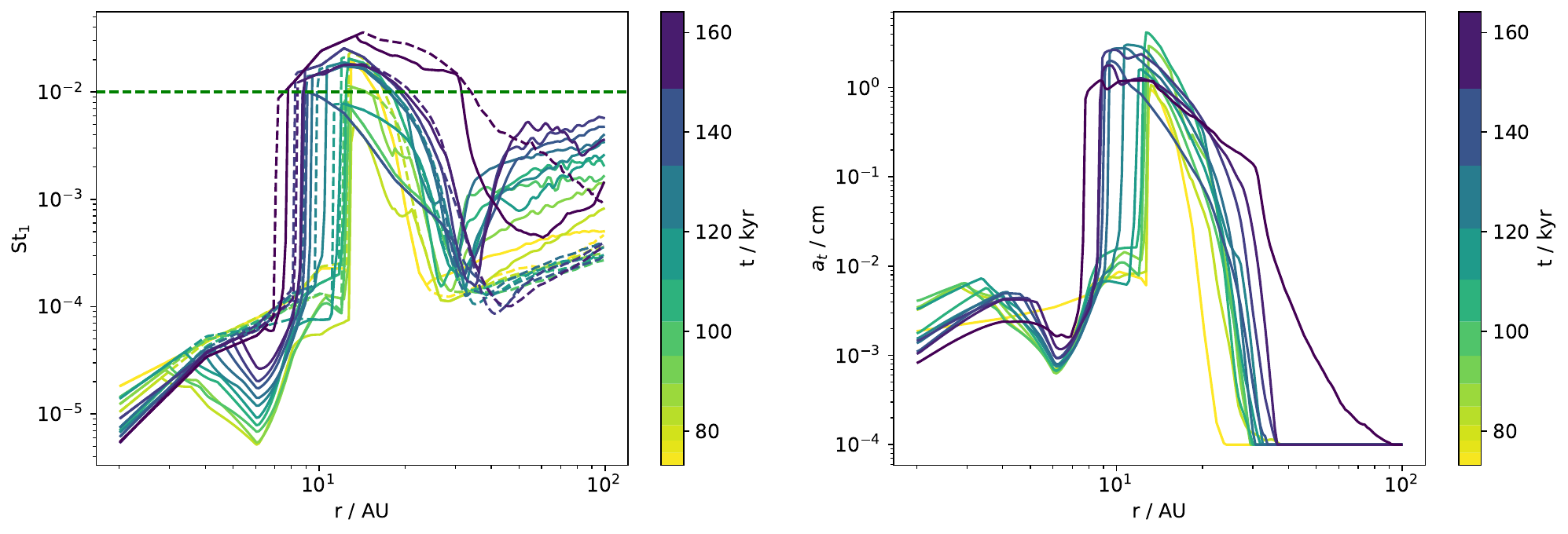}
    \caption{\textit{Left}: Pebble stokes number $\mathrm{St}_1$ as a function of radial distance. \textit{Right}: Corresponding particle size $a_t$. In this simulation, the temperature was scaled by $\xi=5$, and the velocity fluctuations reduced by $\chi=10$. The line coloring corresponds to the simulation time according to the color bar. A darker color depicts the radial profile at a later time. The dashed lines indicate the value $f_f\mathrm{St}_\mathrm{frag}$ for each time. The horizontal green dashed line shows the limiting Stokes number $\mathrm{St}_1=\num{e-2}$ for planetesimal formation.}
    \label{fig:R2_chi10_St}
\end{figure*}%
The resulting pebble Stokes number $\mathrm{St}_1$ is shown in Fig. \ref{fig:R2_chi10_St}, using the same parameters as shown in Fig. \ref{fig:R2_xi5_chi1} but with $\chi=10$. Here, pebbles are able to reach $\mathrm{St}_1\geq\num{e-2}$ required to form planetesimals via the streaming instability. The maximal Stokes number is $\mathrm{St}_1=\num{3e-2}$, which is only reached during the last \SI{10}{\kilo\year} of the simulation. In this run, mixing of grown grains with the small infalling grains becomes impactful, particularly in the vicinity of $r\sim\SI{6}{\astronomicalunit}$. If $\chi=100$ is used instead, we find that a maximal pebble Stokes number of $\mathrm{St}_1=\num{e-1}$ is already reached after \SI{20}{\kilo\year}, due to a further reduction of $\delta_t$, which sets the grain size in the fragmentation limit, which is the smallest one at all times and radial distances in R2.

\begin{figure*}[htp]
    \centering\includegraphics[width=\textwidth]{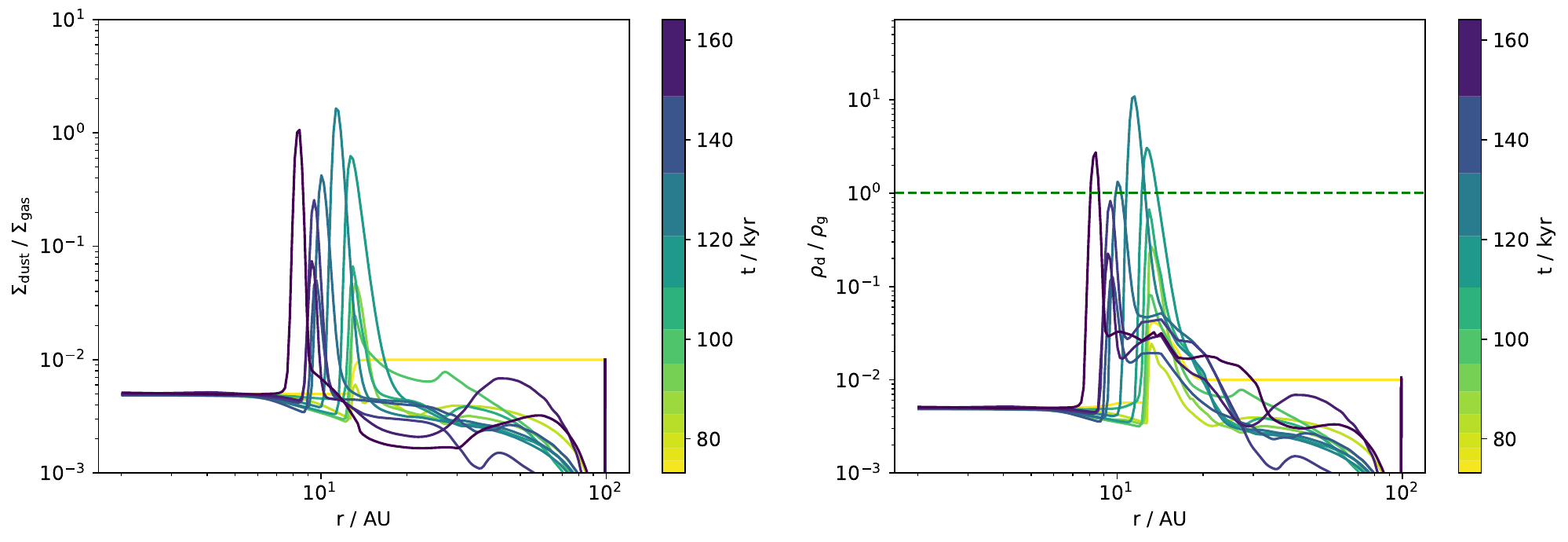}
    \caption{Like Fig. \ref{fig:R2_xi5_chi1}, but the velocity fluctuations experienced by the dust grains were reduced by a factor $\chi=10$.}
    \label{fig:R2_chi10_xi5_dtg}
\end{figure*}%
If dust grains are subject to smaller velocity fluctuations, they can grow to larger sizes. However, radial mixing is reduced, decreasing the effectiveness of the cold finger effect because backward diffusion is less significant and only a smaller fraction of mass is able to cross back to the cold side of the snow line and condense, as detailed in Section \ref{sec:abar_model}. In Fig. \ref{fig:R2_chi10_xi5_dtg}, the vertically integrated and midplane dust-to-gas ratios are shown for turbulence reduced by a factor of $\chi=10$. Now, the maximal vertically integrated dust-to-gas ratio is 1 due to the reduced effectiveness of the cold-finger effect, reached at late times during the simulation. Due to more efficient settling (reduced $\delta_z$) of the now larger grains (reduced $\delta_t$), the midplane dust-to-gas ratio can reach values up to 10, though. Since both criteria for efficient clumping by the streaming instability are fulfilled in this case, planetesimals can form. As an overestimation of the midplane turbulence based on the core-collapse simulation is likely due to poor resolution, we can conclude that planetesimals can form in the Class 0/I stage of protoplanetary disks, under the conditions found in the R2 model and given an adjustment of the temperature. They are formed in the area covered by the movement of the snow line, creating a narrow planetesimals surface density profile around ${\sim}\SI{10}{\astronomicalunit}$ in the case of $\xi=5$.

\begin{figure*}[htp]
    \centering\includegraphics[width=\linewidth]{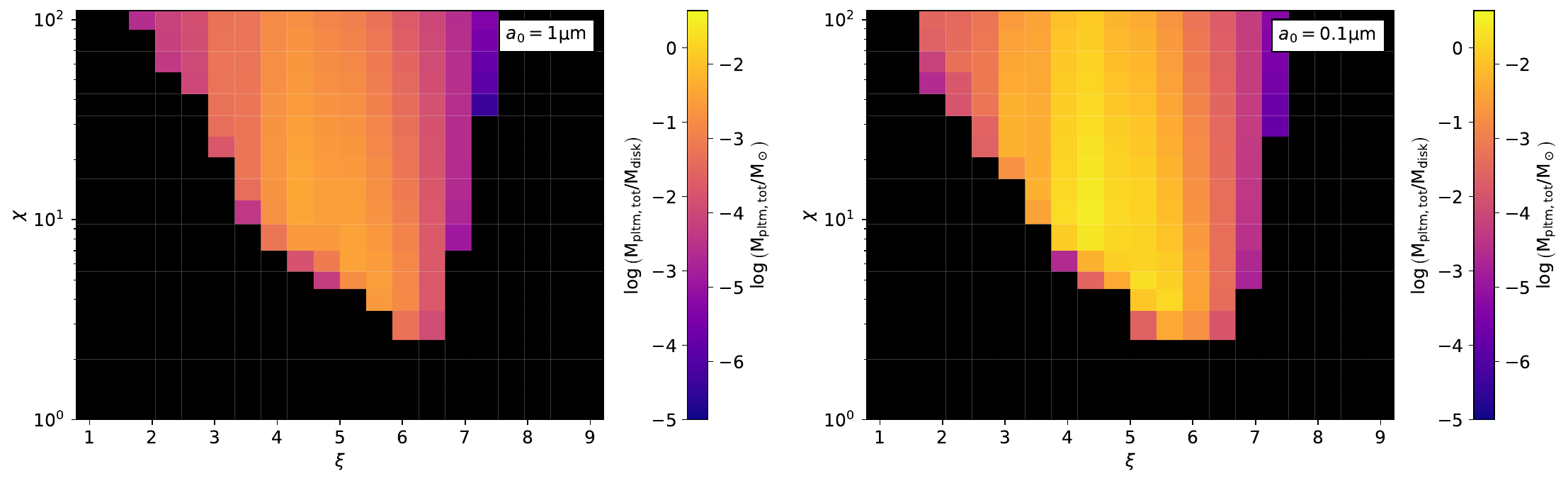}
    \caption{Total mass of planetesimals present at the end of various simulations. The simulations are based on R2 with several parameter adaptations. On the abscissa, the employed temperature scaling parameter $\xi$ is shown, whereas the ordinate indicates the factor $\chi$ by which the velocity fluctuations experienced by the dust grains was reduced. The left panel shows the case of larger dust monomers with size $a_0=\SI{1}{\micro\meter}$ and the right panel the case of a smaller value $a_0=\SI{0.1}{\micro\meter}$.}
    \label{fig:R2_mpla_abar}
\end{figure*}%
The total mass that can be retained in the disk at the snow line due to the cold-finger effect depends on the efficiency of radial mixing and on the location of the snow line, which sets the flux through it. In addition, the average particle size, which is approximately given by the size of the smallest grains, plays a significant role (see Section \ref{sec:abar_model}). In Fig. \ref{fig:R2_mpla_abar} on the left-hand side, the total mass of planetesimals formed during the simulation is shown as a function of temperature and velocity reduction factor for a monomer size of $a_0=\SI{1}{\mu m}$, which is the one we discussed so far. For $\xi<7$, an increase in temperature allows for the formation of planetesimals for lower $\chi$, because a higher temperature promotes dust settling in our model (see Section \ref{sec:dust_model}). At a given $\xi$, an increase in $\chi$ leads to a sharp increase in the planetesimal mass compared to the smallest value that allows formation, but stays constant for even higher $\chi$. For $4\leq\xi\leq6$, the planetesimal mass is slightly diminishing for increasing $\chi$, as the cold-finger effect retains fewer solids for reduced mixing (see Fig. \ref{fig:coldfinger_eff}). When exceeding temperatures $\xi\gtrsim 6.5$, a further increase in $\xi$ requires higher $\chi$ for planetesimals to form as the flux of solids through the snow line is reduced. Formation is not possible for $\xi<1.5$ ($r_\mathrm{snow}\sim\SI{2}{\astronomicalunit}$) and $\xi>7.5$ ($r_\mathrm{snow}\sim\SI{17}{\astronomicalunit}$). This trend does not change qualitatively for a smaller monomer size of $a_0$, shown on the right-hand side of Fig. \ref{fig:R2_mpla_abar}, but the total mass of formed planetesimals increases by a factor ${\sim}5$. Additionally, formation is possible for smaller $\chi$ at the extrema of the temperature space.

\begin{figure*}[htp]
    \centering\includegraphics[width=\linewidth]{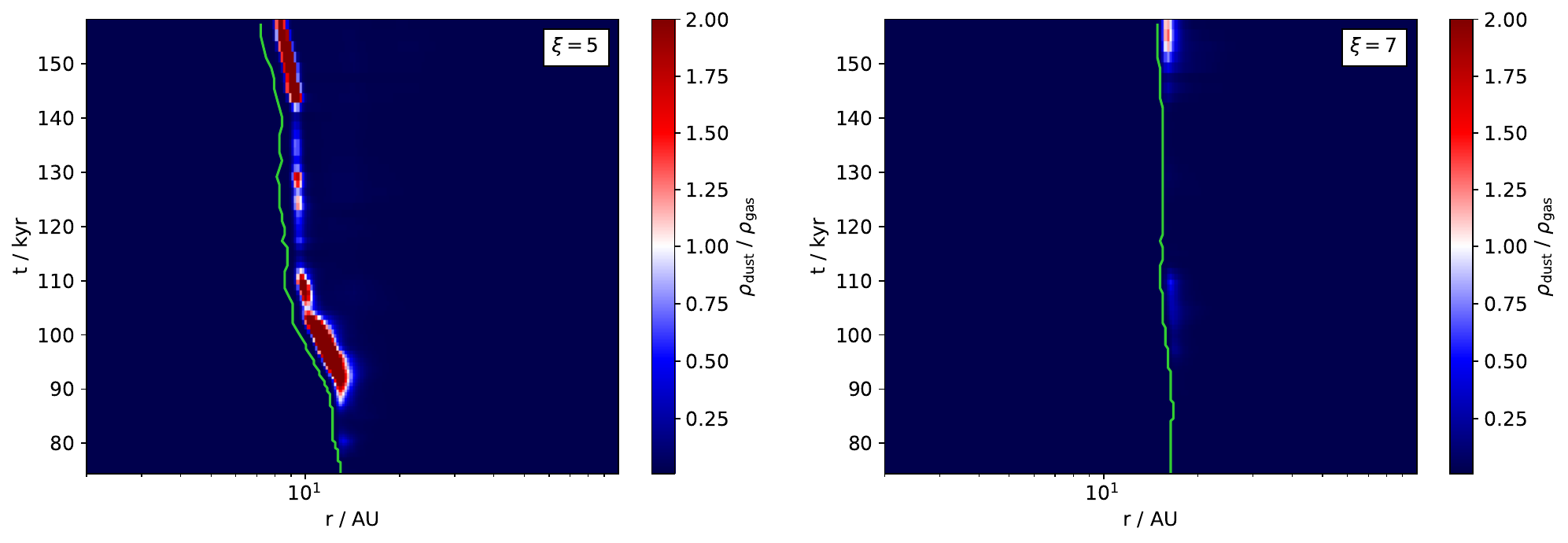}
    \caption{Midplane dust-to-gas ratio as a function of time (ordinate) and space (abscissa). The color map was chosen such that any shade of blue corresponds to a sub-critical value, such that planetesimal formation does not occur, whereas any shade of red corresponds to a super-critical value, creating the opportunity for planetesimal formation. The critical value itself corresponds to a white color. Therefore, all regions of white or red color corresponds to areas and points in time during the simulation where planetesimal formation occurs. Furthermore, the green solid line indicates the position of the snow line. The depicted simulations have the velocity fluctuations reduced by a factor $\chi=10$ compared to the R2 value, and employ a monomer size of $a_0=\SI{1}{\micro\meter}$. The left panel presents the case of a temperature scaling of $\xi=5$ and the right panel of $\xi=7$.}
    \label{fig:R2_chi10_heatmap}
\end{figure*}%
The total mass of formed planetesimals (see Fig. \ref{fig:R2_mpla_abar}) generally depends on two conditions. First, the effectiveness of the cold-finger effect sets the vertically integrated dust-to-gas ratio, but varies over time as disk conditions evolve. Second, changes in the strength of velocity fluctuations set by $\delta_t$, $\delta_r$, $\delta_z$ over time influence the maximal grain size and settling efficiency, in turn influencing the midplane dust-to-gas ratio. In total, this results in the necessary midplane dust-to-gas ratio for planetesimal formation to only be reached episodically in time. In addition, the location where $\epsilon\geq\epsilon_\mathrm{mid,crit}$ changes not only due to the movement of the snow line, but also because the radial profile of the $\delta$ parameters evolves in time. Figure \ref{fig:R2_chi10_heatmap} shows the midplane dust-to-gas ratio for two cases with a more realistic temperature, corresponding to $\xi=5$ and $\xi=7$, where turbulence is reduced by a factor of $\chi=10$. The depicted simulations employ a monomer size of $a_0=\SI{1}{\micro\meter}$. The dependencies on time and radial distance are depicted in the form of a 2D heatmap, where planetesimal formation occurs everywhere a white or red color is seen. With the cold-finger effect being the dominant factor for creating favorable conditions, planetesimal formation occurs just outside the snow line, in a ring of varying width. Compared to $\xi=5$, the run with $\xi=7$ experiences a smaller solid flux at the snow line location, so that planetesimal formation only occurs during times when settling is very efficient. This results in a lower total mass of planetesimals, as shown in Fig. \ref{fig:R2_mpla_abar}. Furthermore, it can be seen that formation of planetesimals is episodic. This is caused by the fluctuations of the $\delta$ parameters in time (see Fig. \ref{fig:mhd_delta}), which not only change the efficiency of settling and the dust grain size, but also the fraction of water vapor that can reach the snow line in the backward diffusion process, which significantly affects the efficiency of the cold-finger effect (see Appendix \ref{sec:appendix_timedep_coldfinger}). Due to this episodicity, we leave the prediction of further planetesimal formation beyond the run time of the core-collapse simulation subject to future work. Therefore, the total mass of formed planetesimals we find in our simulations is only a lower limit, and further formation may be possible at later times during Class I or Class II.

\subsection{Forming planetesimals without the cold-finger effect}\label{sec:a1}

For the parameters considered up to this point, the cold-finger effect is the dominant process allowing favorable conditions for planetesimal formation to be reached. However, its effectiveness drops sharply for an increased average grain size (see Fig. \ref{fig:coldfinger_eff}). The smallest grains dominate the size distribution in number, so that their size is dominant in $\bar a$. However, disk processes might affect the abundance and size of small grains, and observational signatures of larger grains are found around young, embedded disks \citep{galametz2019, valdivia2019, cacciapuoti2023a}. For example, sweep-up processes can create a top-heavy dust distribution \citep{xa2023}. Alternatively, if the bouncing barrier is included in grain size distribution considerations, large grains can efficiently sweep up small grains without replenishment by fragmentation. This results in a narrow grain size distribution with a greatly diminished small grain abundance \citep{dd2024}. While a full consideration of the bouncing barrier would prevent planetesimal formation due to a severely reduced maximal grain size, we investigate the impact of a large average grain size on the formation of planetesimals in infall-dominated disks assuming that the maximal grain size is unaffected. We consider the most extreme scenario where small grains are removed efficiently by some disk process, such that the average grain size equals the pebble size, $\bar a = a_1$.

Under the same conditions that produce a high total mass of planetesimals for $\bar a \approx a_0$ ($\xi=5$, $\chi=10$), we find that no enrichment of solids beyond the initial value of $\epsilon_0$ occurs, and the cold-finger effect does not operate. The remaining impact of the snow line is to slow down the radial velocity of solids located at $r<r_\mathrm{snow}$, which can aid planetesimal formation if it moves further inside and settling becomes efficient at a location that was previously inside the snow line. This phenomenon poses a less significant contribution to creating favorable conditions than the cold-finger effect, though. Thus, one can consider the scenario described in this section as a case where the cold-finger effect is suppressed during the infall stage of the disk. In fact, we find that, if the cold-finger effect is suppressed, no bulk dust-to-gas ratio enrichment takes place for any temperature scaling $\xi$. Even when adopting $\chi=100$ to achieve the highest degree of settling to increase the midplane dust-to-gas ratio, the threshold $\epsilon_\mathrm{mid,crit}=1$ is not met for an initial dust-to-gas ratio of $\epsilon_0=0.01$. Therefore, we conclude that planetesimal formation cannot occur for $\epsilon_0=0.01$ if the cold-finger effect does not operate.

\subsection{Dependence on the initial dust-to-gas ratio}\label{sec:eps0}
\begin{figure}[htp]
    \centering\includegraphics[width=\linewidth]{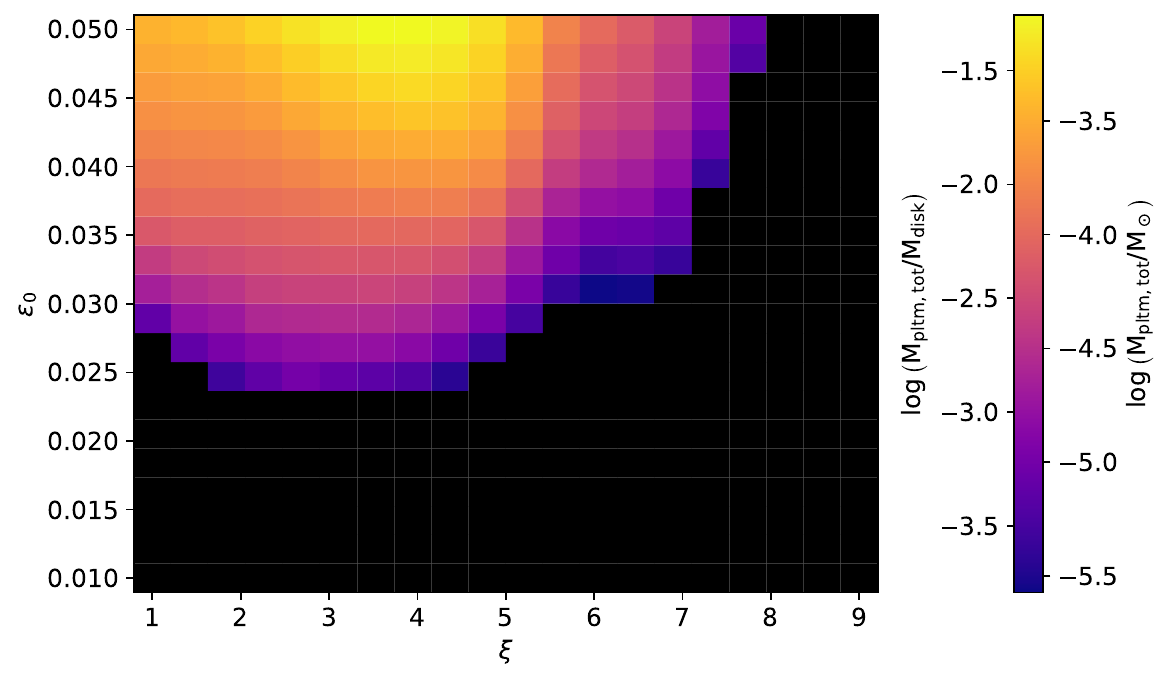}
    \caption{Like Fig. \ref{fig:R2_mpla_abar}, but with changed model assumptions. Here, the case of $\bar a= a_1$, $\epsilon_\mathrm{mid,crit}=0.5$ and a fixed value of $\chi=100$ is shown. Furthermore, the ordinate now shows the initial dust-to-gas ratio $\epsilon_0$ of the simulation, which was fixed previously.}
    \label{fig:R2_a1_mplas_epscrit}
\end{figure}%
Previous studies about the dynamics of dust during core-collapse indicate the possibility of the formation of protoplanetary disks whose vertically integrated dust-to-gas ratio initially exceeds the interstellar medium (ISM) value of 1\% and which is supplied with similarly enriched infalling material \citep{lebreuilly2020,cridland2022}. In these studies, this is caused by the decoupling of dust grains that are ${\sim}\SI{10}{\micro\meter}$ in size from the gas during the collapse of the molecular cloud. Therefore, we extend our study to include cases where the initial dust-to-gas ratio, as well as the dust-to-gas ratio of the infalling material, was increased to $0.01<\epsilon_0\leq 0.05$. Furthermore, we relax the criterion for the midplane dust-to-gas ratio to $\epsilon_\mathrm{mid,crit}=0.5$ \citep{gole2020}. Indeed, we find that simulations where such an increase of $\epsilon_0$ was applied can form planetesimals even in the absence of the cold-finger effect, given that $\chi=100$ and the disk is not too hot. The final planetesimal mass formed under these conditions is shown in Fig. \ref{fig:R2_a1_mplas_epscrit} as a function of $\xi$ and initial metallicity $\epsilon_0$. It can be seen that $M_\mathrm{pltm,tot}\sim\SI{e-4}{\solarmass}$ is reached for $\epsilon_0=0.03$ for the most optimal temperature of $\xi\sim 3.5$. In the more optimistic case of even further increased $\epsilon_0$, the final mass is further increased accordingly. For $\xi\gtrsim 5.5$, the planetesimal mass drops sharply and continues to drop for even higher $\xi$. For $\xi\gtrsim 8$, planetesimal formation no longer occurs.

\begin{figure*}[htp]
    \centering\includegraphics[width=\linewidth]{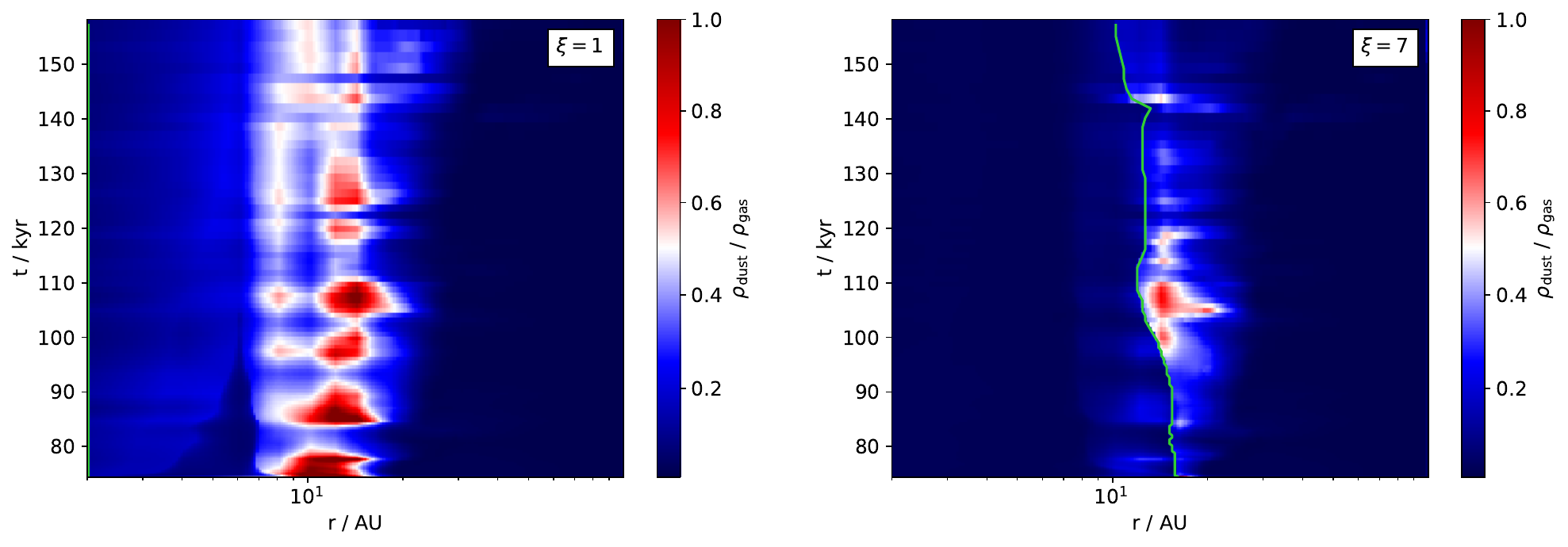}
    \caption{Like Fig. \ref{fig:R2_chi10_heatmap}, but for different model assumptions. Here, we assume the absence of small grains for phase transition purposes, i.e., $\bar a = a_1$, an increased initial dust-to-gas ratio $\epsilon_0=0.05$, a reduced critical midplane dust-to-gas ratio $\epsilon_\mathrm{mid,crit}=0.5$ and velocity fluctuations to be reduced by a factor of $\chi=100$. The left panel depicts the case of $\xi=1$ and the right panel $\xi=7$.}
    \label{fig:R2_a1_eps05_heatmap}
\end{figure*}%
The reason why the mass of formed planetesimals drops sharply for $\xi\gtrsim 5.5$ can be understood by considering the times and locations where the critical midplane dust-to-gas ratio is reached, shown in Fig. \ref{fig:R2_a1_eps05_heatmap} for $\epsilon_0=0.05$. The case of a cold ($\xi=1$) and hot ($\xi=7$) disk is presented. In the absence of the snow line in the simulation grid, a dust-to-gas ratio sufficient for the formation of planetesimals is reached at certain radial distances for some periods of time. Reaching the critical dust-to-gas ratio is not linked to the enrichment of solids at that location; rather, they are location where settling is sufficiently efficient. In fact, unlike the case where the cold-finger effect operates (see Fig. \ref{fig:R2_mass}), we find that infalling solids are not retained in the disk, and the total mass of solids does not increase, keeping the total solid-to-gas ratio at approximately the same order of magnitude, only decreasing in time due to locking of solids in planetesimals.

Settling being sufficient to reach the conditions for planetesimal formation without any enrichment of the vertically integrated dust-to-gas ratio is only possible because the initial $\epsilon_0$ is already high, while the critical dust-to-gas ratio is low. The regions where planetesimal formation is possible are therefore only set by the time and radial profiles of the $\delta$ parameters. However, strong settling is only possible outside the snow line, as grains are too small due to the high fragility of silicate grains inside the snow line. Thus, if the disk temperature is increased to the point where formation locations set by the $\delta$ parameters are inside the snow line, conditions are less favorable for planetesimal formation, as seen in Fig. \ref{fig:R2_a1_eps05_heatmap}. This causes the order-of-magnitude difference in final planetesimal mass for high temperatures in Fig \ref{fig:R2_a1_mplas_epscrit}. Therefore, the scenario most favorable for the early onset of planetesimal formation is a combination of the two cases discussed in the previous sections, which is presented in Appendix \ref{sec:appendix_combined_case}.

In reality, the high dust-to-gas ratio of the infall is not maintained for the entire $\SI{100}{\kilo\year}$ after disk formation, as the reservoir of dust from the natal molecular cloud that is able to reach the disk is eventually exhausted. A drop in infall dust-to-gas ratio is expected to occur earlier for higher $\epsilon_0$. The reason for this is that the enrichment is caused by an increased dust flux, which naturally leads to a faster depletion of the available mass. Consequently, planetesimal formation occurring later in a simulation with high $\epsilon_0$ as indicated in Fig. \ref{fig:R2_a1_eps05_heatmap} should be treated with caution, and a more realistic scenario is described by the combination of a high $\epsilon_0$ case at early times with a lower $\epsilon_0$ case toward the end of the simulation. We do not perform such combinations here in the interest of simplicity.

\section{Dependence on core-collapse initial conditions}\label{sec:othermhd}

In previous sections, all considerations were made for one specific core-collapse simulation, albeit for a family of parameters $\chi$, $\xi$, $\epsilon_0$. In order to also consider the impact of different core-collapse initial conditions, two additional models from H20 were considered: R1 and R7. Compared to the canonical run R2, R1 has a lower ratio of rotational to gravitational energy $\beta_\mathrm{rot}=0.01$ compared to $\beta_\mathrm{rot}=0.04$. On the other hand, R7 has the same value of $\beta_\mathrm{rot}$ as R2, but the initial angle between the magnetic field and the rotation axis is $\theta=\SI{90}{\degree}$, while that angle is \SI{30}{\degree} for R2. In addition, the total simulation runtime differs. Compared to $t_\mathrm{run}=\SI{82.7}{\kilo\year}$ of R2, the runtime of R1 is $t_\mathrm{run}=\SI{107.3}{\kilo\year}$, and the runtime of R7 is shorter with only $t_\mathrm{run}=\SI{27.1}{\kilo\year}$. The changes of the initial conditions of the core-collapse simulations are reflected in the change of the corresponding 1D model parameters. We will briefly present the profiles of the parameters most impactful to planetesimal formation in this section. Profiles of quantities not discussed here can be found in Appendix \ref{sec:appendix_other_profiles}.

\begin{figure*}[htp]
    \centering\includegraphics[width=\linewidth]{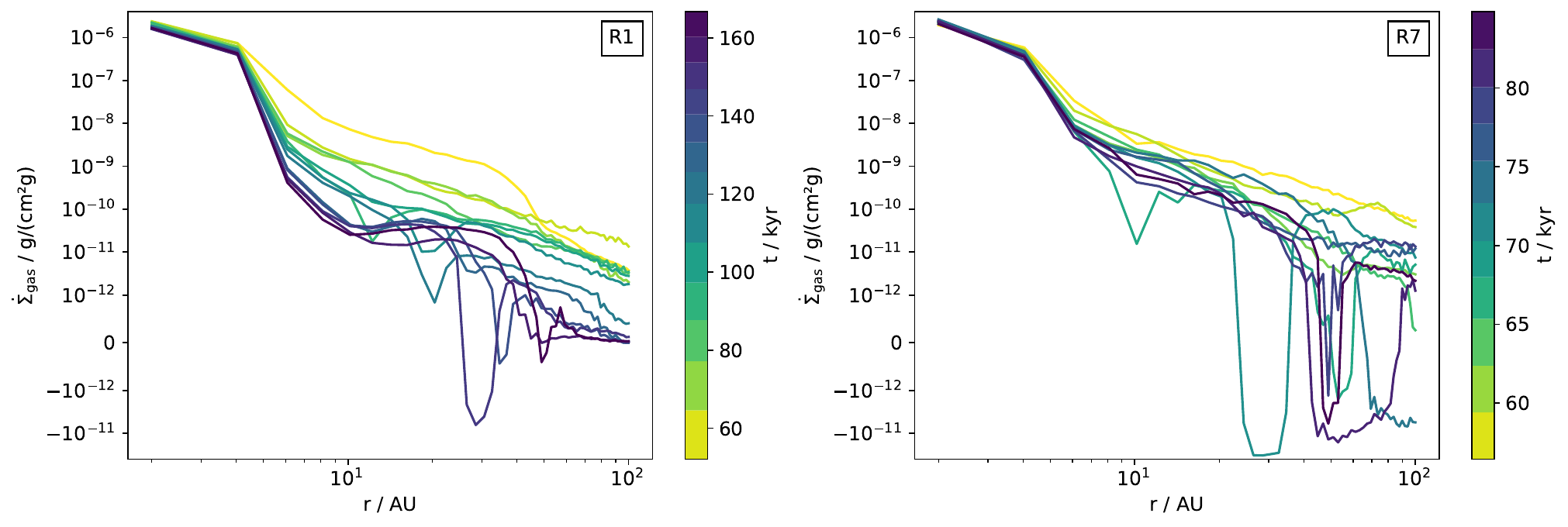}
    \caption{Like the left panel of Fig. \ref{fig:mhd_infall}, but for the core-collapse simulations R1 (left panel) and R7 (right panel) of H20 instead.}
    \label{fig:othermhd_S}
\end{figure*}%
\begin{figure*}[htp]
    \centering\includegraphics[width=\linewidth]{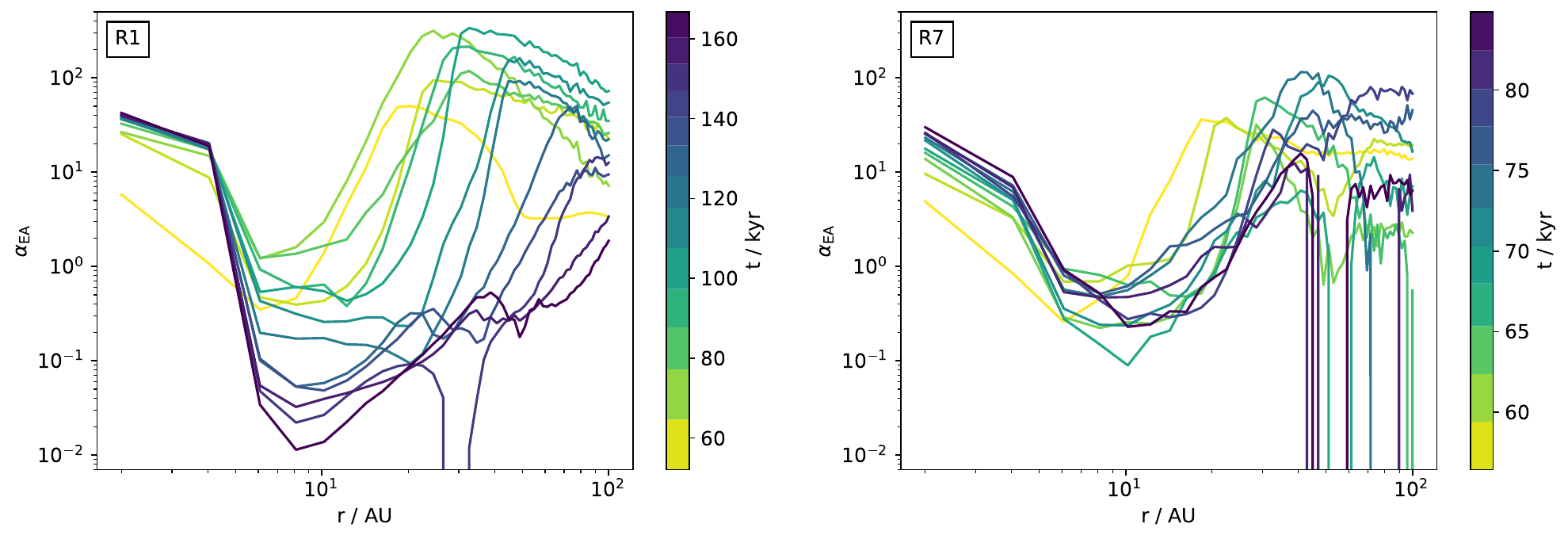}
    \caption{Like the right panel of Fig. \ref{fig:mhd_alpha}, but for the core-collapse simulations R1 (left panel) and R7 (right panel) of H20 instead.}
    \label{fig:othermhd_alphaEA}
\end{figure*}%
In Fig. \ref{fig:othermhd_S}, a comparison of $\dot\Sigma_\mathrm{gas}$ and $\dot M_\mathrm{gas}$ between R1 and R7 is shown. The total infall rate onto the disk is of the same order of magnitude for all three core-collapse simulations, because it is dominated by the strong infall onto the inner disk region, which is in turn set by the sink particle accretion. The radial profile outside the sink particle accretion-dominated area is affected by the choice of the initial core-collapse conditions, though. It is especially noticeable that the infall rate drops much more sharply with radius for R1 than for the other cases. This is related to a similarly sharp drop in $\alpha_\mathrm{EA}$ for R1, especially at late times, as presented in Fig. \ref{fig:othermhd_alphaEA}. On the other hand, the change in $\theta$ between R2 and R7 seems to have a less significant impact on the 1D model parameters. However, it remains to be investigated whether this also holds true at later times.

A change in core-collapse initial conditions also has an impact on the three parameters $\delta_r$, $\delta_z$ and $\delta_t$ related to the mean velocity fluctuations in the gas. We find that, despite the sharp drop-off of the infall profile of R1 compared to R2, the vertical mixing strength does not decline to the same extent. While $\delta_z$ reaches values as low as $\delta_z\sim\num{e-3}$ in R2, it does generally not drop below $\delta_z\sim\num{e-1}$ in R1, except toward the end of the simulation. Similarly, $\delta_t$ does not drop below $\delta_t\sim\num{e-1}$ in R1, whereas values of a few \num{e-2} are reached in R2. In contrast, in R7, the radial diffusion parameter is much larger than in R2, generally increasing with radial distance and reaching values close to 1 within the disk region, but $\delta_z$ and $\delta_t$ exhibit values generally similar to R2. This mismatch is unexpected, as the main driver of the mixing is the infalling mass, so that the mass infall rate and mixing parameters should be related. While the difference in disk size, and therefore the difference in the ratio between infalling and disk mass may contribute to this mismatch, a detailed analysis of the origin of mixing in young Class 0/I disks and its relation to mass inflow require a substantially higher resolution and remain subject to future work. Detailed profiles of these parameters are shown in Appendix \ref{sec:appendix_other_profiles}.

\begin{figure*}[htp]
    \centering\includegraphics[width=\linewidth]{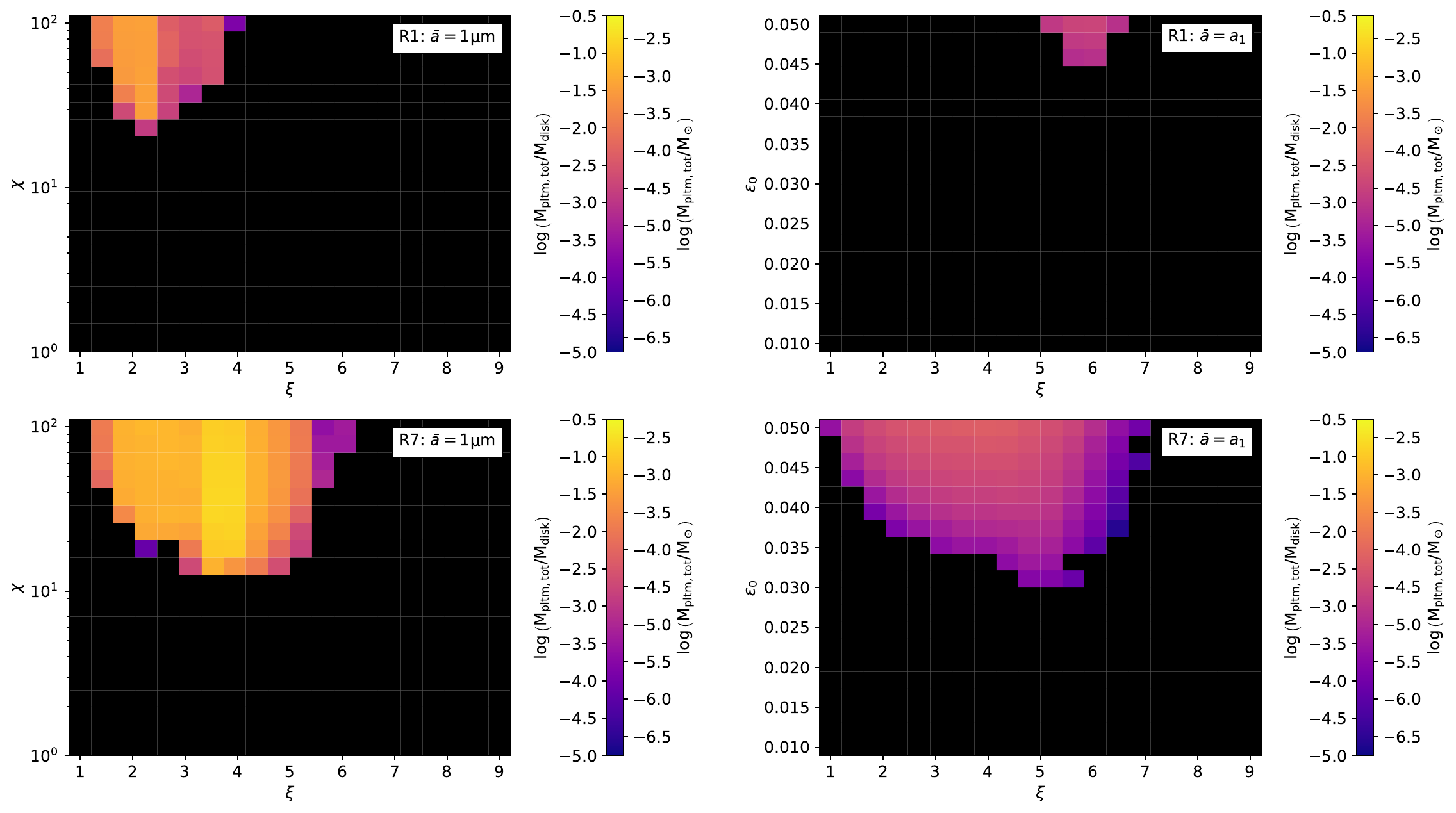}
    \caption{Total mass of formed planetesimals at the end of the simulation for runs based on the core-collapse simulations R1 (top row) and R7 (bottom row). Panels on the left-hand side correspond to Fig. \ref{fig:R2_mpla_abar} for R2, showing the mass for fixed $\epsilon_0=0.01$ and $\bar a = \SI{1}{\micro\meter}$ while varying $\chi$. The panels on the right-hand side correspond to Fig. \ref{fig:R2_a1_mplas_epscrit} for the R2 case, showing the mass for fixed $\chi=100$ and no small grains ($\bar a = \SI{1}{\micro\meter}$) while varying $\epsilon_0$.}
    \label{fig:otherR_mplas}
\end{figure*}%
Naturally, the change of parameters describing the disk in our 1D model impacts the possibility of planetesimal formation and, if applicable, the total mass of planetesimals formed. We note that, due to a difference in total run time of the corresponding core-collapse simulations, the total mass of formed planetesimals is not directly comparable between them. In Fig. \ref{fig:otherR_mplas}, we present the two scenarios considered for R2 in the previous sections for R1 and R7 each. The panels on the left-hand side present the scenario of Section \ref{sec:chi_a0}, where the cold-finger effect is effective in retaining solids in the disk, but the initial solid-to-gas ratio corresponds to the ISM value of 1\%. For both R1 and R7, only runs with $\chi>10$ and $\xi>1$ are able to produce planetesimals. In the case of R7, runs with disks as hot as $\xi=6.5$ produce planetesimals, whereas $\xi\leq4$ is required for R1. The most optimal parameter configuration results in a total mass of formed planetesimals of ${\sim}\SI{e-3}{\solarmass}$ for R1. On the other hand, for R7, planetesimals with a total mass just below $M_\mathrm{pltm,tot}\sim\SI{e-2}{\solarmass}$ may be formed. The right-hand side corresponds to the scenario where the cold-finger effect is not effective, as discussed in Sections \ref{sec:a1}, \ref{sec:eps0}. In that case, conditions are generally not favorable for the streaming instability to facilitate planetesimal formation for R1, with only a small set of configurations with $5\leq\xi\leq6.5$ and a high initial dust-to-gas ratio of $\epsilon_0>0.045$ forming planetesimals. In contrast, R7 is more comparable to R2 regarding what parameter sets allow the formation of planetesimals. Here, $\epsilon_0\geq 0.03$ allows planetesimal formation. The highest total mass is produced for $3\leq\xi\leq 5$, with $M_\mathrm{pltm,tot}\sim\SI{e-5}{\solarmass}$ over the short run time of R7 for $\epsilon_0=0.05$. In summary, velocity fluctuations have to be reduced more for R1 and R7 compared to R2 to reach favorable conditions if small grains are present for evaporation and condensation. The disk produced in the R1 MHD simulation is not favorable for planetesimal formation if the cold-finger effect is not effective in increasing the local dust-to-gas ratio at the snow line. Like R2, no runs based on the disks found in R1 and R7 lead to planetesimal formation if the cold-finger effect does not operate and $\epsilon_0 = 0.01$.

\begin{figure*}[htp]
    \centering\includegraphics[width=\linewidth]{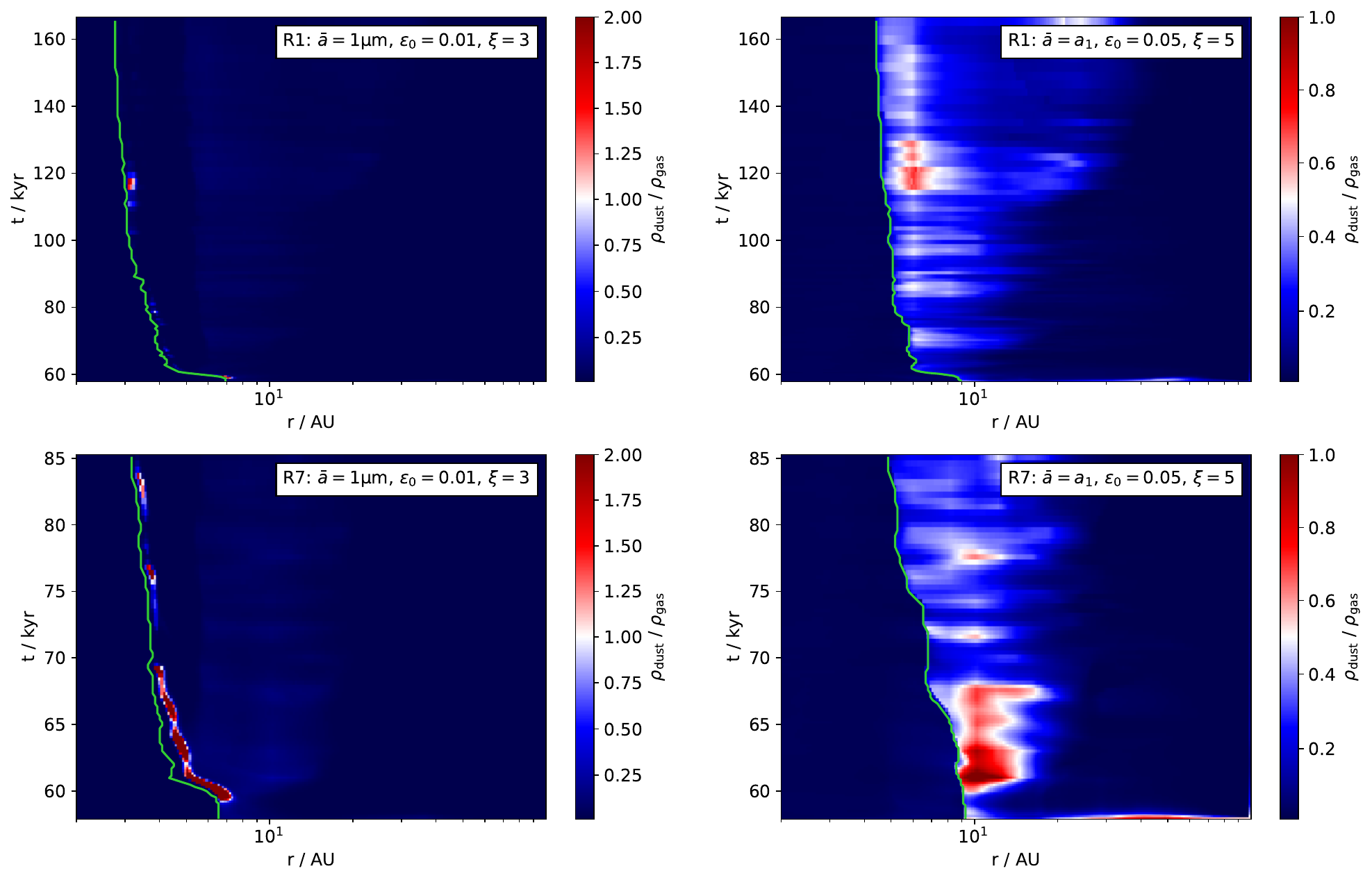}
    \caption{Midplane dust-to-gas ratio as a function of time (ordinate) and space (abscissa). The color map was chosen like in Fig. \ref{fig:R2_chi10_heatmap}, with the green solid line indicating the position of the snow line. The top row shows simulations based on the R1 core-collapse simulation and the bottom row shows simulations based on R7. Panels on the left-hand side present simulations with $\bar a = \SI{1}{\micro\meter}$, $\epsilon_\mathrm{mid,crit}=1$, $\epsilon_0=0.01$, $\chi=100$ and $\xi=3$. This corresponds to the case discussed in Section \ref{sec:chi_a0} for R2. The panels on the right-hand side show simulations with $\bar a = a_1$, $\epsilon_\mathrm{mid,crit}=0.5$, $\epsilon_0=0.05$, $\chi=100$ and $\xi=5$, which corresponds to the case discussed in Section \ref{sec:eps0} for R2.}
    \label{fig:otherR_heatmap}
\end{figure*}%
Finally, Fig. \ref{fig:otherR_heatmap} shows the midplane dust-to-gas ratio for the two aforementioned scenarios for core-collapse simulations R1 and R7. The left-hand side shows the temperature that leads to the highest total mass of planetesimals for both R1 and R7, $\xi=3$. The right-hand side of Fig. \ref{fig:otherR_heatmap} employs $\xi=5$, as this is the best-case temperature for the scenario of an absent cold-finger effect. The qualitative behavior is the same as already discussed in Section \ref{sec:eps0}. However, it can be seen that conditions for planetesimal formation are only met very briefly in the $\bar a \approx \SI{1}{\micro\meter}$ case for R1. Furthermore, for the $\bar a = a_1$ case, the regions where planetesimal formation is possible due to efficient settling are different compared to R2, which is mainly caused by the change in the dependence of $\delta_z$ on time and radial distance for those two simulations. In particular, $\delta_z$ found in R1 is much higher on average than in the other cases (see Fig. \ref{fig:othermhd_delta}), so that planetesimal formation solely caused by settling is achieved only for brief times.

\section{Dependence on the planetesimal formation criterion}\label{sec:pl_form_crit}
Up to this point, we have considered the simple criterion by \citet{da2018} to determine whether planetesimals can form under the conditions present at any given location in the simulated domain. Here, the effect of turbulence on the strength of clumping achievable through the streaming instability is only considered indirectly through a requirement on the midplane dust-to-gas ratio $\epsilon$, which is reduced by the turbulence-induced vertical mixing. However, recent works on the streaming instability in an environment with external turbulence have highlighted the importance of considering it for the study of the degree of clumping that is believed to facilitate planetesimal formation.

On one hand, a turbulent environment has been shown to limit the growth rates of the streaming instability \citep{umurhan2020} and the clumping efficiency \citep{lim2024}. Even if significant clumping occurs, turbulent diffusion might impact the gravitational collapse of the clumps into planetesimals \citep{gerbig2020, gk23}. Given the high degree of mixing induced by strong velocity fluctuations that are found in the H20 simulations, reflected by the large magnitude of the parameters $\delta_r$, $\delta_r$ and $\delta_t$ (see Fig. \ref{fig:mhd_delta}), these findings might imply that that mass of plantesimals that form is overestimated in our model, especially at low turbulent reduction factors $\chi$. On the other hand, turbulence can also aid particle concentration and thereby aid the planetesimal formation process. So-called zonal flows have been found to be produced by the MRI \citep{johansen2007,sa14}, which can lead to dust enhancements that are significant enough to trigger planetesimal formation (e.g., \citealt{xb2022}). Similar findings have been obtained for the Vertical Shear Instability (VSI) \citep{sj22}. Furthermore, turbulent clustering can lead to planetesimal formation \citep{hc20}.

\begin{figure*}[htp]
    \centering\includegraphics[width=\linewidth]{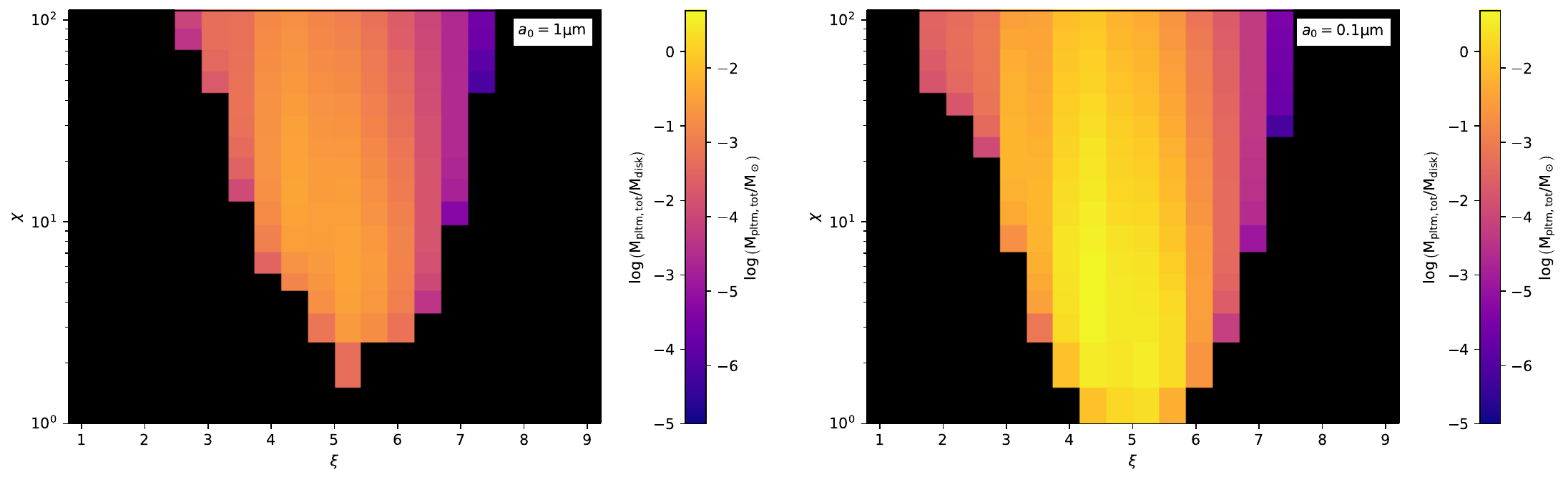}
    \caption{Like Fig. \ref{fig:R2_chi10_xi5_dtg}, but employing the planetesimal formation criterion by \citet{lim2024} (see Eqs. \ref{eq:lim1}, \ref{eq:lim2}) instead.}
    \label{fig:R2_mpla_abar_newcrit}
\end{figure*}
To investigate the impact of external turbulence on our model, we consider a planetesimal formation criterion by \citet{lim2024} in this section. They consider a 3D stratified shearing box with turbulent forcing and derive a criterion for strong clumping based on the midplane dust-to-gas ratio $\epsilon$ and the bulk dust-to-gas ratio $\Sigma_d/\Sigma_g$,
\begin{align}
    \log\epsilon_\mathrm{crit} &\simeq A\left(\log\tau_s\right)^2 + B\log\tau_s+C,\label{eq:lim1}\\
    \begin{split}
        \log\left(\Sigma_d/\Sigma_g\right)_\mathrm{crit} &\simeq A'\left(\log\delta_t\right)^2+B'\log\tau_s\log\delta_t\\
        &+C'\log\tau_s+D'\log\delta_t,\label{eq:lim2}
    \end{split}
\end{align}
with constants given by
\begin{equation*}
    \begin{split}
        A &= 0.42,\ B = 0.72,\ C = 0.37,\\
        A' &= 0.15,\ B' = -0.24,\ C' = -1.48,\ D' = 1.18. 
    \end{split}
\end{equation*}
We apply this criterion to the setup discussed in Section \ref{sec:chi_a0}, i.e., where the cold-finger effect leads to a significant accumulation of solids just outside the snow line due to the choice of small monomer sizes $a_0=\SI{1}{\micro\meter}$ and $a_0=\SI{0.1}{\micro\meter}$, with disk parameter profiles based on the R2 run of H20. Again, we investigate the total mass of formed planetesimals as a function of the temperature scaling factor $\xi$ and the turbulence reduction factor $\chi$. The results are shown in Fig. \ref{fig:R2_mpla_abar_newcrit}.

For the larger monomer size of $a_0=\SI{1}{\micro\meter}$, the difference to the model shown in Fig. \ref{fig:R2_mpla_abar} is not highly significant. When employing the \citet{lim2024} criterion, planetesimal formation is suppressed for comparatively lower temperature factors $\xi$, even at high $\chi$. Runs with $\chi\approx 10$ exhibit a total mass of planetesimals that is slightly lower than for the other criterion, but it is of the same order of magnitude. However, at $\xi\approx 5$, even runs with more turbulent conditions form planetesimals, down to $\chi=2$ compared to the previous case of $\chi=5$. For the smaller monomer size of $a_0=\SI{0.1}{\micro\meter}$, the difference to the other criterion is more substantial. More parameter sets form planetesimals with a total mass of $M_\mathrm{pltm}>\SI{e-2}{\solarmass}$, and planetesimals can form even in runs with no turbulence reduction, $\chi=1$, for $4\leq\xi\leq 6$.

A small reduction in the total mass of planetesimals for runs with $\chi\approx 10$ is expected due to restrictions on $\Sigma_d/\Sigma_g$ that were not previously considered. However, these results also show that, even though the criterion by \citet{lim2024} introduces more restrictive limits on planetesimal formation in a turbulent environment, employing it results in an overall increase in planetesimal formation capabilities, especially for $a_0=\SI{0.1}{\micro\meter}$. This is due to the fact that the criterion by \citet{da2018} is more restrictive on the size of the dust grains, strictly requiring $\mathrm{St}>0.01$, which excludes runs with small $\chi$. On the other hand, small dust grains can still be clumped strongly as long as $\Sigma_d/\Sigma_g$ is large enough according to Eq. \eqref{eq:lim2}. More turbulent runs and a smaller monomer size both imply a higher cold-finger effect efficiency, resulting in high values of $\Sigma_d/\Sigma_g$ and planetesimal formation in previously excluded runs. Despite this, we note that there are two important caveats to consider when interpreting these results. First, even though runs with a highly efficient cold-finger effect have $\Sigma_d/\Sigma_g\gg 1$, we neglect dust back-reaction and the fact that dust self-gravity will impact the scale height, so that Eq. \eqref{eq:dust_scaleheight} is no longer valid. Second, even though strong clumping by the streaming instability has been found to occur also for $\mathrm{St}<0.01$ \citep{carrera2015,yang2017,ly2021} given a high enough value for $\Sigma_d/\Sigma_g$, the smallest Stokes number data point considered by \citet{lim2024} is $\mathrm{St}=0.01$. Because their criterion is based on a fit, it remains to be confirmed whether it also holds true for $\mathrm{St}<0.01$.

We note that, if the cold-finger effect does not operate, as described in Section \ref{sec:a1}, there is no mechanism that could provide an enhancement in the bulk dust-to-gas ratio, so that $(\Sigma_d/\Sigma_g)_\mathrm{crit}$ (cfr. Eq. \ref{eq:lim2}) is never reached. Therefore, if turbulence is considered as being detrimental to the clumping of dust and subsequent planetesimal formation, planetesimals cannot form through settling alone in our model, and the efficient operation of the cold-finger effect is a necessity. Furthermore, even though we assume that planetesimals form once the criterion for strong clumping is met, the dust clumps may be prevented from collapsing gravitationally due to the turbulence and thus planetesimals may not form.

\section{Discussion}\label{sec:discussion}
\subsection{Implications for planet formation}

Early planetesimal formation in infall dominated disks has an impact on the initial conditions for planet formation. If a standard Class-II centric view is employed, the initially present mass of solids is first used for the formation of planetesimals, which occurs at disk locations where drifting pebbles are trapped. In the core accretion scenario, these planetesimals then become the seeds for the rapid accretion of pebble-sized solids to eventually form giant planets. However, once planetesimals have formed from solids that have drifted and been trapped from outer disk regions, it is unclear how much solid mass can remain to serve as accretion material in the core accretion scenario. While rapid in-place accretion of left-over pebbles at the planetesimal formation site can offer an explanation \citep{jo2023} in that case, there is no lack of solids in the outer disk if planetesimals already form during the infall stage. As shown in Fig. \ref{fig:R2_mass}, planetesimals are not formed from solids that were initially present in outer disk regions, but rather from accreted material that was retained in the disk by the cold-finger effect. Therefore, they are still readily available in the outer disk to serve as accretion material in the core accretion scenario likely taking place in later stages where infall has largely subsided.

In fact, ${\sim}\SI{e-3}{\solarmass}\approx\SI{3e2}{\earthmass}$ of planetesimals may form during the $\SI{100}{\kilo\year}$ considered in this work (see Fig. \ref{fig:R2_mpla_abar}), while still leaving ${\sim}\SI{3e3}{\earthmass}$ of dust in the disk (see Fig. \ref{fig:R2_mass}). This is up to 10\% of the disk mass in planetesimal bodies formed via the streaming instability, which is believed to form objects 10 -- $\SI{100}{\kilo\meter}$ in size \citep{simon2016,schaefer2017,ks2020}. Objects of this size are not affected by gas drag and can therefore remain in the disk to serve as the initial condition for subsequent evolution. Assuming that planets grow by accreting pebbles \citep{ok2010,lj2012} and a high accretion efficiency of ${\sim}10\%$ \citep{ormel2017}, planetary cores with a total mass of up to ${\sim}\SI{600}{\earthmass}$ may be formed, greatly exceeding the total mass of the cores of the giant planets in the Solar System and the ${\sim}20$ -- $\SI{30}{\earthmass}$ planetesimal population that is believed to have evolved into the Kuiper Belt \citep{tsiganis2005}. However, these values have a strong dependence on the choice of temperature scaling factor $\xi$ and the reduction of velocity fluctuations by $\chi$. For a less ideal choice of parameters, only ${\sim}\SI{3}{\earthmass}$ of planetesimals may form, and the cold-finger effect may not be effective at retaining infalling solids, leaving as little as ${\sim}\SI{70}{\earthmass}$ of solids available to be accreted as pebbles. In such a case, the existence of the Solar System and giant exoplanets cannot be explained by our model.

However, population studies focusing on the Class II evolutionary stage exhibit an apparent population-wide increase of the dust mass found in the disks between $0.5$ -- $\SI{1}{\mega\year}$ and $\SI{2}{\mega\year}$ \citep{testi2022}. This can be explained by the formation of a secondary generation of dust though fragmentation during planetesimal-planetesimal collisions, which requires the early formation of planets that dynamically excite the planetesimal population. Notably, the dust increase observed by \citet{testi2022} is consistent with fast planetesimal formation followed by core formation at ages $<\SI{500}{\kilo\year}$ \citep{bernabo2022}. This scenario matches well with our findings of fast and efficient planetesimal formation, given a favorable choice of parameters.

\subsection{Comparison to observational findings}

Observations of dust emission in protoplanetary disks suggest an early occurrence of substructure related to planet formation. In particular, the extensively studied disk of HL Tau \citep{hltau} shows an abundance of rings while still being embedded in a remnant envelope. The same holds true for Oph IRS 63, where multiple rings have been observed \citep{seguracox2020}. In addition, the recent ALMA Large Program "eDisk" \citep{ediskI} is focused specifically on the detection of substructure of young, embedded disks, observing 12 Class 0 and 7 Class I sources. They find that substructures like rings and spirals are absent from the Class 0 sources, but not from all Class I sources. If there are truly no substructures in Class 0 disks, i.e., they are not obscured by the high optical depth of the emission, they need to start forming during the transition between the Class 0 and I stage. This relates with our findings that planetesimals start forming only at later times, when infall induced velocity fluctuations diminish and conditions become favorable for settling. As the ages of Class I sources with observed substructures are a few $\SI{100}{\kilo\year}$, the planetesimals formed during the $\SI{100}{\kilo\year}$ in the simulations presented here can serve as the building blocks for the first planetary cores that can carve structures such as a gap, which strengthens the idea of an early onset of planet formation further.

Observations find indications of high temperatures at disk scales for young solar-type protostars. In several of these young disks, the dust brightness temperatures reach values of a few hundreds of Kelvin in the inner \SI{20}{\astronomicalunit} \citep{vanthoff2018, zamponi2021, vanthoff2024}, while the observation of hot corinos suggest the water snow-lines could be located at radii as large as ${\sim}\SI{100}{\astronomicalunit}$, beyond the disk boundary \citep{belloche2020}. Resolved observations of complex organic molecules (COMs) have shown a dependency between the spatial extent of the COMs and their excitation temperatures \citep{maury2014}, in agreement with expectations from thermal desorption in a warm medium \citep{busch2022}. The dust temperatures measured in these young disks are not consistent with the sole passive heating, however, and suggests the presence of accretion or viscous heating \citep{maureira2022, takakuwa2024}. Neither of these heating sources are accounted for in the H20 simulations, which may explain the fact that the disks that arise in these simulations are colder than what observations suggest (see Fig. \ref{fig:mhd_T}). However, the impact of accretion heating is uncertain, so that theoretical modeling of the disk temperature in young disks remains challenging \citep{hennebelle2020b,lebreuilly2024}. We take the uncertainty of the disk temperature into account by introducing a temperature scaling factor $\xi$. Obtaining better constraints for the accretion luminosity in the future would, in turn, place constraints on this factor from the theoretical side.

These observational findings regarding the temperature are matched by our $\xi=9$ model in order of magnitude, where the snow line is located at ${\sim}\SI{100}{\astronomicalunit}$ (see Fig. \ref{fig:R2_snowline}), which is beyond the disk boundary located at ${\sim}\SI{20}{\astronomicalunit}$. For a disk of this temperature, our model implies that planetesimal formation does not occur. The reason for this is that the cold-finger effect cannot lead to a dust-to-gas ratio enhancement if the snow line is not located in the disk, as shown in Fig. \ref{fig:R2_mpla_abar}, and silicate grains are too fragile to grow to sizes where settling alone can trigger the streaming instability even for high metallicity, as shown in Fig. \ref{fig:R2_a1_mplas_epscrit}. Conversely, high resolution observations of HL Tau reveal a significant reservoir of water vapor, where the lower mass limit is estimated to be ${\sim}\SI{7.8e-4}{\earthmass}$ \citep{faccini2024}. These observations suggest an upper limit to the snow line location of ${\sim}\SI{17}{\astronomicalunit}$, which approximately matches the location found in our $\xi=7$ model. While detailed models of the snow line in HL Tau do not exist at the moment, it being at a radial distance that is favorable for planetesimal formation is plausible given these constraints. Consequently, even if young disks are too hot initially to form planetesimals, they could cool down to favorable conditions within the time frame considered in our model.

The typical mass accretion rates for solar-type Class I protostars, when measured at disk scales using infrared spectroscopy, are highly uncertain, but are typically found to be of the order of a few ${\sim}\SI{e-7}{\solarmass\per\year}$ at most \citep{fiorellino2021}. However, the mass infall rates measured for the gas infalling from the envelope to the disk scales are usually an order of magnitude larger, which is in line with the values found in our model (see Fig. \ref{fig:mhd_infall}). Our model indicates that the infalling mass is advected toward the central star rapidly, which is in agreement with higher resolution models, where infalling gas is accreted onto the star through an accretion channel at several pressure scale heights \citep{lee2021}. However, this is caused by the lack of angular momentum of the infalling gas, which is caused by magnetic braking in the envelope. If the braking was weaker, the gas may be retained in the disk, and the high infall rate could suggest that these young disks grow in mass. This could lead to the development of favorable conditions for planetesimal formation in the form of a very optically thick midplane where temperatures could be much lower than the ones measured from the millimeter dust continuum emission, so that the snow line would be located inside the disk. On the other hand, such mass loading could also suggest that most disks quickly become too massive and thus gravitationally unstable.

\subsection{Potential role of other chemical species}

Our work only considers the water snow line as a potential site of solid enrichment due to the cold-finger effect. In principle, including additional chemical species and their evaporation and condensation could create secondary solid enhancements at the location of their evaporation fronts. The impact on the midplane dust-to-gas ratio enhancement is then limited by two aspects.

First, the cold-finger effect retains part of the fraction of the dust mass that is made of the respective chemical species. Therefore, solid enhancement is only significant for species with an abundance comparable to water. From an observation perspective, the abundances of volatile molecules in young disks are not well constrained and are influenced by a complex interplay of physical and astrochemical processes in Class II disks (for reviews, see \citealt{ob2021,oberg2023}). Only recently has the James Webb Space Telescope (JWST) opened a new window into the ice inventory of these disks. In addition to H$_2$O, the main volatile carbon carriers, CO and CO$_2$, have been systematically observed in the ice phase of edge-on disks for the first time (e.g., \citealt{sturm2023}). However, uncertainties remain regarding the implications of these measurements. For instance, \citet{bergner2024} suggest that CO could survive in the ice phase even within its snowline, due to its trapping in H$_2$O and CO$_2$ ices. We can gain some insights by assuming that the molecular composition
of Class II disks is inherited from the earliest phases of star formation, in the so-called inheritance scenario. The composition of the protostellar envelope and the stellar abundance after disk dissipation are not identical, as the stellar composition is altered by the accretion of disk material during the disk's lifetime. However, this effect is only minor \citep{hb2023}. Various astrochemical models (e.g., \citealt{eistrup2016,ow2019}) and observations of ices in the envelopes of low-mass YSOs (e.g., \citealt{boogert2015}) predict similar CO and CO$_2$ abundances in disks under the assumption of chemical inheritance. On average, these abundances are only about a factor of 4 lower than that of H$_2$O, so that an enhancement of the dust-to-gas ratio due to the cold-finger effect operating at the respective snow lines may be non-negligible. We note that the chemical composition of gas and ices in disks can vary throughout the system's evolutionary history, for instance, due to episodic accretion events, which can result in a total or partial reset of the chemistry (e.g., \citealt{eistrup2016}), with significant implications for planet formation \citep{pacetti2022}. Additionally, the relative abundances of chemical species vary across stars of different metallicities \citep{bb2020}. Addressing the impact of these effects, and exploring alternative scenarios to chemical inheritance, requires dedicated investigation.

Second, the resulting solid enhancement depends on the location of the corresponding evaporation front, and the solid flux through that location. Depending on the assumed density and binding energy in mixed ices, the condensation temperature for CO is between 17 and 30 K \citep{miotello2023}. Therefore, for disk temperatures that allow for the formation of planetesimals at the water snow line, the CO snowline would fall outside the disk. However, CO could play a role in colder disks, where the water snow line lies at a radial distance smaller than the inner edge of our simulation grid. The same applies to other highly volatile species, such as N$_2$ and O$_2$, which have similar condensation temperatures. For CO$_2$ on the other hand, the snow line may be located in warmer regions, situated between those of H$_2$O and CO. For example, \citet{eistrup2016} indicate condensation temperatures of \SI{88}{\kelvin} for CO$_2$, while \citet{zhang2015} calculate temperatures ranging from \SI{60}{\kelvin} to \SI{72}{\kelvin} for CO$_2$. This means that it could lie inside the disk and our computational domain in the cold case of $\xi=1$ and the case of a temperature better suited for the formation of planetesimals at the water snow line, like $\xi=5$ (see Fig. \ref{fig:R2_mpla_abar}). In the case of a warm disk like the one described by $\xi=5$, the solid flux through the corresponding snow line may be too low for considerable solid enhancement, though.

Volatile species may not be the only candidate for a cold-finger-like effect. For the Solar System, it is found that comets and the ISM are more enriched in carbonaceous material than the rocky bodies in the inner Solar System. Consequently, it has been suggested that this carbonaceous material must have returned to the gas phase within the so-called soot line (\citealt{vanthoff2020,vanthoff2024}, see also review by \citealt{pontoppidan2014}). While the exact composition, and thereby the condensation temperature, of this material is still under investigation, recent studies indicate that the soot line is located within the water snowline, between 300 and \SI{400}{\kelvin} (e.g., \citealt{gt2017}). If young, embedded disks are very hot, evaporation and subsequent condensation at the soot line could provide a solid enhancement and lead to planetesimal formation similar to the cold-finger effect at the snow line. Unlike the planetesimals formed at the water snow line, which are potentially very water-rich if formed through the cold-finger effect, these planetesimals would then be enriched in carbonaceous material.

\subsection{Core-collapse simulation caveats}\label{sec:discussion_mhd}
The core-collapse simulations performed by H20 integrating nonideal magnetohydrodynamics are computationally expensive and were designed to investigate global properties of protoplanetary disks. While the adaptive mesh refinement they employed produces a highly resolved disk region when compared to the non-refined cell size of $\SI{256}{\astronomicalunit}$, a resolution of $\SI{1}{\astronomicalunit}$ introduces challenges when one is concerned with more detailed models of quantities that are not of global nature, but vary across the disk. The resolution cannot be considered sufficient for a detailed study of planet formation.

The resolution is too low to resolve the injection scale of the MRI, which is one of the typical sources of turbulence considered in the literature and leads to outward angular momentum transport in Class II disks, corresponding to a positive value for $\alpha_\mathrm{SS}$ (for a review, see \citealt{lesur2023}). Due to the absence of this instability, the velocity fluctuations that contribute to the Reynolds part of $\alpha_\mathrm{SS}$ arising in the disks are a combination of the accretion shock and high numerical viscosity introduced by the large Cartesian grid cells. At higher resolution, there is a possibility that the MRI would cause turbulence ultimately leading to a positive value of $\alpha_\mathrm{SS}$, differently from what we find at the current resolution. Higher resolution simulations of core-collapse have been performed by \citet{mauxion2024}, reaching a radial resolution as low as \SI{e-2}{\astronomicalunit} and a vertical resolution of 10 cells per scale height at the inner boundary. Employing a lower mass-to-flux ratio than H20, but a lower rotational over gravitational energy ratio, they initially find the same very vigorous infall of ${\sim}\SI{e-5}{\solarmass\per\year}$ onto a small disk with a radial extent of ${\sim}\SI{10}{\astronomicalunit}$. This infalling material is also deficient in angular momentum and is advected to the star in the upper disk layers. However, they find the dynamics in the bulk of the disk to be dominated by gravitational instability removing angular momentum at small radii. Additionally, they observe a second phase in the disk evolution, where the disk expands in size by an order of magnitude due to the accretion of material with high specific angular momentum, and gravitational instability seizes. Therefore, studies should be conducted in the future, aimed at providing a better and more comprehensive understanding of the early dynamics of protoplanetary disks, depending on the initial conditions of the natal cloud.

Moreover, because the central star itself cannot be resolved at such resolution, a sink particle has to be used. As a consequence, disk quantities do not capture accurate physics within the sink particle accretion radius of $\SI{4}{\astronomicalunit}$, so that we cannot extend our investigation to the inner disk regions. The effects of the sink particle are not limited to the inner disk, though. The specifics of the employed accretion model affect global properties of the disk like total mass and cutoff radius \citep{hennebelle2020a}. Furthermore, with the gas pressure scale height being barely resolved, no distinction can be made between the dynamics in the disk midplane, where planetesimal formation takes place, and the upper layers mostly affected by the accreted material. However, this distinction might be crucial for the consideration of vertical mixing and maximal dust grain sizes, which is only approximated here by introducing a reduction factor $\chi$ to the velocity fluctuations felt by the dust. Additionally, the snow line is not located at a constant radius if a vertical temperature gradient is considered. Magnetically braked material entering the disk and being rapidly advected would do so at the upper disk layers. Therefore, it evaporates its ice mantle and is subject to the cold-finger effect at a different radial distance than material being accreted through the disk midplane, where radial speeds could be much lower, reducing the efficiency of the cold-finger effect. This affects the relationship between the temperature scaling parameter $\xi$ and the location where solid enrichment can occur.

The runtime of our 1D simulations is limited to the runtime of the core-collapse simulations, which is short compared to the lifespan of a protoplanetary disk. This limits the evolutionary stages we can apply our model to. In particular, we cannot investigate the possibility and efficiency of planetesimal formation in older Class I disks, which would be characterized by a decrease in the infall rate. A decrease in the infall rate would, in turn, create more favorable conditions due to smaller velocity fluctuations allowing dust grains to grow larger and settle more efficiently. A treatment of disks during this evolutionary stage requires extrapolating the mass infall rate and the angular momentum of the infalling mass for an accurate description of $\dot\Sigma_\mathrm{ext}$ and $\alpha_\mathrm{EA}$. Furthermore, the time evolution of $\alpha_\mathrm{SS}$ would need to be considered as it plays a more significant role for older disks. This is subject to future work.

Lastly, the core-collapse simulation themselves employ model assumptions that rely on poorly constrained parameters. The main uncertainties are related to the magnitude of the accretion luminosity, that is, the fraction of gravitational energy that is radiated away and contributes to that luminosity (e.g., \citealt{kenyon1995,evans2009}). It has large effects on the disk temperature, structure and gravitational stability \citep{krumholz2007,offner2009,lebreuilly2024}. The dust grain size distribution during the collapse is also crucial as it greatly influences Ohmic resistivities, but dust is not included in the simulation itself. Growth processes can alter the grain size distribution during the collapse \citep{lebreuilly2023,valluccigoy2024}, which a full treatment of nonideal MHD would need to consider. However, the impact of growth on the dust size distribution in the envelope is still highly debated \citep{guillet2020,bate2022,silsbee2022}.

\subsection{1D Model caveats}\label{sec:discussion_1D_caveats}
In the interest of simplicity and computational feasibility, we use a simplified treatment to consider the dust evolution and potentially resulting planetesimal formation. Our model does not contain direct treatment of the magnetic field, and we only include its effects on dust evolution through the calculation of the model parameters $\alpha_\mathrm{EA,SS}$. While the impact is likely minor, this neglects a component in the force balance relating to the azimuthal gas velocity, which sets the drift speed of the dust and can create dust traps. On the time scales investigated here, dust drift does not play a significant role, but it could be enhanced in the magnetically supported region directly exterior to the Keplerian disk due to the slower rotation speed of gas in this region. This could result in an accumulation of dust. We do not consider this effect here.

Our dust model considers only the radial dimension, so that vortices and spirals that can serve as dust traps cannot be modeled. Spirals are a phenomenon found in marginally stable, self-gravitating disks that might present an alternative pathway to planetesimal formation in Class 0/I protoplanetary disks and their growth \citep{rice2004}. However, their consideration would require a 2D treatment and more detailed modeling of dust growth and evolution. In fact, \citet{bhandare2024} find, using a 2D treatment, that meridional flows can retain and efficiently mix dust grains in the inner disk regions and the outer disk edge. Additionally, outflows can transport dust up to \SI{100}{\micro\meter} in size from the inner to the outer disk. Furthermore, the limit created by the necessity for gravitationally stable disks prevents us from extending our study to a wider range of core-collapse initial conditions. In particular, we could not extend our study to simulations with a higher initial mass-to-flux ratio, because the disks produced in such simulations are generally not stable against self gravity. However, a higher mass-to-flux ratio impacts the angular momentum of infalling material as magnetic braking is less severe, which may have an impact on the possibility and efficiency of planetesimal formation.

Ultimately, how representative gravitationally stable Class 0/I disks are of a general sample is an open question that has also not been answered definitively by observations. Indications of gravitational instability have been found in the late Class I disk around HL Tau \citep{bi20}, and surveys suggest that young disks are massive and compact \citep{tychoniec2018,maury2019}. Despite that, while some structures have been observed in some Class I disks \citep{seguracox2020,ediskI} that suggest the existence of potential instabilities to create such dust features, the examples are very scarce and most young disks are featureless at the spatial resolution we can observe. This may suggest that gravitational instability is not a ubiquitous phenomenon. Moreover, most protostellar disk masses inferred from observations are very poorly constrained, as the dust properties are oversimplified or unknown (e.g., the size distribution, see \citealt{lebreuilly2020}), and radiative transfer effects are neglected when estimating the dust masses, despite young embedded disks being largely optically thick at millimeter wavelengths. Therefore, disk masses extrapolated from dust continuum observations must be interpreted with care (see \citealt{tung2024}) if measurements of the kinematic mass are not possible.

In our parameter study, we investigate the impact of the core-collapse simulation uncertainties (see Section \ref{sec:discussion_mhd}) using scaling factors for temperature ($\xi$) and velocity fluctuations ($\chi$). This allows us to obtain first-order results and investigate the impact of temperature and turbulence on planetesimal formation at low computational cost, but only serves as a rough approximation. As we find that the resulting mass of planetesimals are sensitive to the disk temperature, more detailed radiative physics including heating terms from the central star and accretion should in principle be applied to the core-collapse simulations themselves. Additionally, a more self-consistent treatment of the gas scale height and vertical settling is necessary. Overall, this would lead to more accurate radial profiles of the temperature, as the change of the radial profile is not taken into account by the parameter $\xi$. In simulations where the cold-finger effect can operate, planetesimal formation takes place in a narrow region outside the snow line, so that for a given snow line location, a change in the radial profile is likely not significant for the fragmentation barrier of the planetesimal forming grains. While the radial profile has an impact on the global picture of the fragmentation barrier and resulting dust grain sizes, drift is not significant on the time scales we consider in this work, so that the planetesimal formation region would be unaffected. However, the radial profile may affect simulation where planetesimals form through settling alone (see Sect. \ref{sec:a1}), as the formation region is wider in that case, and further work is required to investigate this.

Including more detailed radiative physics will not only affect the radial temperature profile and resulting scale height, but is likely to have an impact on the overall structure of the formed disk. For example, a higher temperature generally stabilizes disks against gravitational instability (see Eq. \ref{eq:toomre_Q}), so that a range of MHD parameters that were excluded from this work due to forming gravitationally unstable disks could lead to stable disks instead. On the other hand, the inclusion of more physics comes at significant computation cost. For example, \citet{lebreuilly2024} model both radiative transfer and stellar radiative feedback in the \texttt{RAMSES} code while investigating the collapse of a molecular cloud, but the simulated time of the individual sinks does not exceed \SI{30}{\mega\year} for the majority of sinks, which is insufficient for an analysis like presented in this work.

From an observational point of view, high resolution observations like those performed by \citet{faccini2024} for the HL Tau disk, obtaining information about the spatial distribution of water vapor, could provide information about the location of the snow line and the disk temperature. If such studies are extended to young Class 0/I disks, they could provide insight on the effectiveness of the cold-finger effect in these disks. However, this may prove difficult given that most young disks are optically thick at millimeter wavelengths in the inner regions. Furthermore, higher resolution simulations need to be performed to investigate whether velocity fluctuation in the midplane of the disk are indeed a factor 10 or 100 lower in magnitude than in the disk upper layers.

We employ several simplifications in our treatment of dust evolution and planetesimal formation. We assume spherical grains of homogeneous density and composition-independent size, neglecting how porous grains could grow larger and settle more efficiently \citep{krijt2016}, which could alleviate the need for significantly reducing the velocity fluctuations felt by the dust. Instead of considering the size change of grains due to evaporation and condensation, we change the fragmentation velocity based on grain composition, which also reduces the size of the silicate grains compared to grains with an ice mantle. Furthermore, the two-population model we employ utilizes several constants ($f_f$, $f_d$, $f_m$; see \citealt{birnstiel2012}) that were calibrated to match this simplified model to more detailed dust coagulation simulations. However, these simulations considered a disk with $M_\mathrm{disk}=\SI{0.1}{\solarmass}$ around a solar-mass star with no external infall and parameterize the turbulence with a single viscous $\alpha$ parameter. While our model takes into account the mass variability of the central star for the calculation of the growth limit of the dust grains, it was assumed that the numerical values found for these constants still serve as a good approximation for the Class 0/I disks we consider. Given that in our case, the host star has only grown to $M_\star\sim\SI{0.4}{\solarmass}$ at the end of the R2 simulation and that the disk is subject to a significant influx of mass, detailed simulations of dust growth and dynamics in such an environment are required to confirm these assumptions in the future.

Even though we find $\epsilon=\Sigma_\mathrm{dust}/\Sigma_\mathrm{gas}\gg 1$ in some runs where the cold-finger effect is highly efficient at retaining infalling dust, we do not consider the dust back-reaction on the gas. While a high bulk dust-to-gas ratio can be indicative of considerable back-reaction slowing down gas accretion rates, dust grains need to be large enough to result in a significant impact on gas dynamics \citep{garate2020}. In fact, it is required that $\mathrm{St}\epsilon/(\epsilon+1)\sim\alpha$ \citep{kanagawa2017,dipierro2018}. Employing $\alpha\approx\alpha_\mathrm{EA}\sim\num{e-1}$ and $\epsilon\sim 10$ to match the scenario in Fig. \ref{fig:R2_xi5_chi1} at $r=\SI{10}{\astronomicalunit}$, dust grains would be required to have a Stokes number of $\mathrm{St}\sim\num{e-1}$ to change gas dynamics, which is two orders of magnitude higher than the Stokes number of the largest grains in this scenario. While this may indicate that the dust back-reaction on the gas does not change our results significantly, more involved hydrodynamical calculations would be required to understand how it would affect the accumulation of solids at the snow line due to the cold-finger effect in detail.

We form planetesimals in our 1D framework by using a parameterized criterion for the streaming instability in Class II disks, rather than modeling it self-consistently. It is based on simulations without infall or strong turbulence, and it remains to be investigated how the conditions in early stage disks impact the clumping of solids by the streaming instability and how the minimal initial solid concentration and size for efficient planetesimal formation is affected by it. Furthermore, in order to form planetesimals, we find that, among other necessary disk conditions, the magnitude of midplane velocity fluctuations needs to be reduced by a factor of 10 or 100 compared to the values found directly in the core-collapse simulations. If such low velocity fluctuations do not reflect the reality during the first $\SI{100}{\kilo\year}$ after disk formation, they are likely to still be reached at some point before Class II is reached, so that early substructures can be formed by potential early planetary cores. This is because the infalling material is the main source of velocity fluctuations here, and it must diminish over time to the point when it eventually ceases. Therefore, the conditions that correspond to an arbitrary reduction here might be reached naturally at later times, which remains subject to future work.

\section{Conclusions}\label{sec:conclusions}
We performed 1D protoplanetary disk simulations investigating the dust distribution found in the disks arising in the core-collapse simulations performed by H20. The model we used was designed to investigate the possibility of planetesimal formation during the Class 0/I evolutionary stage of protoplanetary disks, i.e. young disks subject to significant accretion of material from their envelope. We draw the following conclusions.
\begin{itemize}
    \item Velocity fluctuations found at the resolution employed in the core-collapse simulations from H20 do not lead to outward angular momentum transport, i.e., $\alpha_\mathrm{SS}<0$. The evolution of the disks are governed by strong inward advection, which is caused by the infalling mass being strongly deficient in angular momentum.
    \item Employing disk quantity profiles obtained directly from core-collapse simulations performed by H20, the resulting disk conditions do not lead to the formation of planetesimals. However, adapting the temperature to match core-collapse simulations where more realistic temperature models were used, $\xi\sim 5$, as well as reducing the velocity fluctuations felt by the dust by a factor $\chi=10$ to mitigate effects likely caused by the poor resolution, we find that planetesimals with a total mass of up to $M_\mathrm{pltm,tot}\sim\SI{e-3}{\solarmass}$ may be formed in Class 0/I disks if strong clumping of dust leads to gravitational collapse. This holds true even if the disks are not initially enriched in solids, and an ISM-motivated initial dust-to-gas ratio of $\epsilon_0=0.01$ is used, and for both strong clumping criteria we investigated. This has strong implications for the onset of planet formation and the creation of giant planet cores.
    \item The assumption of the average particle size $\bar a$ for evaporation and condensation plays a significant role due to the rapid advection of disk material caused by the strong infall. While planetesimals form for a range of parameters for $a_0\lesssim\SI{1}{\micro\meter}$, the same does not hold true if disk processes lead to a sweep-up and effective removal of small dust grains. In such a case, planetesimal formation is suppressed fully for ISM-motivated initial dust-to-gas ratio of 1\%. Planetesimal formation can then only occur to a significant extent if the velocity fluctuations are reduced significantly ($\chi=100$), the disk is not too hot ($\xi\leq 5$) and the disk is initially enriched in solids ($\epsilon_0>0.03$) and strong clumping due to the streaming instability already occurs at a midplane dust-to-gas ratios of 0.5. Even then, planetesimals only form under the assumption that the strong turbulence found in young disks does not prevent clumping by the SI.
    \item The highest total planetesimal mass that can be formed during the ${\sim}\SI{100}{\kilo\year}$ run time of the core-collapse simulation under the most optimistic conditions is $M_\mathrm{pltm,tot}=\SI{e-2}{\solarmass}$, which is approximately one disk mass at the end of the core-collapse simulation.
    \item Solids used for the build-up of plantesimals do not decrease the dust content within the disk, as infalling material from the natal molecular cloud continuously replenishes them. Formation of planetesimals is episodic and occurs generally via two mechanisms, corresponding to different formation locations. First, the cold-finger effect can lead to efficient planetesimal formation near the snow line given a sufficiently $\bar a$ even for moderately strong settling. If settling is highly efficient and the disk initially enriched in solids, planetesimals can be formed solely due to settling, which leads to a broader area of planetesimal formation, whose location is then set only be the velocity fluctuations arising during core-collapse
    \item A smaller rotation to gravitational energy fraction, corresponding to a more compact disk, creates much less favorable conditions for planetesimal formation, mainly due to a greatly reduced settling efficiency. On the other hand, a change in the initial angle between the magnetic field and the disk rotation axis does not impact the results significantly.
\end{itemize}
\begin{acknowledgements}
The authors acknowledge financial support from the European Research Council via the ERC Synergy Grant ECOGAL (grant 855130). Part of this work is
funded by the DFG via the Heidelberg Cluster of Excellence STRUCTURES
in the framework of Germany's Excellence Strategy (grant EXC-2181/1 --
390900948). R.S.K. also acknowledges funding from the German Ministry for Economic Affairs and Climate Action in project ``MAINN'' (funding ID 50OO2206). The authors acknowledge support by the High Performance and Cloud Computing Group at the Zentrum für Datenverarbeitung of the University of Tübingen, the state of Baden-Württemberg through bwHPC and the German Research Foundation (DFG) through grants INST 35/1134-1 FUGG, 35/1597-1 FUGG and 37/935-1 FUGG. Additionally, R.S.K. is grateful for data storage at SDS@hd funded through grants INST 35/1314-1 FUGG and INST 35/1503-1 FUGG. And R.S.K. thanks the Harvard-Smithsonian Center for Astrophysics and the Radcliffe Institute for Advanced Studies for their hospitality during his sabbatical, and the 2024/25 Class of Radcliffe Fellows for highly interesting and stimulating discussions. A.M. acknowledges support the funding from the European Research Council (ERC) under the European Union's Horizon 2020 research and innovation programme (Grant agreement No. 101098309 - PEBBLES). G.R. acknowledges funding from the Fondazione Cariplo, grant no. 2022-1217, and the European Research Council (ERC) under the European Union's Horizon Europe Research \& Innovation Programme under grant agreement no. 101039651 (DiscEvol). Views and opinions expressed are however those of the author(s) only, and do not necessarily reflect those of the European Union or the European Research Council Executive Agency. Neither the European Union nor the granting authority can be held responsible for them.
\end{acknowledgements}
\bibliography{references}
\begin{appendix}

\section{1D model boundary conditions}\label{sec:appendix_boundaries}
When choosing the inner boundary condition for the background gas, we considered multiple factors. First, the inner edge of the simulation grid does not correspond to the physical inner edge of the protoplanetary disk. Therefore, the condition has to be chosen in a way that is consistent with the existence of the inner disk inside the inner edge of the computational domain. Furthermore, most of the infalling material leaves the computational domain through the inner edge. Thus, we choose the boundary condition such that the mass flux is not impeded and mass is not artificially accumulated at the inner edge. In the following, to simplify notation, $\Sigma=\Sigma_\mathrm{H+He}$, unless specified otherwise.

For the background gas, the viscous contribution to the radial velocity at the inner boundary can be inferred from Eq. \eqref{eq:1D_adv_diff} to be
\begin{equation}
    v_0 = -\frac{3}{\Sigma_0\sqrt{r_0}}\left(\frac{\nu_0}{\sqrt{r_0}}\frac{\Sigma_1r_1-\Sigma_0r_0}{r_1-r_0}+\Sigma_0r_0\frac{\frac{\nu_1}{\sqrt{r_1}}-\frac{\nu_0}{\sqrt{r_0}}}{r_1-r_0}\right),\label{eq:inner_boundary_v}
\end{equation}
where the lower indices $0,1$ refer to the values at the inner boundary cell and the one adjacent to it, respectively, and $\nu$ denotes the kinematic viscosity. Imposing the common zero gradient boundary condition $\partial(\Sigma r)/\partial r$ reduces the radial velocity at the inner boundary, because the first term in Eq. \eqref{eq:inner_boundary_v} is zero. To alleviate that, we instead impose
\begin{equation}
    \left.\frac{\partial}{\partial r}\left(\Sigma^{i+1}\nu^i\sqrt{r}\right)\right|_0 = \frac{\nu_0^i}{\sqrt{r_0}}\frac{\Sigma_1^ir_1-\Sigma_0^ir_0}{r_1-r_0}+\Sigma_0^{i+1}\frac{\frac{\nu_1^i}{\sqrt{r_1}}-\frac{\nu_0^i}{\sqrt{r_0}}}{r_1-r_0},\label{eq:inner_boundary}
\end{equation}
where the high indices indicate the time step, i.e., high index $i+1$ indicates the time step to be calculated, whereas $i$ indicates the previously calculated time step. Equation \eqref{eq:inner_boundary} is equivalent to
\begin{equation}
    \left.\frac{\partial}{\partial r}\left(\Sigma^{i+1}r\right)\right|_0 = \left.\frac{\partial}{\partial r}\left(\Sigma^{i}r\right)\right|_0.
\end{equation}
The outer boundary lies in the diffuse envelope surrounding the young protoplanetary disk. We impose
\begin{equation}
    \Sigma_{N-1} = \Sigma_\mathrm{env},
\end{equation}
with $\Sigma_\mathrm{env}$ describing the envelope surface density. It is obtained directly from the MHD simulations by H20.

For the water vapor, we assume that there is no concentration gradient at the inner boundary,
\begin{equation}
    \left.\frac{\partial}{\partial r}\left(\frac{\Sigma_\mathrm{vap}}{\Sigma_\mathrm{H+He}}\right)\right|_0 = 0,
\end{equation}
whereas the cold outer boundary cell contains no water vapor,
\begin{equation}
    \Sigma_\mathrm{vap}(r_{N-1}) = 0.
\end{equation}

Analogously, we assume that the concentration gradient vanishes for icy and silicate solids at the inner boundary,
\begin{equation}
    \left.\frac{\partial}{\partial r}\left(\frac{\Sigma_\mathrm{ice,sil}}{\Sigma_\mathrm{\Sigma_\mathrm{H+He}}}\right)\right|_0 = 0,
\end{equation}
whereas the right boundary is given by the envelope density,
\begin{equation}
    \Sigma_\mathrm{ice,sil}(r_{N-1}) = \frac{1}{2}\epsilon_0\Sigma_\mathrm{env}.
\end{equation}

In the interest of numerical stability, we change the inner boundary condition for the background gas in the R7-based simulation to
\begin{equation}
    \left.\frac{\partial}{\partial r}\left(\Sigma^{i+1}\sqrt{r}\nu^i\right)\right|_0 = 0.
\end{equation}
This choice of boundary condition limits the mass outflow to the radial velocity contribution caused by advection, as detailed above. However, since the advective term provides the dominant contribution to the radial velocity, this does not have a significant impact on the resulting surface density evolution. 

\section{Time-dependent effectiveness of the cold-finger effect}\label{sec:appendix_timedep_coldfinger}
\begin{figure*}[htp]
    \centering\includegraphics[width=\linewidth]{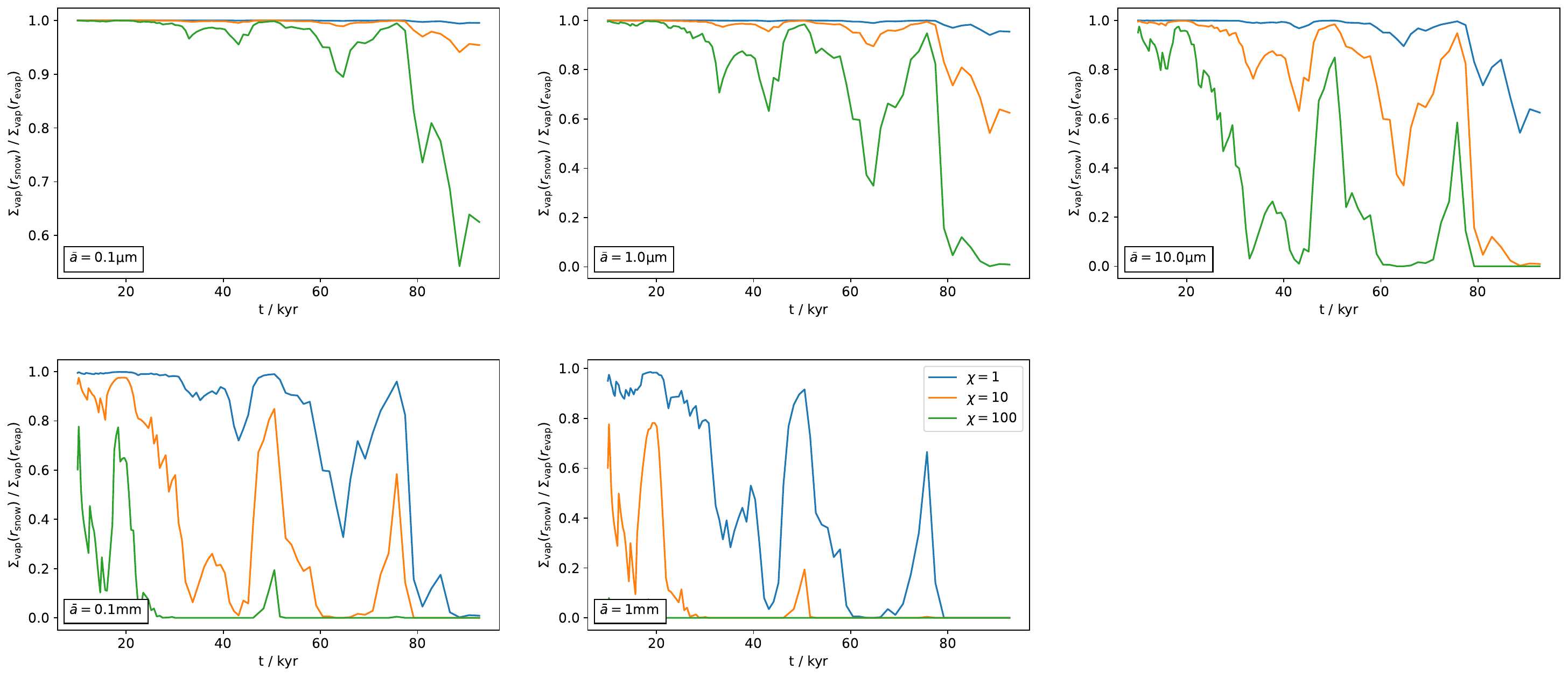}
    \caption{Mass fraction of water vapor reaching the snow line via backward radial diffusion from $r_\mathrm{evap}=r_\mathrm{snow}+v_\mathrm{gas}\tau_\mathrm{evap}$ as a function of time. The panels show the cases of different $\bar a$, with values spanning from $\bar a = \SI{0.1}{\micro\meter}$ to $\bar a =\SI{1}{\milli\meter}$. All shown cases use $\xi=5$, and the line color indicates the velocity fluctuation reduction factor $\chi$ that was used. If a time average is applied to each panel, the result is Fig. \ref{fig:coldfinger_eff}.}
    \label{fig:timedep_coldfinger}
\end{figure*}%
The fraction of water vapor that can reach the snow line after evaporating from the grain surface, resulting in condensation, is an indicator for the effectiveness of the cold finger effect. It is shown in Fig. \ref{fig:coldfinger_eff} averaged over the entire runtime of the simulation. In more detail, disk conditions vary greatly over time, leading to fluctuations of the effective Schmidt number $\Sctilde=\alpha_\mathrm{EA}/\delta_r$. Figure \ref{fig:timedep_coldfinger} shows this time dependence for the considered values of $\bar a$. Even if a large fraction of vapor can reach the snow line on average, indicating a high effectiveness of the cold-finger effect, it can still fail in providing a sufficient increase in the vertically averaged dust-to-gas ratio for planetesimal formation. This is because planetesimal formation episodically occurs during times of and in regions where $\delta_z$ and $\delta_t$ are small. If the formation episodes do not coincide with the times when substantial backward diffusion is taking place, planetesimal formation might be hindered as a result.

\section{Cold-finger effect in a dust-enriched disk}\label{sec:appendix_combined_case}
\begin{figure}[htp]
    \centering\includegraphics[width=\linewidth]{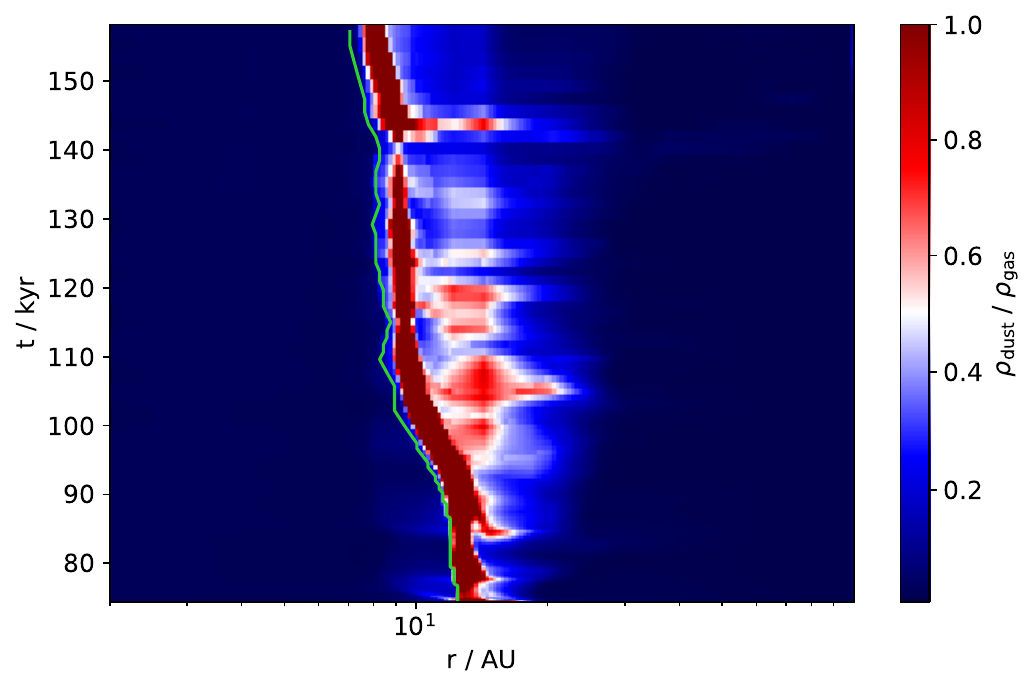}
    \caption{Like Fig. \ref{fig:R2_a1_eps05_heatmap}, but with $\bar a = \SI{1}{\micro\meter}$ and $\xi=5$.}
    \label{fig:R2_comb_heatmap}
\end{figure}%
\begin{figure}[htp]
    \centering\includegraphics[width=\linewidth]{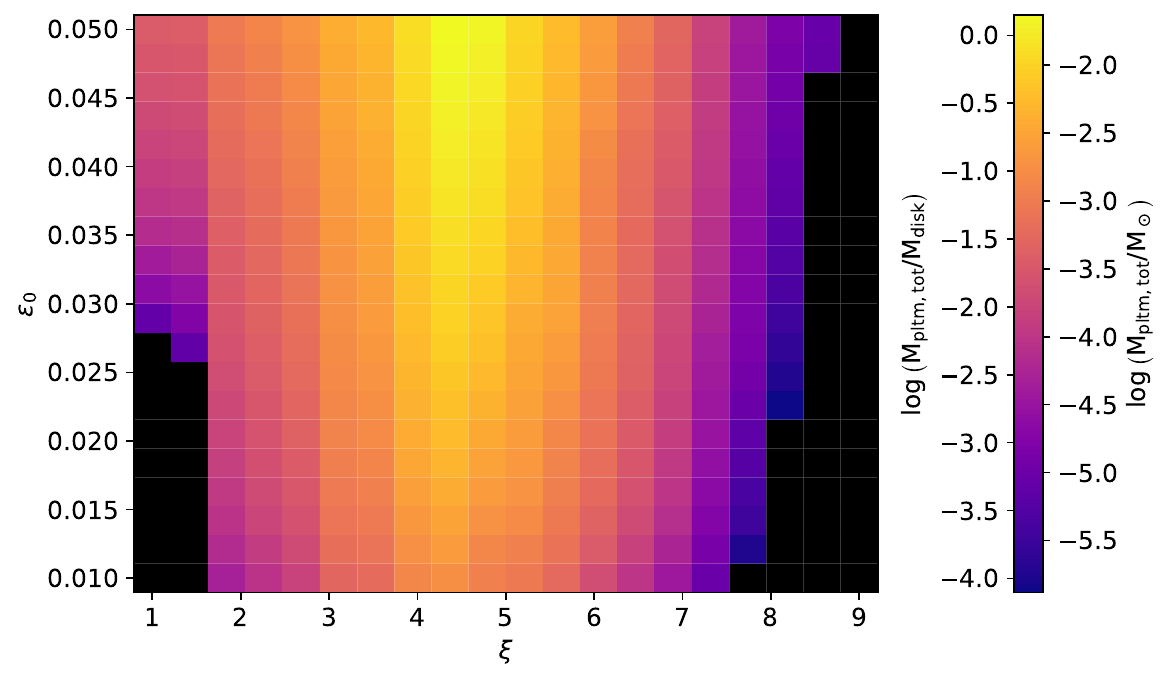}
    \caption{Like Fig. \ref{fig:R2_a1_mplas_epscrit}, but with $\bar a = \SI{1}{\micro\meter}$.}
    \label{fig:R2_comb_mplas}
\end{figure}%
A protoplanetary disk that is initially enriched in solids and exhibits comparatively low velocity fluctuations while also retaining the smallest grains for efficient operation of the cold-finger effect would enable a high midplane dust-to-gas ratio in a considerable area of the disk and over most of the simulation time. This combined case is presented in Fig. \ref{fig:R2_comb_heatmap}, showing the midplane dust-to-gas ratio for the previously discussed ideal case $\chi=100$, $\epsilon_0=0.05$ and $\epsilon_\mathrm{mid,crit}$ for various temperatures where small grains are now available for evaporation and condensation such that $\bar a = \SI{1}{\micro\meter}$. It can be seen that the cold-finger effect, as expected, strongly raises the midplane dust-to-gas ratio close to the snow line, while the higher value of $\epsilon_0$ additionally allows planetesimals to form in places unrelated to the snow line solely caused by strong settling. The resulting total mass of formed planetesimals is depicted in Fig. \ref{fig:R2_comb_mplas}. In this combined scenario, most planetesimals are formed for $\xi\sim 4.5$ within the simulation time, and the total mass increases with $\epsilon_0$ as expected. However, these results are limited by the simulation run time. Planetesimal formation is still commencing at the end of the simulation in many runs where planetesimal formation is efficient. Furthermore, due to the episodic nature of planetesimal formation we find in this work, this cannot be excluded for the other cases. We note that the total mass of formed planetesimals is comparable to the final gas disk mass in the most efficient cases, greatly exceeding the total mass of all available solids in a typical Class II disk model. As discussed above, the reason for this lies in the fact that material accreted from the envelope is converted into planetesimals, rather than material initially present in the disk.

\section{R1 and R7 disk quantity profiles}\label{sec:appendix_other_profiles}
\begin{figure*}[htp]
    \centering\includegraphics[width=\linewidth]{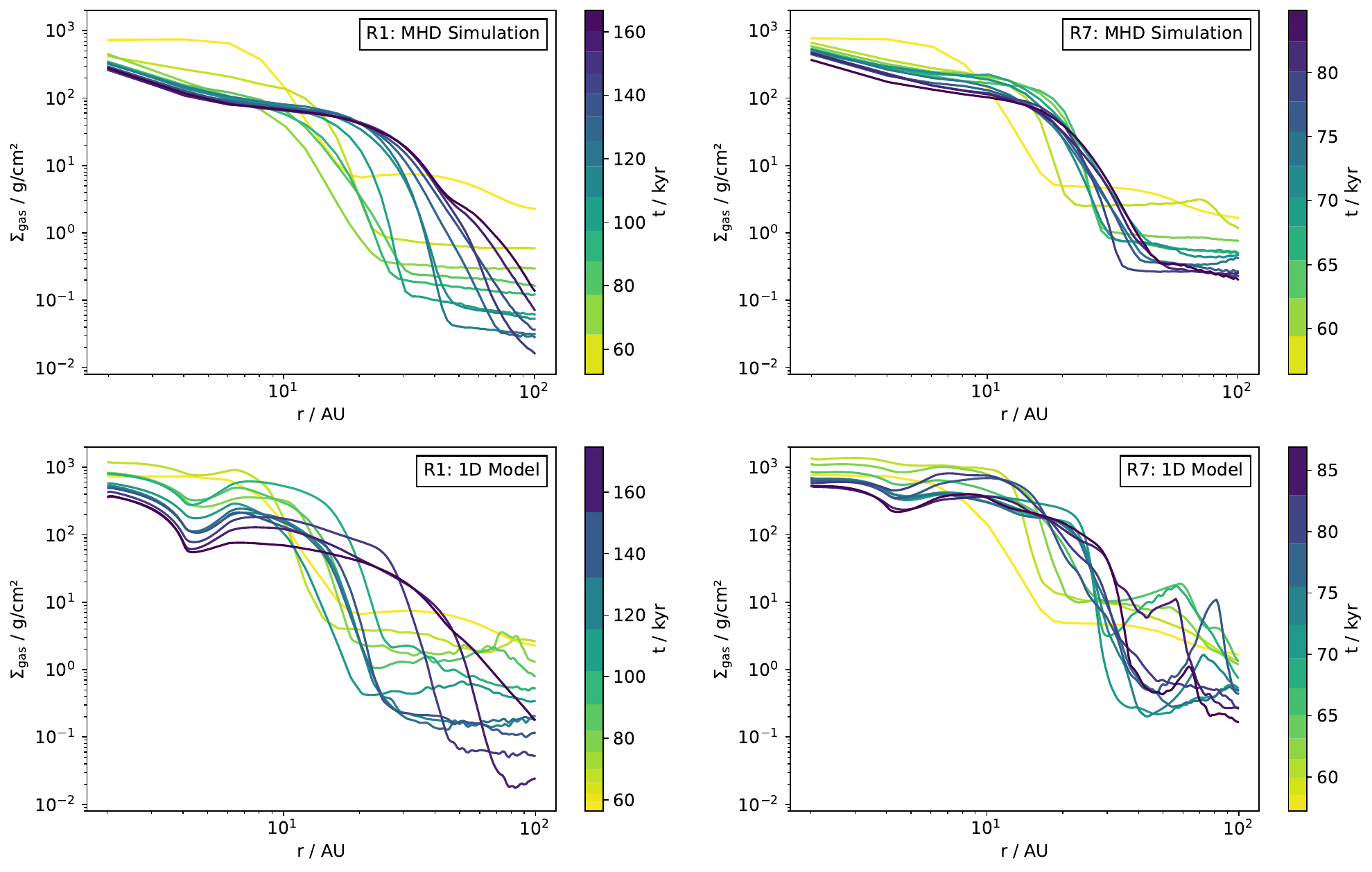}
    \caption{Gas surface density $\Sigma_\mathrm{gas}$ as a function of radial distance. The line color denotes the elapsed simulation time as indicated by the color bar, where a darker color indicates later times. \textit{Top}: Calculated directly from the R1 (left) or R7 (right) core-collapse simulation. \textit{Bottom}: Obtained in our 1D model when starting from the first core-collapse snapshot and employing the previously extracted model parameters.}
    \label{fig:othermhd_sigma}
\end{figure*}%
To verify the continued validity of our 1D model reproducing the disk evolution of the core-collapse simulations, the time evolution of the radial profile of the gas disk surface density is shown in Fig. \ref{fig:othermhd_sigma} and compared to the time evolution obtained in the 1D framework, analogously to R2 (see Fig. \ref{fig:mhd_sigma}). Similar to the case of R2, the time evolution of $\Sigma$ matches the original evolution order-of-magnitude. Most notably, the disk found in R1 is more compact than in R2 initially, but the disk expands to a cut-off radius comparable to that of R2 for later times. Like for R2, the $\Sigma$ radial profiles of R1 show a kink close to the sink particle accretion radius, likely related to unphysical parameters caused by the sink particle accretion. We note that it is less pronounced for R7, but this is related to the shorter run time of that simulation, as the kink becomes more pronounced over time. Additionally, the surface density in the envelope area of the radial grid is reproduced worse for R7 than for the other core-collapse simulations. However, since this region is largely devoid of mass, there is no significant impact on the creation of conditions necessary for planetesimal formation.

\begin{figure*}[htp]
    \centering\includegraphics[width=\linewidth]{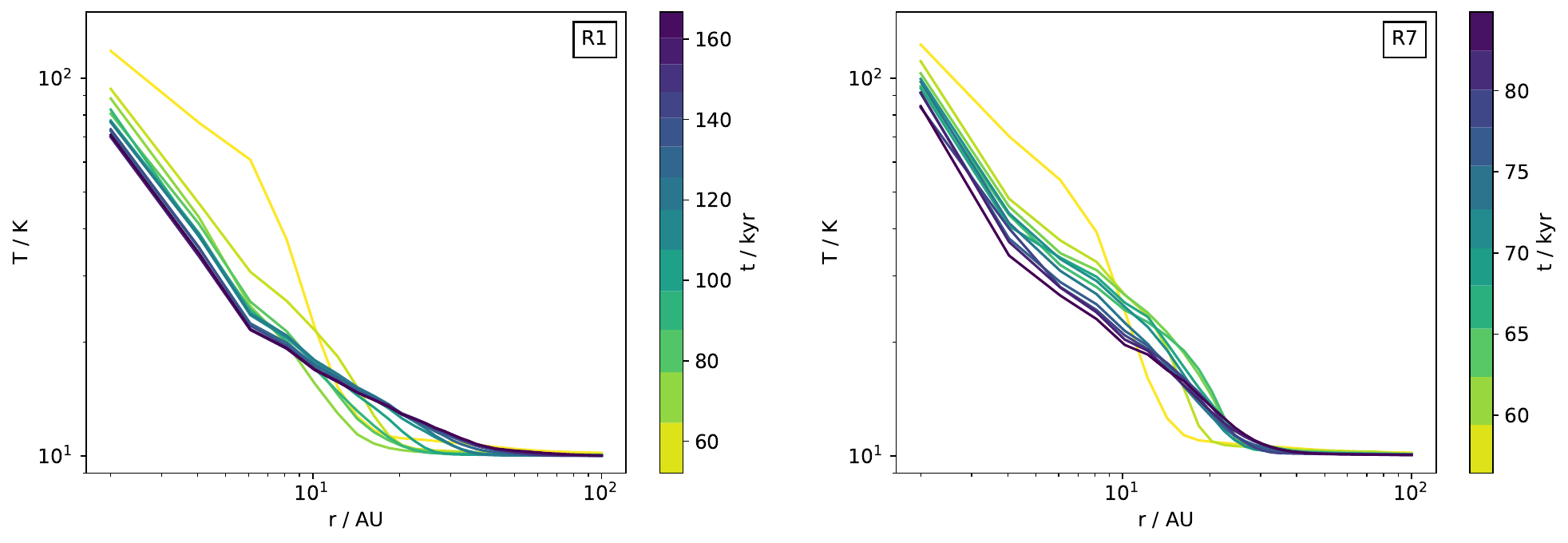}
    \caption{Average temperature as a function of radial distance found in R1 (left) and R7 (right). Different line colors indicate different simulation times according to the color bar, like Fig. \ref{fig:mhd_sigma}.}
    \label{fig:othermhd_T}
\end{figure*}%
\begin{figure*}[htp]
    \centering\includegraphics[width=\linewidth]{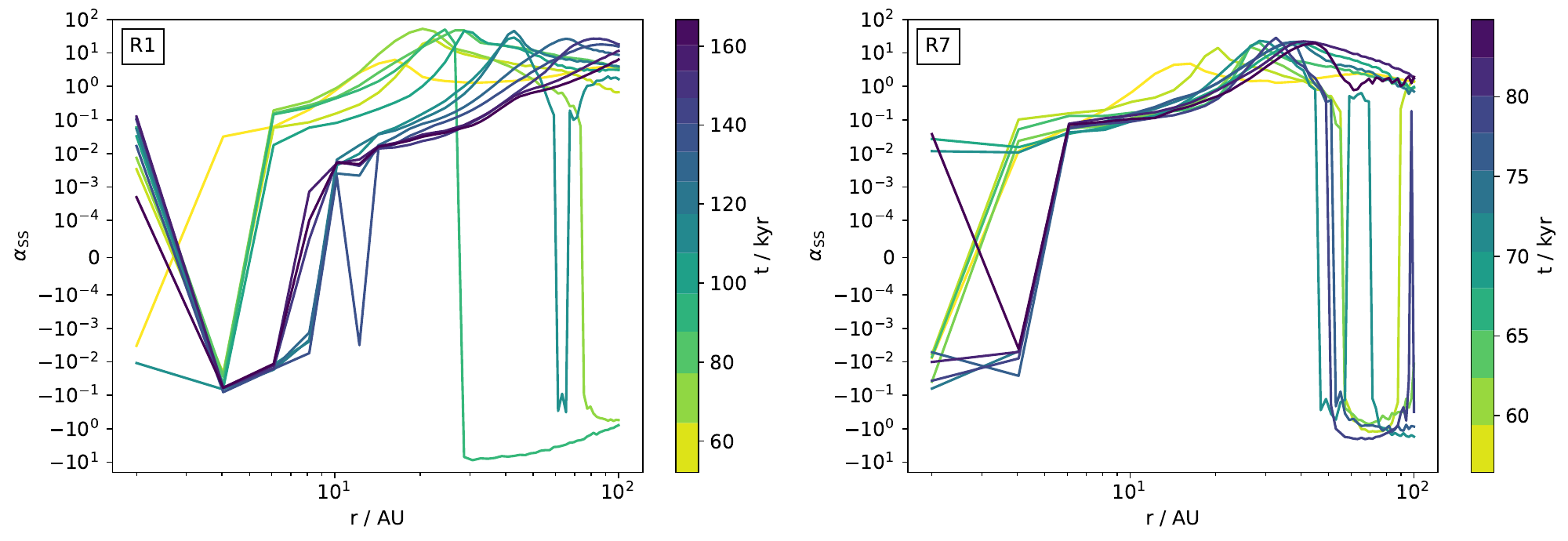}
    \caption{Like Fig. \ref{fig:othermhd_T}, but instead of the temperature, the radial profile of the internal torque parameter $\alpha_\mathrm{SS}$ (cfr. Eq. \ref{eq:alpha_SS}) is shown, which is scaled linearly for values with absolute value less than \num{e-4}, but logarithmically otherwise.}
    \label{fig:othermhd_alphaSS}
\end{figure*}%
\begin{figure*}
    \centering\includegraphics[width=\linewidth]{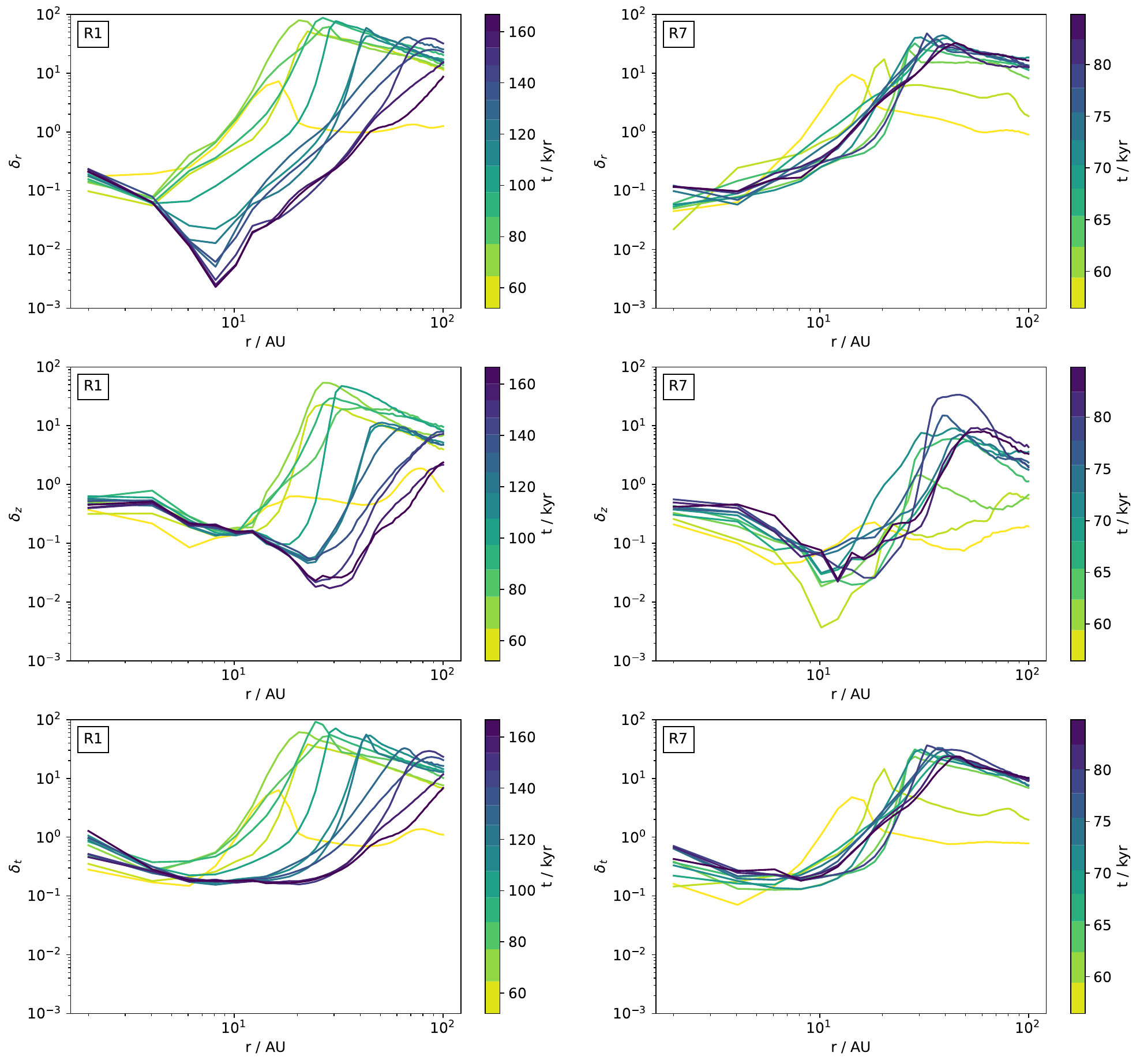}
    \caption{Radial profiles of model parameters describing dust properties and evolution, calculated from the disk in R1 (left-hand side) and R7 (right-hand side) and with line coloring like in Fig. \ref{fig:mhd_sigma}. \textit{Top}: Radial diffusion parameter $\delta_r$ (cfr. Eq. \ref{eq:delta_r}). \textit{Middle}: Vertical mixing parameter $\delta_z$ (cfr. Eq. \ref{eq:delta_z}). \textit{Bottom}: Turbulence parameter $\delta_t$ (cfr. Eq. \ref{eq:delta_t}).}
    \label{fig:othermhd_delta}
\end{figure*}%
In the interest of brevity, the radial profiles of some disk quantities from the R1 and R7 core-collapse simulations were not shown in Section \ref{sec:othermhd}. They are shown in this Appendix instead. The temperature is presented in Fig. \ref{fig:othermhd_T} and the Shakura-Sunyaev parameter $\alpha_\mathrm{SS}$ in Fig. \ref{fig:othermhd_alphaSS}, whereas the velocity fluctuation parameters $\delta_r$, $\delta_z$ and $\delta_t$ are shown in Fig. \ref{fig:othermhd_delta}.
\end{appendix}

\end{document}